\documentstyle[12pt,epsfig]{article}
\begin{document}
 
%%%%%%%%%%%%%%%%%%%%%
\newdimen\picraise
\newcommand\picbox[1]
{
  \setbox0=\hbox{\input{#1}}
  \picraise=-0.5\ht0
  \advance\picraise by 0.5\dp0
  \advance\picraise by 3pt      % correct for height of `=' above baseline
  \hbox{\raise\picraise \box0}
}
%%%%%%%%%%%%%%%%%%%%%

\title{Quark structure of hadrons and high energy collisions}
\author{J. Nyiri \\
 Research Institute for Particle and Nuclear Physics \\
 Budapest, Hungary }
\maketitle
\tableofcontents
\newpage
\section*{Foreword}

There exists a large field for phenomenological models in which the
knowledge of the structure of hadrons in terms of QCD constituents
obtained from deep inelastic scatterings is related to their behaviour
in soft processes. One of the simplest and oldest models is the additive
quark model, with the rules of quark statistics following from it.
Originally, the relations of quark combinatorics for hadron yields
were based on the qualitative description of a multiparticle
production process as a process of the production of non-correlated
quarks and antiquarks followed by their subsequent fusion into hadrons
[20],[21]. An analogy with the statistical description of multinucleon
processes was used, including the fact that there are final state
interactions both in multinucleon reactions and in multiquark reactions.
In the latter case it was natural to suppose [9] that $q\bar{q}$ and
$qqq$ final state interactions lead to the dominant contribution of
comparatively low-mass mesons and baryons into the spectra of the
produced hadrons.

The additive quark model turned out to be rather successful in describing
different experimental data. The model for hadronization suggested in
[20],[21] which is based on the hypothesis of soft colour neutralization
is used up to now when considering hadron production in jets [36]. As a
large amount of new precision measurements appear, and, on the other
hand, our understanding of QCD becomes deeper, a new level of
understanding of quark-gluon physics in the region of soft interactions
forces us to review the relations of quark combinatorics. To do so,
an especially good possibility is provided by the experimental data for
hadronic $Z^0$ decays which allow us to check the relations of quark
combinatorics for a new type of processes: quark jets in the decays
$Z^0 \rightarrow q\bar{q} \rightarrow hadrons$ [32].

\section*{Introduction}

All the recent conceptions of strong interactions are based on the
notion of quarks [1] which appeared in the early sixties as a
mathematical expression of the SU(3) symmetry properties of hadrons.
Since then, it had gone through a long way of evolution. We are sure
today that all the complexity of hadronic phenomena is due to the fact
that hadrons are composite systems built up from a small number of
"elementary" particles, quarks and gluons. Quarks as objects inside the
hadrons appeared first in the "classical" models of constituent quarks
([2]-[4]). The quark model gave a comprehensive tool for classifying the
hadronic states. Its success turned out to be not just descriptive, but
quantitatively accurate; in many cases the the consequences of the model
were known before data were available, and new hadrons were found due
to these predictions.

Strong evidence for the quark structure of hadrons was provided by
the investigation of hard processes, such as deep inelastic scatterings
of high-energy electrons, muons and neutrinos on hadrons (in practice,
nucleons), $\mu^+ \mu^-$ production with large effective masses in
hadron collisions, $e^+ e^-$ annihilation into hadrons.
Indeed, the quantitative description of these processes on the basis
of the parton model required the introduction of point-like objects,
the symmetry properties of which coincides with those of the
constituent quarks. Also, it turned out, that in the framework of
the quark model hadron-hadron collision processes at high energies
could also be handled ([5]-[7]).

The problem is that while in the theory we deal with microscopic
dynamics of quarks and gluons, we want to understand the spectrum
and interactions of hadrons. Quantum chromodynamics is the microscopic
theory of hadrons and their interactions, the success of the QCD-based
phenomenology leaves no doubt about that. It contains the free quarks
supplied with the required colour degrees of freedom. Being an
asymptotically free theory, i.e. a theory in which interactions at short
distances (at large momentum transfers) are small, QCD gives a description
of hard processes which is in accordance with the prediction of the
parton model. At large distances the interaction increases, and we face
the phenomenon of quark confinement.

There exist different approaches to explain it within QCD. According
to Gribov ([8] and references therein), the confinement is a property of
our world and it is largely determined by the existence of practically
massless quarks. QCD is here formulated as a quantum field theory
containing both perturbative and non-perturbative phenomena, and the
confinement is based on the supercritical binding of light quarks.
The theory of the supercritical confinement seems to be, at present,
the only possibility to give a natural explanation for the dynamical
mechanism of the interaction.

If the proposed theory proves to be successful, it means that we can go
down to the small momentum scale which implies understanding and
describing the physics of confinement essentially perturbatively. Still,
up to now, when considering soft processes, one has to deal with all
the problems connected with strong interactions. It is reasonable,
therefore, to describe soft processes in a different,
semi-phenomenological way, which is in agreement with the experimental
data and at the same time does not contradict the theory, moreover,
gives some indications to the character of the confinement. Hence,
one can hope that even if the problem of the confinement is solved,
the results of this semi-phenomenological description remain valid.

We review here an approach that enables us to handle soft
processes\footnote{A detailed description of this approach is given in
[9].}. This is based on a hadron picture due to which the baryons
(and mesons) are formed by three (or, respectively, two) constituent
quarks which are separated in space, i.e. the sizes of the quarks are
much less than those of the hadrons ([10],[11]). The presence of three
or two discrete objects in a hadron can be reconciled with the parton
picture assuming that a fast moving hadron is a system of three (or
two) spatially separated clouds of partons, each containing a valence
quark, a sea of quark-antiquark pairs and gluons. In the case of such
a hadron, which, like a nucleus, is characterized by two different
sizes, the impulse approximation can be applied for hadron collisions
at high energies.

There are several experimental facts which seem to support this hadron
picture. In elastic hadron-hadron scattering processes at high energies
the shrinkage of the diffraction cone was observed. The parameter
$\alpha'_p$ (the slope of the Pomeron) characterizing the shrinkage
is small compared to the slope of the diffractional cone itself. This
is an indication for the existence of a second, characteristic size
inside the hadron (in addition to the size of the hadron itself) which
could be the small radius of the constituent quark [10]. Theoretical
arguments have also been expressed in favour of such a double structure
of hadrons. Due to [12], the confinement region of gluonium states
might be much less than that of the quarks, i.e. the "coat" of the
constituent quark is consisting mainly of gluons. Another possibility
to explain the existence of two sizes comes from instanton-type
fluctuations in the QCD vacuum [13].

The considered hadron picture is, of course, a simplified one. Still,
apart from describing well the soft processes at high energies, it gives
a possibility to connect the results of investigations in hard processes
[14],[15] with the "old" quark physics.

\section{The quark structure of hadrons}\label{I}

The first serious success of the quasi-nuclear quark model was the
description of static properties of hadrons. Hadron spectroscopy gave
and continues to give results which confirm the quark structure of
hadrons. Also, the calculated magnetic moments of baryons as well as the
decay widths of vector mesons were in reasonably good agreement with the
experimental data.

There is another field where the additive quark model is rather
successful: the description of hadron collision processes
at moderately high energies.

Let us remind now the well-known arguments which support the impulse
approximation\footnote{The notion of impulse approximation comes
from the deep inelastic scattering of an electron on a nucleus, when
the momentum transfer is much larger than the average internal momentum
of the nucleons in the ground state. Under these conditions the
interaction between the electron and the nucleons is so sudden, and the
change of the momentum is so large that one can neglect the binding
forces between the constituents during the collision. In the first
approximation, the constituents behave like free particles. This is
called the impulse approximation; the electron can be considered as
being scattered by one of the "free" nucleons.

We can apply this analogy to the scattering process by an isolated
nucleon. The r\^ole of the electrons is played by one of the leptons,
that of the nucleus by the nucleon, and that of nucleons within the
nucleus by quarks. There is, of course, an important difference: because
of confinement, quarks cannot be ejected from the nucleon. This, however,
is not significant from the point of view of the scattering cross section.
On the other hand, the properties of the final states depend on the
mechanisms that appear when the highly energetic quark attempts to
leave the hadron. At this stage quark-antiquark pairs are produced which
eventually materialize into new hadrons.} in hadron collision processes
at high energies. Comparing theoretical predictions with experimental
data, it turned out that the processes \\
\[ 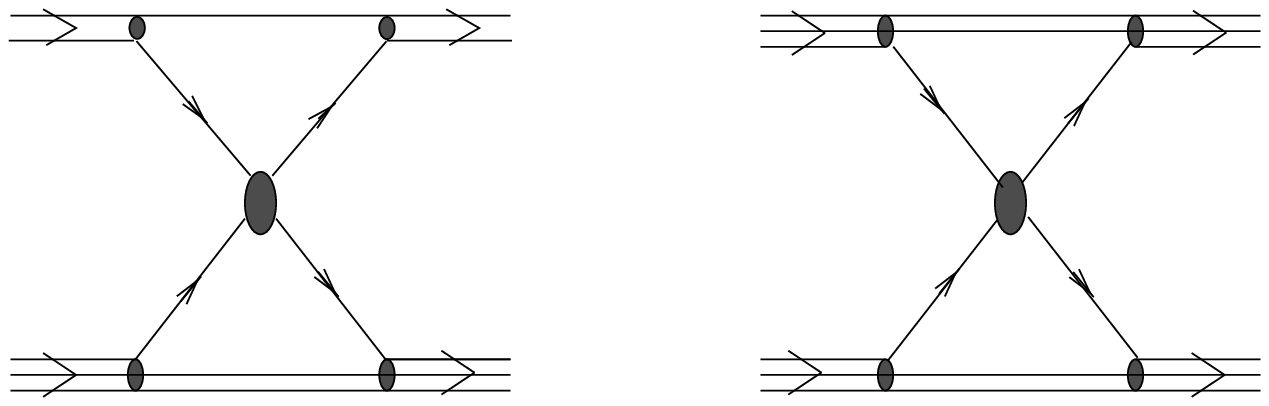  \]
\centerline{Fig.1}
describe sufficiently well the ratio of the total cross sections in
$NN$ and $\pi N$ scattering ([5]-[7])
\begin{equation}
\label{1}
\frac{\sigma_{tot}(NN)}{\sigma_{tot}(\pi N)} = \frac{3}{2}
\end{equation}
as well as the decrease of the elastic $pp$ - cross section with the
increase of the momentum transfer [6]
\begin{eqnarray}
\label{2}
\frac{d\sigma(\pi p \rightarrow \pi p)}{dt} & = & |a_{qq}(t)|^2
 F^2_p(t)F^2_{\pi}(t), \\
\frac{d\sigma(pp \rightarrow pp)}{dt} & = & F^4_p(t)
\end{eqnarray}
where $F_p(t)$ and $F_{\pi}(t)$ are the proton and pion form factors,
respectively, and $a_{qq}(t)$ is the elastic quark scattering amplitude.

Impressive arguments in favour of the additiveness of the interaction
between dressed quarks are given by hadron-nucleus collisions. The
quasi-nuclear hadron structure allows us to calculate the ratio of
multiplicities of secondary hadrons in the central region for high
energy $NA$ and $\pi A$ scatterings. At $A \rightarrow \infty$ this
gives
\begin{equation}
\label{2a}
\frac{<n_{ch}>_{NA}}{<n_{ch}>_{\pi A}} = \frac{3}{2} \; .
\end{equation}
Accepting the hadron picture with two radii, we assume that hadrons are
similar to light nuclei: the meson, consisting of a quark and an
antiquark sufficiently far from each other reminds the deuteron while
the baryon contains three constituent quarks in the same way as $H_3$
or $He_3$ is built up. The constituent quarks are surrounded by their
"coat" of virtual particles. The radius of this "coat" is in fact the
radius of the constituent quark. The mean distances between the
constituent quarks determine the size of the hadron ([10]-[11], [16]).

The radius of the constituent quark can be estimated from the total
hadron-hadron cross-section, which, as follows from Fig.1, can be
expressed in terms of the total quark-quark cross-section. At moderately
high energies $ \sigma_{tot}(qq) \simeq 4,5 mb $. Assuming that the
total quark-quark cross section is determined by the geometrical sizes
of the colliding quarks $\sigma_{tot}(qq) \simeq 2\pi (2r_q)^2 $, we
get
    \[   r^2_q \simeq 0,5 GeV^{-2} . \]

There is another way of obtaining the radius of the constituent quark
in the framework of the parton hypothesis. Without going into details,
we give here only the results: $\pi p$-scattering experiments [17] at
moderately high energies $\sqrt{s} \simeq 40 GeV $ lead to
      \[  r^2_q \simeq 3\alpha'_p \simeq 0,45 GeV^{-2} . \]
Hence, having $R^2_h \simeq 17 GeV^{-2}$,
\[   \frac{r^2_q}{R^2_h} \simeq \frac{1}{30} .\]
(According to calculations compared to recent experimental data [18],
the radius of the constituent quark turns out to be even smaller:
$r^2_q \approx 0,1 GeV^{-2}$).
We consider here, naturally, coloured quarks. Since the quark confinement
is due to the colour forces, we are bound to accept the following hadron
picture. (We will consider here a nucleon). At large momenta (but
$P < 10^8 GeV/c $) the nucleon contains three clouds of quarks-partons
(Fig.2a). \\
\vspace{1.cm}
\[   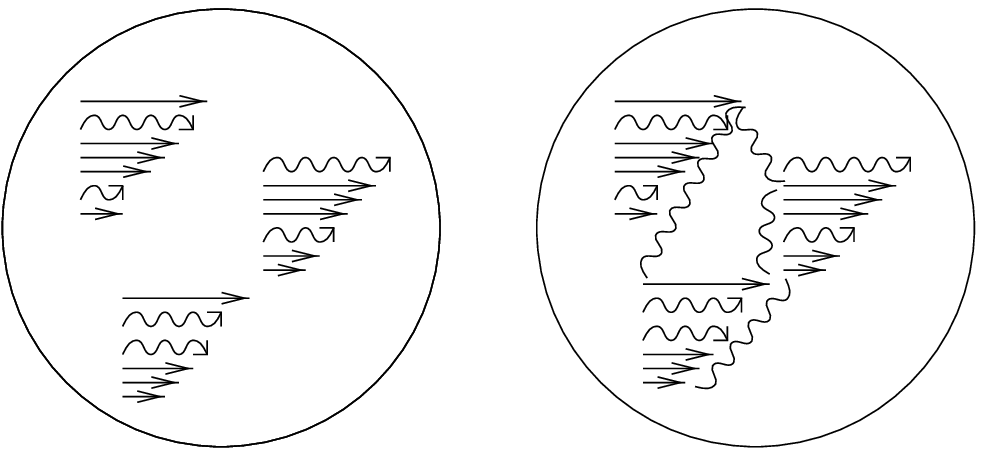  \]
\centerline{Fig.2}
Each of the clouds contains a coloured quark-parton which carries the
quantum numbers of the constituent quark, and a sea of quark-antiquark
pairs and gluons, which is colourless and has zero quantum numbers. The
gluon interaction which keeps the constituent quarks inside the hadrons
is taking place between the fast components $I$ [19] (Fig.2b).
The gluon exchange is improbable between the partons $II$ carrying a
relatively small fraction of the momentum.

The transverse dimension of a cloud increases with the energy as $\sqrt{
\alpha'_p\ln P/P_0} $, $P_0 \sim 10 GeV/c $. Up to $P \ll 10^8 GeV/c $,
$r_q$ remains essentially less than $R_h$ and, practically, the three
(or, in the case of a meson, two) clouds do not overlap\footnote{The
case of overlapping quark-parton clouds may become important at RHIC
energies. The presented approach is not suitable for handling this energy
region}. When a fast hadron collides with the target, only one of the
constituent quarks participates in the interaction; the other constituent
quarks, or quark-parton clouds, remain spectators. The situation is
different in the case of a hadron-nucleus interaction, i.e. when the
target is large, and not only one, but two or three constituent quarks
of the incident hadron can interact. We will come to this question
later. As soon as $r^2_q \ll R^2_h$, repeated collisions of the quarks
are improbable. The interaction with the target is due to the slow
components of the partons (a parton carrying energy $E$ needs a time
of the order of $\tau \sim {E}/{\mu^2} $ to interact). The
quark-parton cloud the slow component of which participated in the
interaction breaks into partons. These partons then, interacting with
each other, obtain their own "coats" and become constituent (dressed)
quarks, giving rise to the production of new particles (Fig.3). \\
\vspace{1.cm}
\[   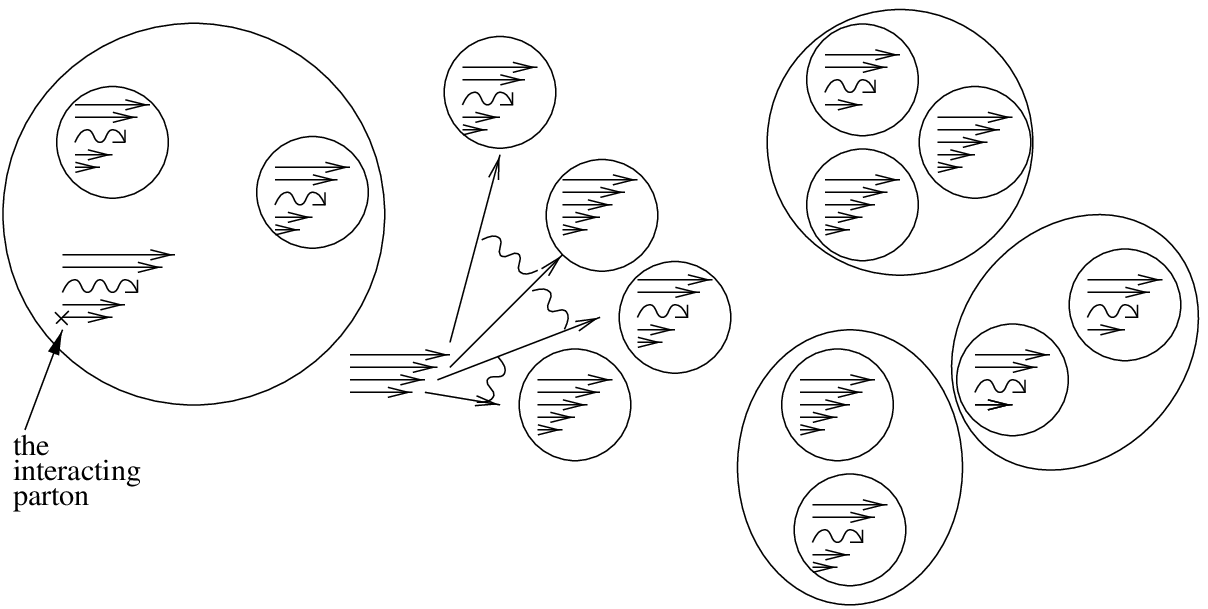  \]
\centerline{Fig.3}
In the framework of the model, the three stages of the multiparticle
production process are sufficiently well separated in time.
The approach we are presenting deals in fact with the second and the
third steps: the interactions of the partons and the gluons with each
other which lead to the formation of constituent quarks, and the
transition of these constituents into hadrons (mesons, baryons, meson
and baryon resonances) in such a way that the set of hadron states
corresponds to the states of the constituent quarks in the
multiperipheral ladder.

The formation time of the constituent quark in its rest frame has to be
determined by the quark radius $r_q$. A crucial point of our consideration
is the assumption that the dressed quark exists long enough to interact
repeatedly producing additional pairs of dressed quarks. One can
assume that the resulting cloud of constituent quarks is "prepared"
for hadronization. This means that every quark has suitable colour and
flavour partners sufficiently close on the rapidity scale. Indeed, if
there is a given quark for which partners suitable to form a hadron are
too far on the rapidity scale, some additional quark-antiquark pairs
will be produced so that the hadronization of all quarks becomes possible
without any suppression. The same is true for colour excitations: if a
quark has no suitable partners to form a white hadron state, then new
quarks are produced, and the wrong colour is transferred to another
rapidity region. This means that again, the transition of quarks into
white hadron states takes place without any suppressing factors, and
does not depend, e.g., on the probability of finding suitable partners
for the quark. We call this property the soft hadronization and soft
colour neutralization of the dressed quarks.

This approach is by no means the only possibility to handle the problem
of the quark-hadron transfer. Our task is to present it and to
demonstrate, why such a rough model may be justified and successful
even now.

\section{Multiparticle production processes}\label{II}
\subsection{General description of the approach. The spectator mechanism}
\label{2.1}

Considering a picture with quark confinement, one assumes the existence
of two equivalent descriptions of the physical processes, namely: the
description in terms of quark states and that in terms of real particles,
since each quark state corresponds to a set of hadron states.

Our aim is, in a sense, to translate the quark language into the hadron
language. Dealing with soft processes (i.e. processes with small momentum
transfer) and especially with inelastic scatterings at high energies,
which lead to the production of many particles, we expect to have a
large field for comparison with experiment.

The quark combinatorial calculus which has been proposed in [20],[21]
provides a good possibility to handle the multiparticle production
processes. Apart from the usual hypothesis about the quark structure
of hadrons, two main assumption have been made. The first one concerned
the spectator mechanism, which was based on the picture of spatially
separated quarks. As we have told before, practically only one constituent
quark of the incident hadron (and of the target) is taking part in the
collision process, the other ones remain spectators. As a result of the
collision, many new quarks are produced, which afterwards join the
quark-spectators and form fast secondary hadrons, observable in
experiment. Fig.4 shows a picture of meson-baryon and baryon-baryon
collisions of this kind. \\
\vspace{1.cm}
\[  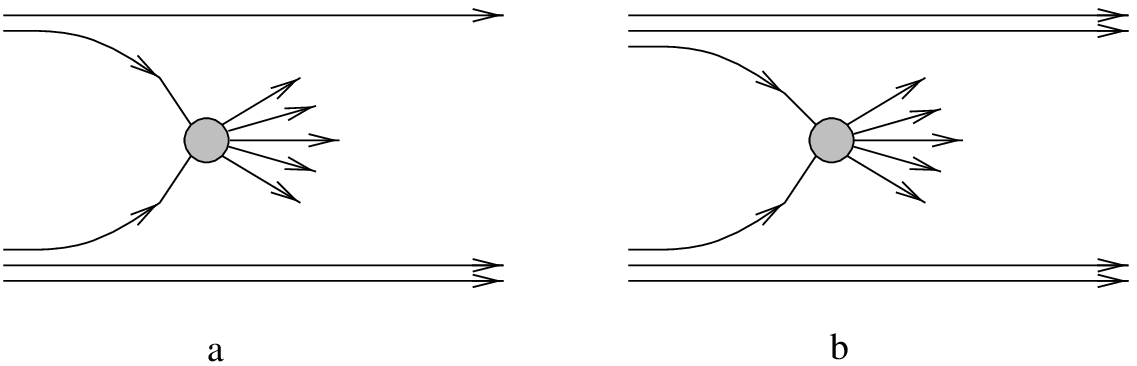  \]
\centerline{Fig.4}
If the hadron consists of discrete "dressed" quarks, then inside a fast
baryon each of them has to carry about $1/3$ of the total baryon momentum,
while inside a meson -- about half of the meson momentum. Consequently,
multiparticle production processes in hadron-hadron collisions can be
divided into two, energetically different, regions: the central and
the fragmentation ones ($I$ and $II$ in Fig.5). \\
\vspace{1.cm}
\[  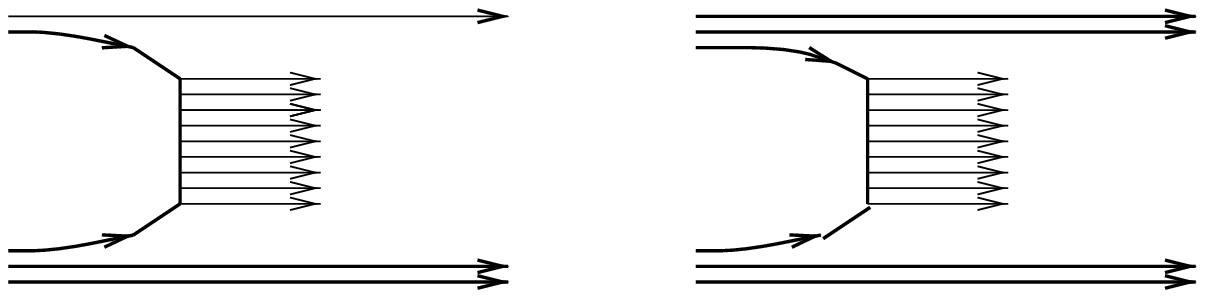  \]
\centerline{Fig.5}
The quarks in the central region are sea-quarks, carrying a small
fraction of the incident momentum. Joining each other, they form
the spectrum of slow hadrons.

The quarks-spectators of the colliding particles ($q_i$, $q_j$ and
$q_{i'}$, $q_{j'}$ in Fig.5) join quarks (or antiquarks) of the sea
forming the hadrons in the fragmentation region. The pair of quarks
$q_k$ and $q_{k'}$ produced in the interaction "remember" their
origin and have to be considered as belonging to the fragmentation
region.

Let us see now, what processes are possible in the fragmentation
region. (For the sake of simplicity we consider a baryon fragmentation
process). The interacting  quark $q_k$ can join the spectators, forming
a baryon state containing the same quarks as the incident one (Fig.6a).
If the collision of $q_i$, $q_j$ and $q_k$ is coherent, then the
produced hadron $B_{ijk}$ is analogous to the initial state (in the
case of an incident proton that means $p \rightarrow p$ transition). If
the collision is not coherent, the produced $B^*_{ijk}$ state is a
superposition of possible real hadrons (e.g. $p \rightarrow p$,
$p \rightarrow \Delta^+$ etc.).

The spectators $q_i$, $q_j$ can join a sea quark; in this case a baryon
state $B_{ij}$ is formed (Fig.6b). At the same time $q_k$ and a sea
antiquark form a meson state $M_k$.

The baryon states $B_{ijk}$ and $B_{ij}$ carry about $2/3$ of the
momentum of the initial hadron. The interacting quark $q_k$ carries
away $x \sim 1/3$ (where $x = {p_L}/{p_{max}}$ ; $p_L$ is the
longitudinal momentum of the constituent quark, $p_{max}$ that of the
incident hadron). The longitudinal momentum of the newly produced
quark $q_k$, which comes from the central region after the interaction,
can be estimated assuming that quarks produced in the central region
are distributed homogeneously in $\log x$, i.e. their longitudinal
momenta follow the geometrical progression law. This is the so-called
comb regime which leads to a Regge-pole exchange in elastic scattering.
If so, the fastest produced quark has a momentum equal to one half of
the incoming quark momentum, the next one $1/4$ of it etc. This means
that the meson state $M_k$ is produced in the $x \leq 0,15 $ region.

If one spectator joins two sea quarks, a baryon state $B_i \left(x \sim
1/3\right)$ is formed; the other spectator joining a sea antiquark forms
a meson state $M_j \left(x \sim 1/3\right)$ (Fig.6c). There are also
cases when only meson states are produced (Fig.6d,e).

The meson fragmentation process can be considered in the same way
(Fig. 6f, g, h). \\
\vspace{1.cm}
\[   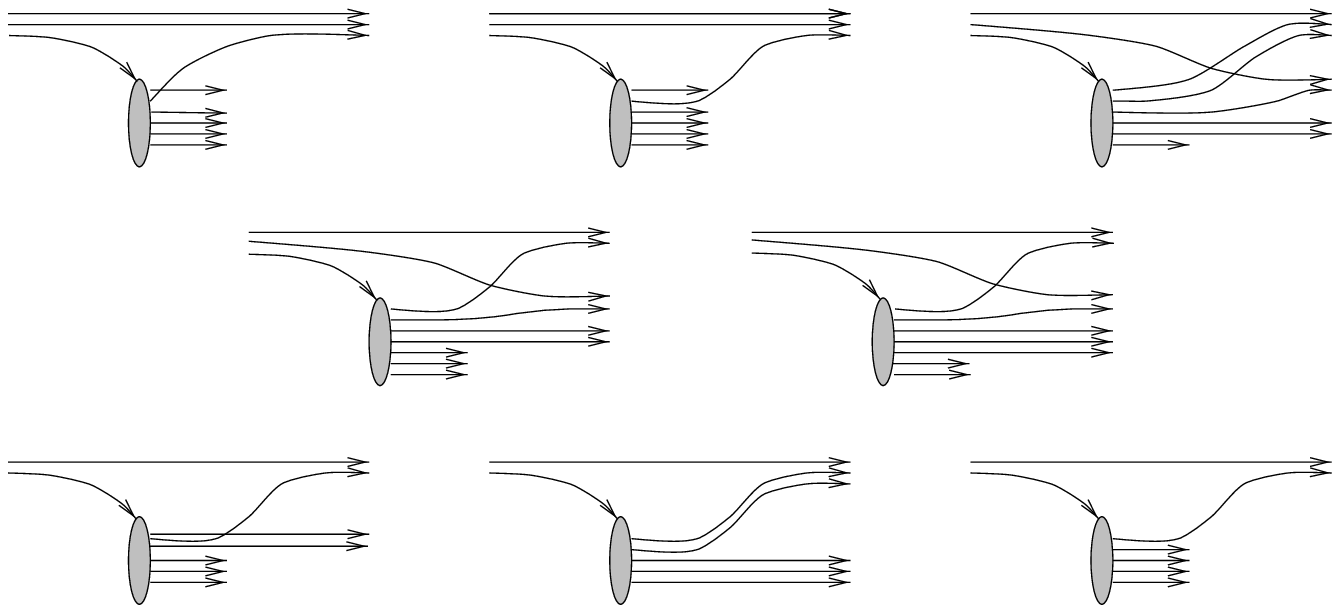 \]
\centerline{Fig.6}
As it is seen, the spectator mechanism leads to the production of
hadrons with a very definite momentum distribution. The comparison
of the theoretical predictions with the experimental data shows a good
agreement in different fields, such as resonance production in the
region of secondaries with large momenta [30] or inclusive spectra of
secondaries in $pp$ and $pA$ collisions [22],[23],[31].

\subsection{Quark combinatorics}\label{2.2}

The second assumption made in the quark combinatorial calculus is
connected with the newly produced particles. The classical quark model
is $SU(6)$ symmetric. Hence, it is natural to assume that $SU(6)$ holds
also for the production processes of secondary particles. This means
that in the multiparticle production processes not only stable particles
appear but also resonances, and the production probabilities of all
hadron states belonging to one $SU(3)$ multiplet are equal\footnote{
There are some particles which do not to fit into this scheme; their
properties seem to be due to the dynamics of the quark confinement [24].
Due to [25], there are two sorts of light quarks: constituent quarks
with masses about 300 MeV and much lighter relativistic current quarks.
This means that we have to take into account two different types
of bound states. If a constituent quark is combined into a meson, then
due to its small mass difference compared to the strange quark we find
an approximate $SU(3)$ flavour symmetry in the corresponding mesonic
spectrum. If the current quark is used to build a meson, then scalar
or pseudoscalar mesons are created that have nothing to do with $SU(3)$
symmetry}. Hence, the probability of the hadron production within one
$SU(6)$ multiplet is proportional to the number of of spin states of
these hadrons, i.e. $2J+1$.

In the framework of quark combinatorics it is assumed that hadrons are
formed by quarks with small relative momenta, i.e. by neighbours on the
rapidity axis. The quarks join each other with equal probability
independently of their quantum numbers and of the fact if they are quarks
or antiquarks.

In the central region, where the hadrons are formed by sea quarks only,
an arbitrary particle might be with equal probability a quark or an
antiquark: $1/2q\, +\, 1/2 \bar{q}$. The nearest neighbour is again
either a quark, or an antiquark. The probability of the states $qq$,
$\bar{q}\bar{q}$ and $q\bar{q}$ is then
\[ \left(\frac{1}{2}q + \frac{1}{2}\bar{q}\right)\left(\frac{1}{2}q +
\frac{1}{2}\bar{q}\right)
\rightarrow \frac{1}{4}qq +\frac{1}{4}\bar{q}\bar{q} +\frac{1}{2}q\bar{q}
\rightarrow \frac{1}{4}qq +\frac{1}{4}\bar{q}\bar{q} + \frac{1}{2}M , \]
where $M=q\bar{q}$ is a meson state. Taking into account a third
possible quark or antiquark, we get
\[ \left(\frac{1}{4}qq +\frac{1}{4}\bar{q}\bar{q} + \frac{1}{2}M\right)
\left(\frac{1}{2}q + \frac{1}{2}\bar{q}\right)
 \rightarrow \frac{1}{8}B + \frac{1}{8}\bar{B} +
\frac{3}{4}M \left(\frac{1}{2}q + \frac{1}{2}\bar{q}\right) ,\]
where $B=qqq$, $\bar{B}=\bar{q}\bar{q}\bar{q}$. Further iterations lead
to the following multiplicity of particles produced in the central
region:
\begin{equation}
\label{3}
(q,\bar{q} - sea ) \rightarrow 6N\cdot M + N\cdot B + N\cdot \bar{B} .
\end{equation}
The number $N$ depends on the total energy of the colliding particles,
and increases with the growth of $s$. Supposing that the multiplicity
$N(s)$ is increasing logarithmically, it is convenient to write
\[ N(s) = b\ln \frac{s}{s_0} \]
at asymptotic energies. The parameters $b$ and $s_0$ cannot be determined
by quark combinatorics, but have to be the same for all processes. Hence,
the relation between the produced mesons $M$, baryons $B$ and antibaryons
$\bar{B}$ is [20]
\begin{equation}
\label{4}
M : B : \bar{B} = 6 : 1 : 1  .
\end{equation}
In the same way one can get relations between baryons and mesons in the
fragmentation region [26]. In this case we consider an incident quark
$q_i$ which, joining a quark or an antiquark of the sea, forms mesons
or baryons containing this quark with the probability $2:1$:
\begin{equation}
\label{5}
(q_i + q,\bar{q} - sea ) \rightarrow \frac{1}{3}B_i + \frac{2}{3}M_i +
\frac{1}{3}M +N(s)(6M+B+\bar{B}).
\end{equation}
Here $B_i=q_i qq$, $M_i = q_i\bar{q}$ ; $N(s)$ is a large number which
is characterized by the number of quarks in the sea.

A similar relation is valid for the case when a pair of quarks $q_i q_j$
transforms into hadrons:
\begin{equation}
\label{6}
(q_i + q,\bar{q} - sea ) \rightarrow \frac{1}{2}B_{ij} +\frac{1}{12}(B_i
+ B_j) + \frac{5}{12}(M_i + M_j)+ \frac{1}{6}M +N(s)(6M+B+\bar{B}).
\end{equation}
The baryon state $B_{ij}$ contains both incident quarks:
$B_{ij}=q_i q_j q$.

Supposing that quarks $q_i,q_j$ and the quark $q_k$ (Fig.5) form hadrons
in an independent way, the relations (\ref{5}) and (\ref{6}) provide
a possibility to find the relative weight of the fragmentation
processes in Fig.6b, 6c and 6e: $1/2 : 1/12 : 1/3$ . The probability
of the process 6a can not be obtained in the framework of quark
combinatorics.

Hence, if a quark $q_k$ belonging to the baryon $B_{ijk}$ hits the
target, fast particles are produced with the following probabilities
[27]:
\begin{eqnarray}
\label{7}
\lefteqn{B_{ijk} \rightarrow }\nonumber\\
 & & \Delta B_{ijk} + \Delta^* B_{ijk}^* + (1-\Delta-\Delta^*)\left[
 \frac{1}{2}B_{ij} + \frac{1}{12}(B_i+B_j) + \right. \nonumber\\
 & & \left. +\frac{5}{12}(M_i+M_j) + \frac{1}{3}B_k + \frac{2}{3}M_k +
 \frac{1}{2}M \right] + \cdots
\end{eqnarray}
Here $\Delta$ and $\Delta^*$ are the probabilities of the coherent and
incoherent transitions $B_{ijk} \rightarrow B_{ijk}$ and $B_{ijk}
\rightarrow B_{ijk}^*$, respectively; they have to be determined from
the experiment. (The contribution of hadrons produced in the central
region is absent in this expression).

Similarly, the probability of the production of fast hadrons after the
collision of a meson $M_{i\bar{j}}$ with the target is
\begin{eqnarray}
\label{8}
\lefteqn{M_{i\bar{j}} \rightarrow }\nonumber\\
 & & \delta M_{i\bar{j}} + \delta^* M_{i\bar{j}}^* + \nonumber\\
 & & + (1-\delta-\delta^*)\left[ \frac{1}{3}(B_i+\bar{B}_{\bar{j}}) +
 \frac{2}{3}(M_i+M_{\bar{j}}) + \frac{2}{3}M \right] + \cdots
\end{eqnarray}
Here the probabilities $\delta$ and $\delta^*$ of the processes
$M_{i\bar{j}} \rightarrow M_{i\bar{j}}$ and $M_{i\bar{j}} \rightarrow
M_{i\bar{j}}^*$ cannot be defined in the framework of quark combinatorics.
The probabilities $\Delta$, $\Delta^*$ and $\delta$, $\delta^*$ can
depend on the initial hadron and on the type of the collision, thus in
fact one has to write $\Delta_p(pp)$, $\Delta_p(Kp)$, $\delta_K(Kp)$
and so on. For the sake of simplicity, we will neglect this.

Let us mention that the presented relations, and (\ref{4}) and (\ref{5})
in particular, were obtained without taking into account the colour
degrees of freedom of the quarks. Having in mind that quarks are
coloured objects, the method [20]-[21] which leads to these results
can be understood in the following way. It is, practically, supposed
that there are strong colour correlations between the quarks in the
process when they join each other to form hadrons. These correlations
provide automatically the colour combinations necessary for the
production of white states.

Indeed, as we have said already, in quark combinatorics it is assumed
that quarks which join each other to form hadrons appear
accidentally as neighbours on the rapidity axis. If, however, the colour
states are not taken into account, hadron states can be formed only
if these quarks have the proper colours; inside small domains there
occur white groups of quarks, all the other quantum numbers of which
are arbitrary. This additional assumption of colour correlation is a
very strong one, which cannot be a rigid condition for the production
of hadron states. It is therefore interesting to understand what could
be the consequences of the absence of colour correlations.

In [28] we consider the situation when the quarks and antiquarks which
are close to each other on the rapidity axis, have uncorrelated,
arbitrary colours. It turns out that quark combinatorics for coloured
quarks does not determine unambiguously the relation between the number
of baryon and meson states either in the central region, or in the
fragmentation region. This relation is a function of a parameter $\alpha$
characterizing the diffusion of quarks along the rapidity axis and their
formation into hadrons. If, however, the relation (\ref{5}) is satisfied,
we obtain $M:B:\bar{B}\simeq 5,2:1:1$. In other words, the fact that
there is strong colour correlation or no colour correlation has only
little impact on the ratio of mesons and baryons in the central region.
(We interpret $q \rightarrow 1/3 B + 2/3 M $ as a relation expressing
that the baryon number of the quark manifests itself as the probability
of the production of a baryon state by this quark).

Let us see the case of quarks with uncorrelated colours in more detail.
The main feature of quark-hadron transitions in this case is that only a
few white states appear on a small interval of the rapidity axis:
in a set of nine states this is
only one colour singlet $1/{\sqrt{3}}q_i\bar{q}_i$ ($i=1,2,3$ are
the colour indices). In a system of three quarks $qqq$
only one of 27 states is white: the totally antisymmetric colour state
$1/{\sqrt{6}}\varepsilon_{ijk}q_i q_j q_k$. The small relative
probability of a transition to white states is the reason why the quarks
formed in the interaction process cannot be transformed into hadrons
immediately, but only in the course of many stages. Only white states
of quarks combine into hadrons, while the quarks that have not found
suitable partners inside the small domain will diffuse, interacting
with each other, along the rapidity axis. Only after many "collisions"
will these quarks find partners for the formation of white states.
In fact the picture of quarks with uncorrelated colours which form
hadrons as the result of repeated interactions is much closer to the
spirit of "soft" quark confinement.

The peculiarity of this approach is the formation of antisymmetric colour
states of two quarks $\frac{1}{\sqrt{2}}\varepsilon_{ijk}q_j q_k$ --
diquarks (we shall denote these states as $(\hat{qq})_i$). However,
inside a small domain such states may not find a suitable partner to
form a baryon. But they are, in fact, parts of real baryons, thus we have
to assume that they remain "bound" states which also diffuse along the
rapidity axis.

If the quarks are formed as a result of repeated "collisions", this
explains why the quarks must be in a statistical equilibrium before
they can form hadrons. Hence, in our calculations we assume that the
probability to find a quark, an antiquark, a diquark or an antidiquark
in the see is determined by constants. If the probability to find
a diquark or an antidiquark is $\alpha$, that of a quark or an antiquark
is $(1-\alpha)$:
\begin{equation}
\label{8'}
(1-\alpha)\frac{1}{2}(q+\bar{q})+\alpha \frac{1}{2}(\hat{qq} +
\hat{\bar{q}\bar{q}})
\end{equation}
where we omitted the colour indices; in fact
$q=\frac 13\sum\limits_{i=1}^3 q_i$  and
$\hat{qq}=\frac 13\sum\limits_{i=1}^3(\hat{qq})_i$.
The condition of the equilibrium means that after the
collision of two systems of the type (\ref{8'}) hadrons, quark and
diquark states are formed in the same proportions as in (\ref{8'}),
i.e. \[ [(1-\alpha)\frac{1}{2}(q+\bar{q})+\alpha \frac{1}{2}(\hat{qq} +
 \hat{\bar{q}\bar{q}})]
  [(1-\alpha)\frac{1}{2}(q+\bar{q})+\alpha \frac{1}{2}
 (\hat{qq} + \hat{\bar{q}\bar{q}})]  \]
\begin{equation}
\label{9'}
  \rightarrow
  hadrons + a\frac{1}{2}(q+\bar{q}) + \frac{1}{2}(\hat{qq} +
 \hat{\bar{q}\bar{q}}) ;
\end{equation}
here
\begin{equation}
\label{10'}
\frac{b}{a} = \frac{\alpha}{1-\alpha} .
\end{equation}
Further, we will assume that only the nearest neighbours form hadrons.
This does not alter, of course, the result (such an assumption is rather
the definition of what is a "neighbour"), but it simplifies the
calculations. In addition, let us suppose that the diquarks do not form
bound states forever, but dissociate with a probability $X$ into
constituent quarks.

Thus, if two quarks turn out to be the nearest neighbours, the
probability to form a diquark is $1/3$:
\begin{equation}
\label{11'}
qq \rightarrow 1/3\hat{qq}+2/3q\cdot q.
\end{equation}
We denote here by $q\cdot q$ a quark pair which does not form a diquark.

If the nearest neighbours are a quark and an antiquark, then
\begin{equation}
\label{12'}
q\bar{q} \rightarrow 1/9 M+8/9 q\cdot \bar{q},
\end{equation}
where $M$ is a meson (a white state of $q\bar{q}$). If a diquark is
"colliding" with a quark, a baryon state $B$ is formed with a probability
$1/9$:
\begin{equation}
\label{13'}
\hat{qq}q \rightarrow 1/9 B + 8/9 \hat{qq}\cdot q,
\end{equation}
and, similarly,
\begin{equation}
\label{14'}
\hat{qq}\bar{q} \rightarrow 1/9 M\cdot q + 8/9 \hat{qq}\cdot \bar{q}.
\end{equation}
The collisions of two diquarks or of a diquark and an antidiquark result
in the following states:
\begin{equation}
\label{15'}
\hat{qq}\hat{qq} \rightarrow 2/9 B\cdot q + 7/9 \hat{qq}\cdot \hat{qq},
\end{equation}
\begin{equation}
\label{16'}
\hat{qq}\hat{\bar{q}\bar{q}} \rightarrow 1/9 q\cdot M\cdot \bar{q} +
8/9 \hat{qq}\cdot \hat{\bar{q}\bar{q}}.
\end{equation}
The relations (\ref{13'})-(\ref{16'}) lead to
\begin{eqnarray}
\label{17'}
\lefteqn{\left[\frac{1-\alpha}{2}(q+\bar{q}) + \frac{\alpha}{2}(\hat{qq}+
\hat{\bar{q}\bar{q}})\right] \left[\frac{1-\alpha}{2}(q+\bar{q}) +
\frac{\alpha}{2}(\hat{qq}+ \hat{\bar{q}\bar{q}})\right] \rightarrow }
\nonumber\\
& & \frac{1}{2}(q+\bar{q})\frac{1}{9}(-\alpha^2-11\alpha+14) + \nonumber\\
& & \frac{1}{2}(\hat{qq}+\hat{\bar{q}\bar{q}})\frac{1}{18}(\alpha^2+
26\alpha+3) + \frac{1}{18}M + \frac{1}{18}B + \frac{1}{18}\bar{B} .
\end{eqnarray}
The number of baryons (antibaryons) and mesons in the see (in the central
region) is determined by the relation (\ref{17'}):
\begin{equation}
\label{20'}
B/M = \alpha , \qquad \bar{B}/M = \alpha .
\end{equation}
This means that the proportions of baryons (antibaryons) and mesons are
the same as the probability to find a diquark (or an antidiquark) in
the see. Depending on the value of $\alpha$, this can vary between $0$
and $1$. The parameter $\alpha$ can be fixed if we assume that the
fragmentation probability of the quark is determined by (\ref{5}).

Let us consider now the process of meson and baryon formation by a tagged
quark $q'$. If this quark "collides" with coloured quarks and diquarks
of the sea, white states containing this quark are formed: $M'=q'\bar{q}$
and $B'= q'qq$.
We shall denote by $c$ the probability to form a meson state $M'$, by
$d$ that to form a baryon state $B'$. These probabilities depend on
the ratio of the numbers of quarks and diquarks in the sea, i.e. on
$\alpha$:
\begin{eqnarray}
\label{21'}
q'+ \quad\mbox{sea}\quad q,\,\bar{q},\,\hat{qq},\,\hat{\bar{q}\bar{q}}
\rightarrow c(\alpha)M' + d(\alpha)B', \nonumber\\
c(\alpha) + d(\alpha) = 1.
\end{eqnarray}
A single "collision" of $q'$ with quarks and diquarks of the sea leads to
\begin{eqnarray}
\label{22'}
\lefteqn{q'\left[\frac{1-\alpha}{2}(q+\bar{q}) +
 \frac{\alpha}{2}(\hat{qq}+\hat{\bar{q}\bar{q}})\right] \rightarrow
 \frac{1}{18}[2(7+\alpha) + } \nonumber\\
 & & +3X(1-\alpha)]q'+(1-X)\frac{1}{6}(1-\alpha)
 \hat{qq'} + \frac{1}{18}M' + \frac{\alpha}{18}B' .
\end{eqnarray}
As before, the coefficients of $M'$, $B'$, $q'$ and $\hat{qq'}$
correspond to the probabilities of the transition of the quark $q'$ to
these states. Calculating these probabilities, we have used the relations
\begin{eqnarray}
\label{23'}
q'q & \rightarrow & 1/3 \hat{qq'} + 2/3 q'\cdot q \; , \nonumber\\
q'\bar{q} & \rightarrow & 1/9 M' + 8/9 q'\cdot \bar{q} \; , \nonumber\\
q'\hat{qq} & \rightarrow & 1/9 B' + 8/9 q'\cdot \hat{qq} \; , \nonumber\\
q'\hat{\bar{q}\bar{q}} & \rightarrow & 1/9 M'\cdot\bar{q} + 8/9 q'\cdot
 \hat{\bar{q}\bar{q}} \;.
\end{eqnarray}
As we see, after the first "collision" diquark states containing $q'$
appear. The "collision" of these diquarks with quarks and diquarks of the
sea leads to the transition
\begin{eqnarray}
\label{24'}
\lefteqn{\hat{qq'}\left[\frac{1-\alpha}{2}(q+\bar{q}) +
 \frac{\alpha}{2}(\hat{qq}+\hat{\bar{q}\bar{q}})\right] \rightarrow
  \frac{1}{36}[1+\alpha+3X(16-\alpha)]q'+ } \nonumber\\
 & & + (1-X)\frac{1}{18}(16-\alpha)\hat{qq'}
 + \frac{1}{36}M' + \frac{\alpha}{36}(2+\alpha)B' .
\end{eqnarray}
The probabilities $c$ and $d$ are results of repeated "collisions" of
quarks with subsequent transitions of the type (\ref{22'}), and of
"collisions" of the appearing diquarks with subsequent transitions
(\ref{24'}). Fig.7 shows the ratio $c(\alpha)/d(\alpha)$ as a function
of $\alpha$. \\
\begin{figure}
\centerline{\epsfig{file=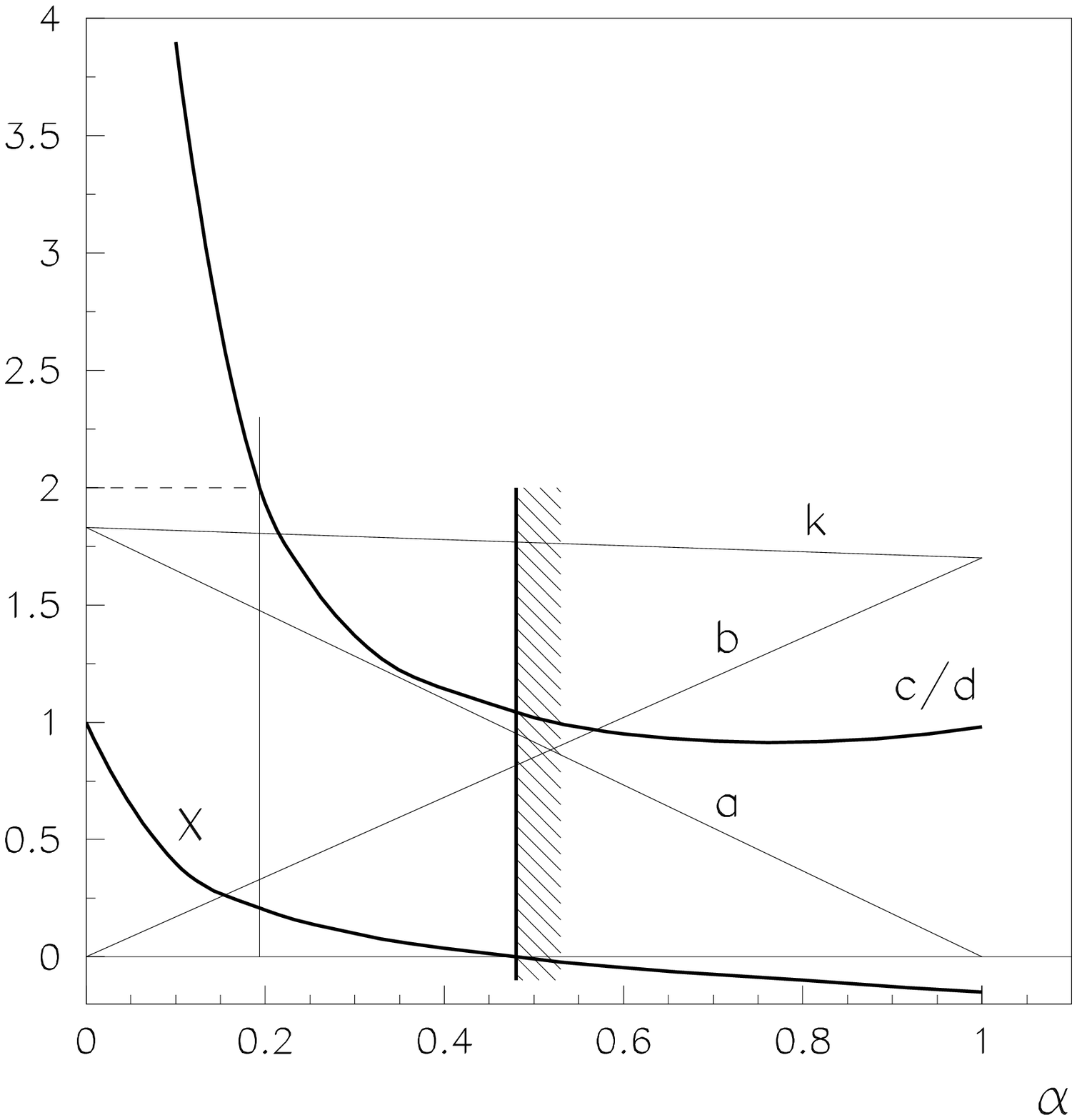,width=9cm}}
\vspace{-1cm}
\centerline{Fig.7}
\end{figure}
At the value $\alpha \simeq 0,194$ the probabilities of baryons $B'$ and
mesons $M'$ are determined by the relation (\ref{5}), and (\ref{17'})
leads to
\begin{equation}
\label{25'}
M:B:\bar{B} \simeq 5,2 : 1 : 1 .
\end{equation}
Thus, if $q' \rightarrow 1/3 B' + 2/3 M'$ is satisfied (which can be
considered as the "local" baryon charge conservation), the ratio of
the number of produced mesons, baryons and antibaryons is pretty
stable.

\subsection{Probabilities of the production of hadron states in the
central and fragmentation regions}\label{2.3}

It is a well-known experimental fact that in multiparticle production
processes the production of strange quarks is suppressed. According to
that, it was proposed [20] to consider a non-symmetrical quark sea
with a relatively suppressed production of strange quarks:
\begin{equation}
\label{27}
     u\bar{u} : d\bar{d} : s\bar{s} = 1 : 1 : \lambda \, .
\end{equation}
This suppression is characterized by a parameter $0 \leq \lambda \leq 1$.
In the case of $\lambda=1$ the symmetry between the quarks $u,d,s$ is
restored.

To be in a position to compare (\ref{7}) and (\ref{8}) with the
experimental data, we face here the question what means $M$ and $B$,
what real hadrons correspond to the mesonic and baryonic states $B_{ij}$,
$B_i$, $M_i$ etc. Indeed, quark combinatorics, while operating with
constituent quark states $q\bar{q}$ and $qqq$ does not answer the
question by what real particles they are saturated. In [20] the dominance
of the lowest $SU(6)$ multiplets was supposed, i.e. the meson 1+35-plet
($ J^P = 0^{-1},1^{-1}$) for the $q\bar{q}$ states and the baryon
56-plet ($ J^P = 1/2^+, 3/2^+ $) for $qqq$, respectively. This is a
rather rough approach, and, of course, a contribution of hadrons
belonging to higher multiplets is quite natural.

The determination of hadrons saturating the meson and baryon states is
in fact an experimental question which, in a sense, characterizes the
quark confinement. The analysis of experimental data shows that the
contribution of hadrons with L=1 is significant, 20-30 \% of the produced
particles ([42],[43]). The share of L=2 multiplets seems to be about
5-10 \% (see [43],[44]).

In the process of hadronization white states $q\bar{q}$ and $qqq$ are
produced. The decomposition of the production amplitude of these white
states into hadronic wave functions determines the content of $M$ and
$B$. In general, we can write
\begin{eqnarray}
\label{9}
M_i & = & \alpha_i(1)M_i(1) + \alpha_i(0)M_i(0), \nonumber \\
M & = & \alpha(1)M(1) + \alpha(0)M(0) .
\end{eqnarray}
The indices $L=0,1$ correspond to the $s$ and $p$-wave states,
respectively. The probabilities $\alpha_i(L)$ and $\alpha(L)$ are fixed
by the conditions $\alpha_i(1)+\alpha_i(0)=1$, $\alpha(1)+\alpha(0)=1$.

In a more general form one can write
\begin{eqnarray}
\label{10}
&& M\ =\ \sum_L\mu_L M_L \; ,\nonumber\\
&&M_i\ =\ \sum_L\mu_L^{(i)}  M_L^{(i)} \;
\end{eqnarray}
The real hadron content of the states $B$ and $B_i$ is defined as
\begin{eqnarray}
\label{11}
&& B\ = \ \sum_L\beta_L B_L \; ,\nonumber \\
&& B_i\ = \ \sum_L\beta_L^{(i)} B_L^{(i)} \, .
\end{eqnarray}
Here $L$ defines the multiplet, while the coefficients $\mu_L$,
$\mu_L^{(i)}$ and $\beta_L$, $\beta_L^{(i)}$ are production
probabilities of mesons and baryons of the given multiplet in the
process of quark hadronization.
These probabilities are determined by characteristic
relative momenta of the quarks which join each other, or rather by their
invariant masses $\sum_i(m_i^2+k_{i\bot})/x_i(1-x_i)$.

Graphically, the colour neutralization of meson and baryon states
can be represented as it is shown on Fig.8.
\vspace{1.cm}
%\begin{figure}
\[ 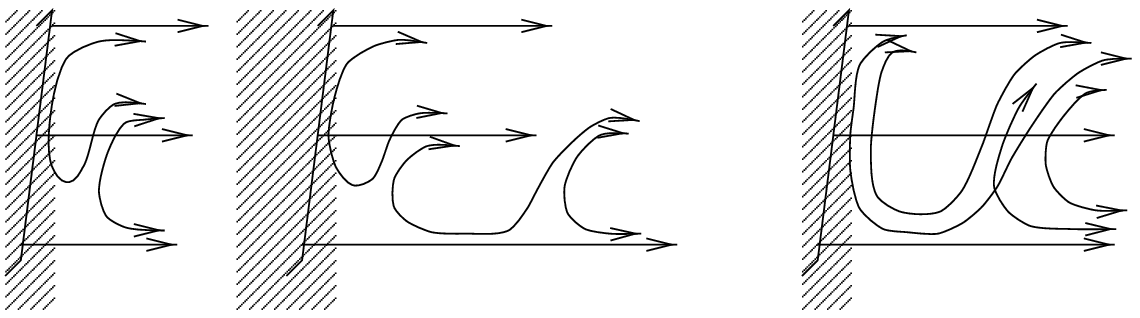 \]
\centerline{Fig.8}
%\end{figure}

\subsection{Multiplicities of the secondary particles in the central
and fragmentation regions}\label{2.4}

We shall give here expressions for the multiplicities of secondary
particles in both the fragmentation and the central regions [27]. We
consider the cases of incident proton, $\Lambda$ and $\Sigma^+$ hyperons,
$\pi^+$ and $K^+$ mesons. Expressions for other incident particles can
easily be obtained from these ones. For example, the case of a neutron
can be obtained from that of a proton by isotopic reflection, i.e.
substituting $ p \leftrightarrow n $, $ \Delta^{++}
\leftrightarrow  \Delta^- $, $\pi^+ \leftrightarrow \pi^- $,
  $K^+ \leftrightarrow K^0$. In the case of an initial antiparticle one
has to carry out charge conjugation $ p \leftrightarrow \bar{p} $,
 $ \Delta^{++} \leftrightarrow  \bar{\Delta}^{--} $ etc.

The relations (\ref{7}) and (\ref{8}) and the expressions of $B_{ij}$,
$M_i$, $B_i$ in terms of the real hadrons allow us to get the
fragmentation multiplicities easily. For this purpose we have to take
the wave function of the incident particle and to consider all possible
interactions of its constituents.

As an example, let us consider in detail the fragmentation of the proton.
We assume that the incident proton is completely polarized (this fact
is of no significance from the point of view of the result). The proton
wave function in this case is
\[  \Psi(p^{\uparrow}) = \sqrt{\frac{2}{3}}\{u^{\uparrow}u^{\uparrow}
 d^{\downarrow}\} - \sqrt{\frac{1}{3}}\{u^{\uparrow}u^{\downarrow}
 d^{\uparrow}\}  . \]
It is implied that the functions are symmetrized with respect to the
$SU(6)$ indices, e.g.
 \[  \{u^{\uparrow}u^{\uparrow}d^{\downarrow}\} =
\sqrt{\frac{1}{3}}(u^{\uparrow}u^{\uparrow}d^{\downarrow} + u^{\uparrow}
d^{\downarrow}u^{\uparrow} + d^{\downarrow}u^{\uparrow}u^{\uparrow}). \]
It can be seen immediately that for the quarks-spectators the probability
of being in a $\{u^{\uparrow}u^{\uparrow}\}$ is $2/9$ (while the quark
$d^{\downarrow}$ is interacting), in $\{u^{\uparrow}u^{\downarrow}\}$
it is $1/9$, in $\{u^{\uparrow}d^{\uparrow}\}$ also $1/9$. When the quark
$u^{\uparrow}$ is interacting, the spectators are in a state
\[ \frac{1}{\sqrt{5}}(2\{u^{\uparrow}d^{\downarrow}\}- \{u^{\downarrow}
d^{\uparrow}\}) \equiv (ud)_p . \]
Thus, we have
\[  B_{ij} = \frac{2}{9}B(u^{\uparrow}u^{\uparrow})+\frac{1}{9}
 B(u^{\uparrow}u^{\downarrow}) + \frac{1}{9}B(u^{\uparrow}d^{\uparrow})
+\frac{5}{9}B_p(ud).\]
The decompositions of $B(u^{\uparrow}u^{\uparrow})$ and $B(u^{\uparrow}
u^{\downarrow})$ into the real hadrons of the 56-plet lead to identical
results, and therefore we write
\[ \frac{1}{9}B(u^{\uparrow}u^{\downarrow})+\frac{2}{9}B(u^{\uparrow}
   u^{\uparrow}) = \frac{1}{3} B(uu) . \]
For the sake of simplicity we introduce the notation
$B(u^{\uparrow}d^{\uparrow})= B_1(ud)$. In the case of an incident proton
the states $B_i$ and $M_i$ are
\[ B_i=\frac{2}{3}B(u)+\frac{1}{3}B(d)\qquad\mbox{and}\qquad
 M_i=\frac{2}{3}M(u)+\frac{1}{3}M(d) ,\]
respectively. As a result, we can write
\begin{eqnarray}
\label{12}
\lefteqn{p \rightarrow \Delta_p\cdot p + \Delta_p^*\cdot B_p^* + }
\nonumber \\
 & + & (1-\Delta_p-\Delta_p^*)\left\{\frac{1}{2}\left[\frac{5}{9}B_p(ud)
 + \frac{1}{3}B(uu) + \frac{1}{9}B_1(ud)\right]\right. + \nonumber\\
 & + & \left.\frac{1}{2}\left[\frac{2}{3}B(u)+\frac{1}{3}B(d)\right]
 \right\} +\frac{3}{2}\left[\frac{2}{3}M(u)+\frac{1}{3}M(d)\right] .
\end{eqnarray}
Expanding the right-hand side in terms of the hadron states $h$ (i.e. the
meson states $h_{M(L)}$ and the baryon states $h_B$) we, finally, obtain
\begin{eqnarray}
\label{13}
\lefteqn{p \rightarrow \Delta_p\cdot p + } \nonumber\\
 & & +\sum_h h_B\bigg\{\Delta_p^*\cdot \beta_h(p)+(1-\Delta_p-\Delta_p^*)
 \bigg[\frac{5}{18}\beta_h(ud_p) + \nonumber\\
 & & +\frac{1}{6}\beta_h(uu) + \frac{1}{18}\beta_h(ud_1) +
 \frac{1}{3}\beta_h(u) + \frac{1}{6}\beta_h(d)\bigg]\bigg\} + \nonumber\\
 & & +(1-\Delta_p-\Delta_p^*)\sum_L \sum_h h_{M(L)} \alpha_i(L)
 \bigg[ \mu_h^L(u) + \frac{1}{2}\mu_h^L(d) \bigg]
\end{eqnarray}
or
\begin{equation}
\label{14}
p \rightarrow \sum_h F_h(p)\cdot h ,
\end{equation}
where $F_h(p)$ denotes the multiplicity of the secondary proton in the
fragmentation region. Similarly, the multiplicities of $\Lambda$ and
$\Sigma^+$ hyperons in the fragmentation region are
\begin{eqnarray}
\label{15}
\lefteqn{\Lambda \rightarrow \sum_h F_h(\Lambda)\cdot h = } \nonumber\\
 & & = \Delta_{\Lambda}\cdot \Lambda + \sum_h h_B\bigg\{\Delta_{\Lambda}^*
 \cdot \beta_h(\Lambda)+(1-\Delta_{\Lambda}-\Delta_{\Lambda}^*)
 \bigg[\frac{\xi}{2(2+\xi)}\beta_h(ud_{\Lambda}) + \nonumber\\
 & & + \frac{1}{4(2+\xi)}\left(\beta_h(us_1)+\beta_h(ds_1)+\beta_h(us_0)+
 \beta_h(sd_0)\right) + \nonumber\\
 & & + \frac{1+2\xi}{6(2+\xi)}\beta_h(s) + \frac{5+\xi}{12(2+\xi)}
 (\beta_h(u)+ \beta_h(d))\bigg]\bigg\} + \nonumber\\
 & & +(1-\Delta_{\Lambda}-\Delta_{\Lambda}^*)\sum_L \sum_h h_{M(L)}
 \alpha_i(L) \bigg[ \frac{5+4\xi}{6(2+\xi)}\mu_h^L(s) + \nonumber\\
 & & \frac{5\xi+13}{12(2+\xi)}(\mu_h^L(u)+\mu_h^L(d)) \bigg]
\end{eqnarray}
and
\begin{eqnarray}
\label{16}
\lefteqn{\Sigma^+ \rightarrow \sum_h F_h(\Sigma^+)\cdot h = } \nonumber\\
 & & = \Delta_{\Sigma}\cdot \Sigma^+ + \sum_h h_B\bigg\{\Delta_{\Sigma}^*
 \cdot \beta_h(\Sigma^+)+(1-\Delta_{\Sigma}-\Delta_{\Sigma}^*)
 \bigg[\frac{\xi}{2(2+\xi)}\beta_h(uu) +  \nonumber\\
 & & + \frac{5}{6(2+\xi)}\beta_h(us_{\Sigma}) +\frac{1}{6(2+\xi)}
 \beta_h(us_1)+ \frac{\xi+5}{6(2+\xi)}\beta_h(u)+\frac{1+2\xi}{6(2+\xi)}
 \beta_h(s)\bigg]\bigg\} + \nonumber\\
 & & +(1-\Delta_{\Sigma}-\Delta_{\Sigma}^*)\sum_L \sum_h h_{M(L)}
 \alpha_i(L) \bigg[ \frac{5\xi+13}{6(2+\xi)}\mu_h^L(u) +
 \frac{5+4\xi}{6(2+\xi)}\mu_h^L(s) \bigg]  .
\end{eqnarray}
Contrary to the proton case, in (\ref{15}) and (\ref{16}) it is taken
into account that the cross section of the interaction is less for the
strange quark than for the non-strange one. Their ratio
$\xi = \sigma_{inel}(sq)/\sigma_{inel}(qq)$ is close to $2/3$.

Formula (\ref{8}) enables us to calculate the fragmentation secondaries
for incident mesons. In the cases $\pi^+$ and $K^+$ we get, as
follows,
\begin{eqnarray}
\label{17}
\lefteqn{\pi^+ \rightarrow \sum_h F_h(\pi^+) \cdot h = \delta_{\pi}\cdot
 \pi^+ } \nonumber\\
 & & + \sum_L \sum_h h_{M(L)}\cdot \alpha_i(L)\bigg\{\delta_{\pi}^*
 \mu_h^L(\pi^+) + (1- \delta_{\pi} - \delta_{\pi}^*)\bigg[\frac{2}{3}
 \mu_h^L(u)+\frac{2}{3}\mu_h^L(\bar{d})\bigg]\bigg\} + \nonumber\\
 & & +\sum_h h_B (1- \delta_{\pi} - \delta_{\pi}^*)\frac{1}{3}\beta_h(u)+
 \sum_h h_{\bar{B}} (1- \delta_{\pi} - \delta_{\pi}^*)\frac{1}{3}
\beta_h(\bar{d}) ,
\end{eqnarray}
\begin{eqnarray}
\label{18}
\lefteqn{K^+ \rightarrow \sum_h F_h(K^+) \cdot h = \delta_{K}\cdot
 K^+ } \nonumber\\
 & & + \sum_L \sum_h h_{M(L)}\cdot \alpha_i(L)\bigg\{\delta_K^*
 \mu_h^L(K^+) + (1- \delta_K - \delta_K^*)\bigg[\frac{2}{3}
 \mu_h^L(u)+\frac{2}{3}\mu_h^L(\bar{s})\bigg]\bigg\} + \nonumber\\
 & & +\sum_h h_B (1- \delta_K - \delta_K^*)\frac{1}{3}\beta_h(u)+
 \sum_h h_{\bar{B}} (1 - \delta_K - \delta_K^*)\frac{1}{3}
\beta_h(\bar{s}) .
\end{eqnarray}
In (\ref{17}, \ref{18}) the parameter $\xi$ does not occur because, due
to (\ref{7}), the secondary hadron content of the quark-spectator is
equal to that of the quark which underwent interaction.

In the central region the multiplicity of secondary particles is
given by (\ref{3}). Due to the additive quark model, the energy which
is used for the production of the new (sea) quarks is determined by
the energy of the colliding quarks. In the pion-nucleon collision the
energy squared is about $1/6$, in nucleon-nucleon collision about $1/9$
of the total energy of hadrons. That means that in the case of
pion-nucleon collision we have
\begin{equation}
\label{19}
N_{\pi N}(s) = b\ln\frac{s}{6s_0} = b\ln\frac{s}{s_{\pi N}^0} ,
\end{equation}
while for the nucleon-nucleon case
\begin{equation}
\label{20}
N_{N N}(s) = b\ln\frac{s}{9s_0} = b\ln\frac{s}{s_{N N}^0} .
\end{equation}
In the collision processes of strange particles one has to remember the
difference between the cross sections of the interaction of strange and
non-strange quarks, and the fact that the heavier strange quark takes
away a larger fraction of the hadron momentum. Hence, for the
kaon-nucleon collision one can write
\begin{equation}
\label{21}
N_{KN}(s) = \frac{b\xi}{1+\xi}\ln\frac{s}{3(1+\mu)s_0} +
 \frac{b}{1+\xi}\ln\frac{s\mu}{3(1+\mu)s_0} = b\ln\frac{s}{S_{KN}^0} ,
\end{equation}
where $\mu=\frac{m_q}{m_s} \approx \frac{2}{3}$ is the ratio of the
strange and non-strange quarks. Finally,
\begin{equation}
\label{22}
N_{\Lambda N}(s) = N_{\Sigma N}(s) = \frac{2b}{2+\xi}\ln\frac{s\mu}{3(1
 + 2\mu)s_0} + \frac{\xi b}{2+\xi}\ln\frac{s}{3(1+2\mu)s_0}
 = b\ln\frac{s}{S_{\Lambda N}^0} .
\end{equation}
The obtained expressions give a possibility to calculate the absolute
values of average multiplicities of secondary particles in hadron-hadron
collisions. The parameters are fitted to the experimental data and
according to them the coefficients in (\ref{19}-\ref{22}) are calculated
[27]. (For example, the value of $\lambda$ is selected to give the best
agreement with the experimental $K/\pi$ ratio in the central region and
is found to be $0,3$). Supposing that the probabilities $\Delta$ and
$\delta$ of the coherent processes $B_{ijk} \rightarrow B_{ijk}$ and
$M_{i\bar{j}} \rightarrow M_{i\bar{j}}$ are mostly of diffractional
origin, the value of these probabilities is estimated using the data on
diffraction scattering. In the additive quark model the cross sections
of diffraction processes in the meson-nucleon and baryon-nucleon
scatterings are determined by the diagrams in Fig.9. \\
\vspace{1cm}
\[  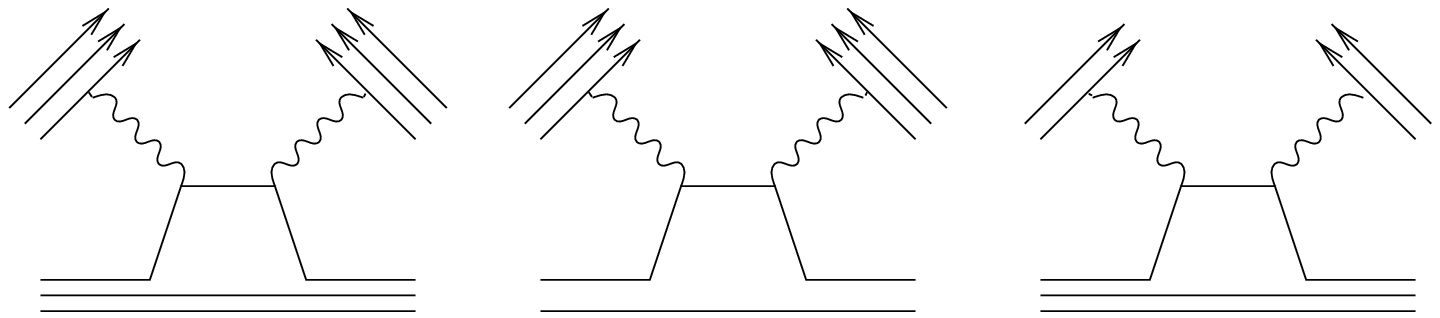  \]
\centerline{Fig.9}
(For the values of $\Delta$ and $\delta$, see [27]).

Experimental data on average multiplicities of secondary hadrons in
the $pp$, $\pi p$ and $Kp$ collisions permit us to prove the basic
statements of quark combinatorics.

\newpage
\begin{figure}
\centerline{\epsfig{file=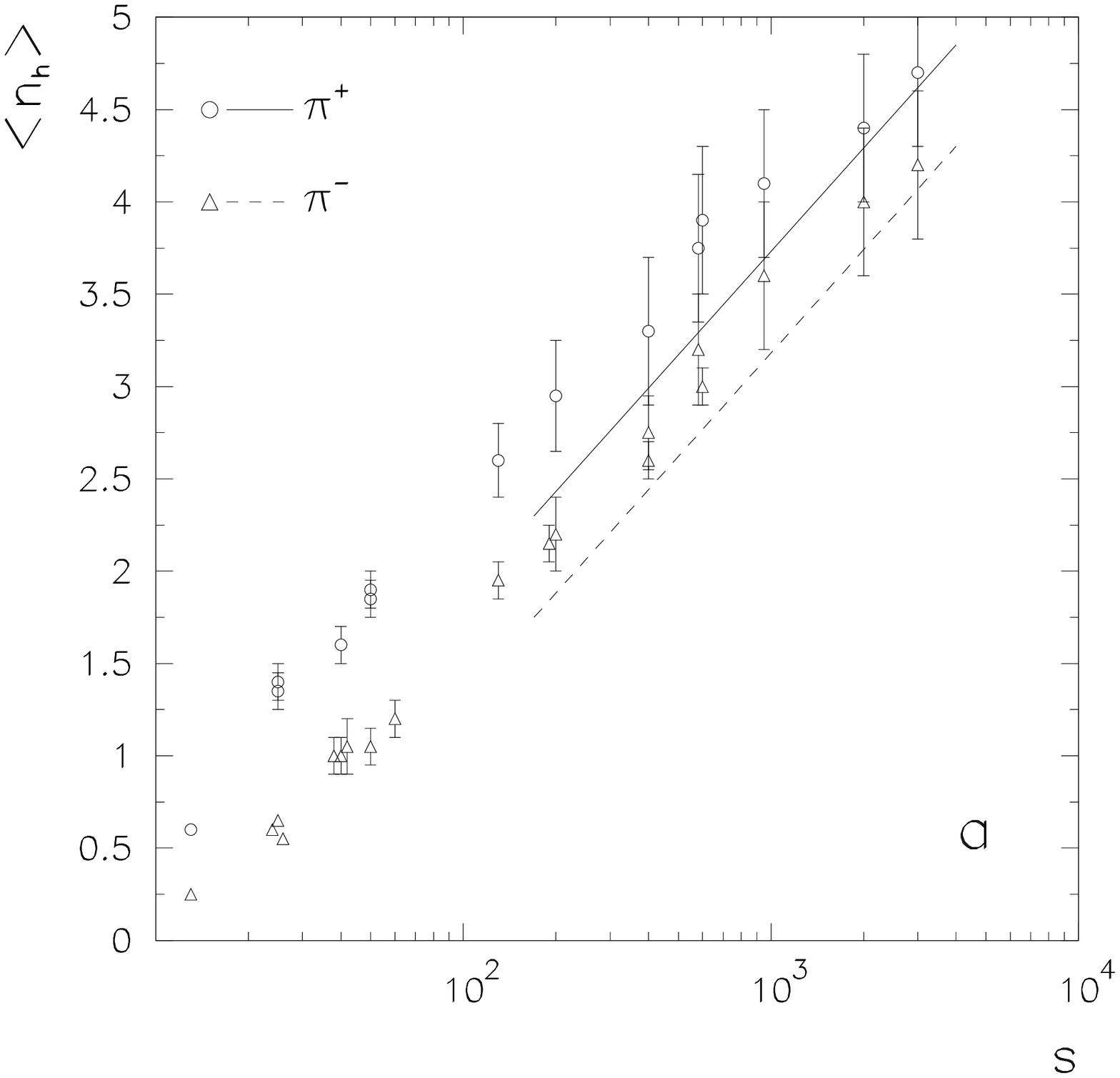,width=9.0cm}
            \epsfig{file=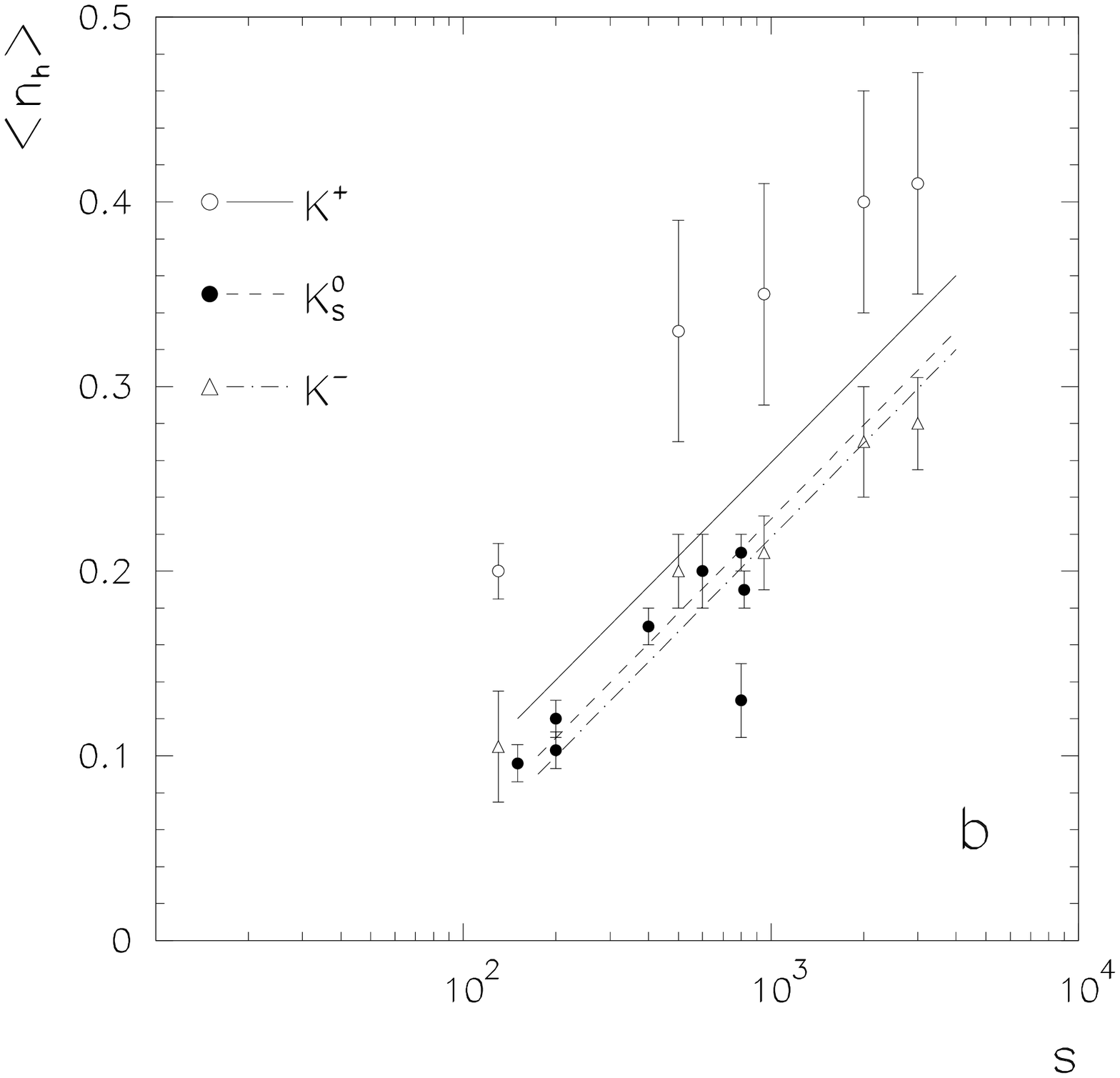,width=9.0cm}}
\centerline{\epsfig{file=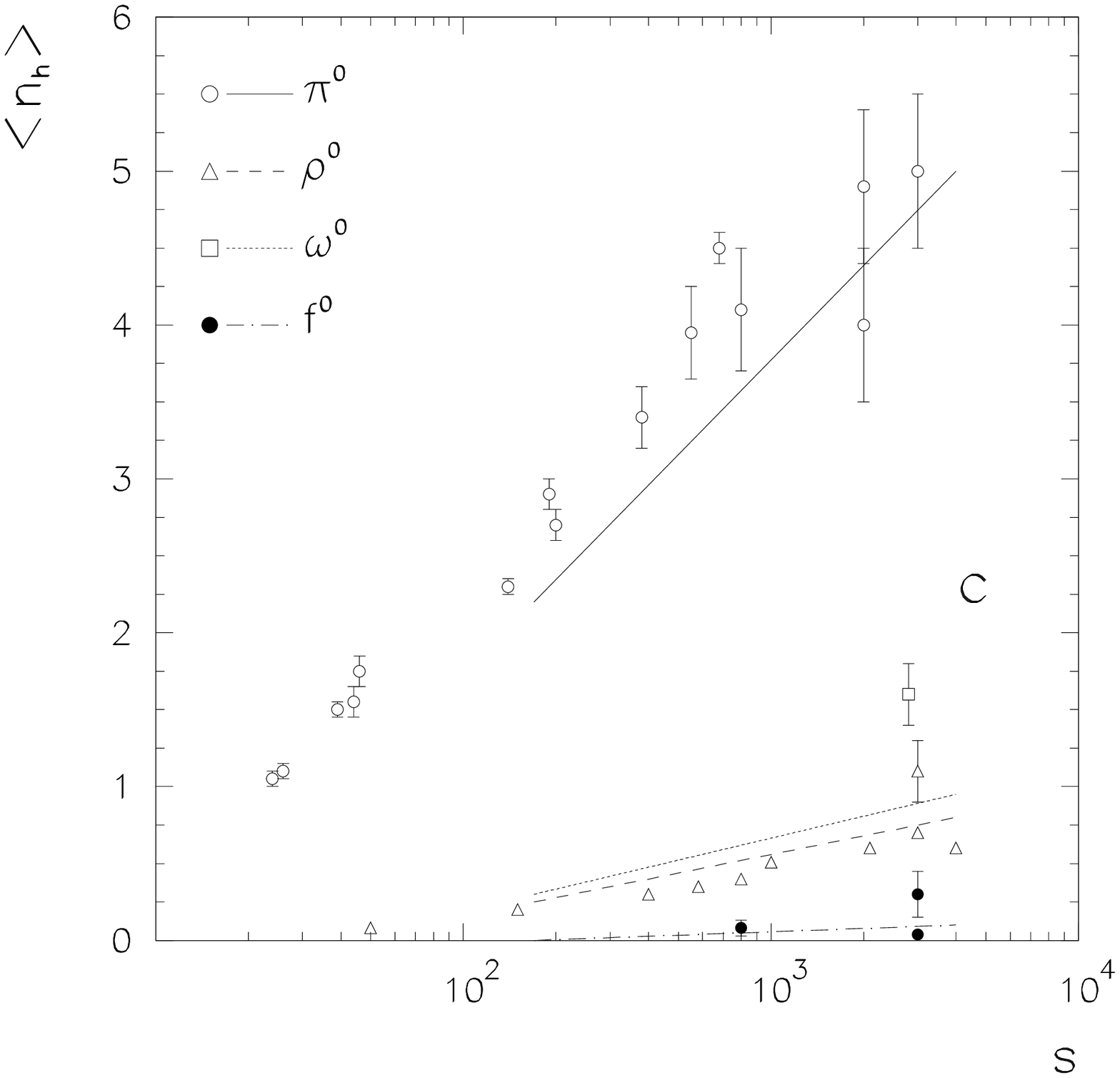,width=9.0cm}
            \epsfig{file=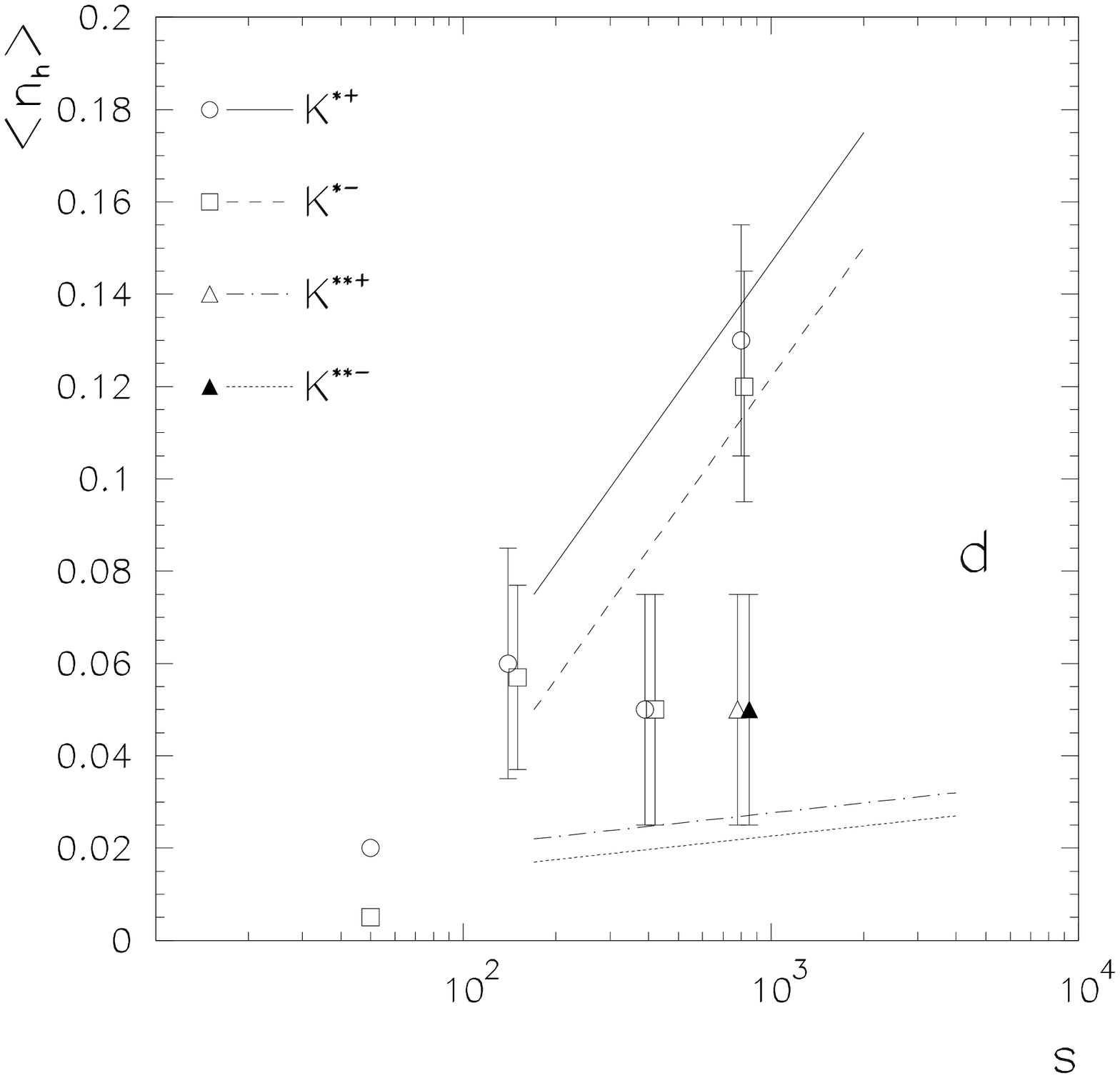,width=9.0cm}}
\end{figure}
\begin{figure}
\centerline{\epsfig{file=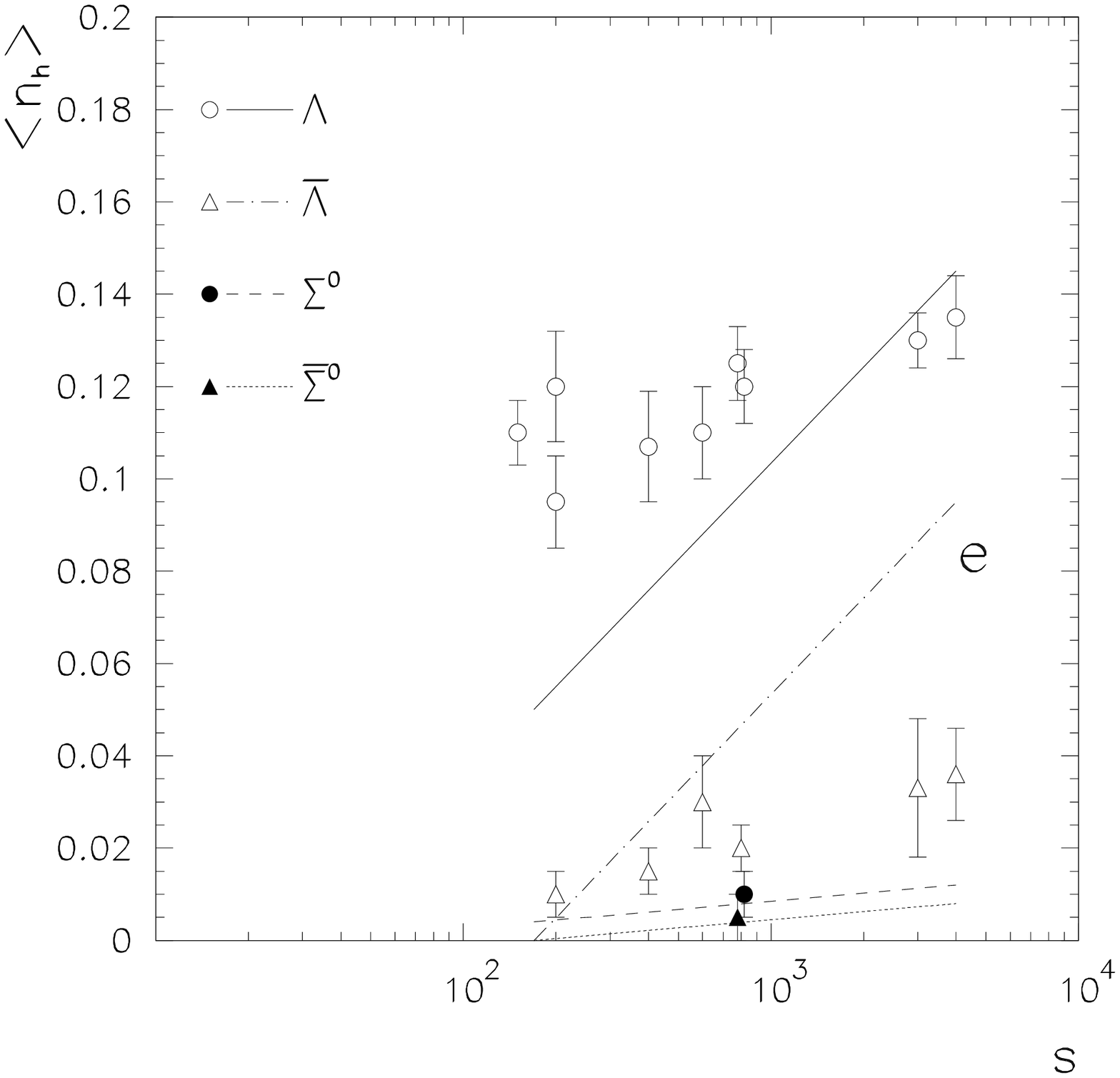,width=9.0cm}
            \epsfig{file=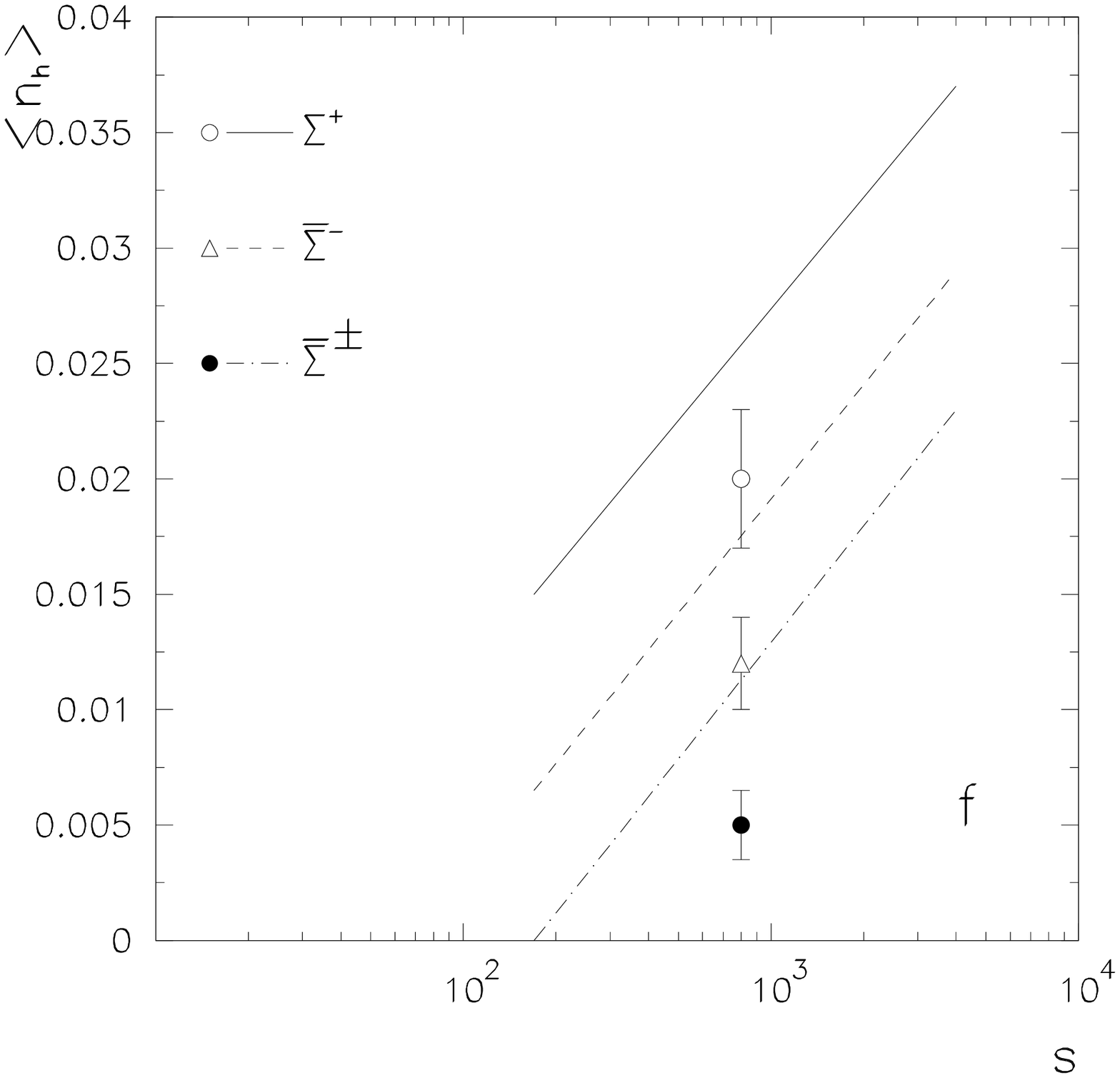,width=9.0cm}}
\centerline{\epsfig{file=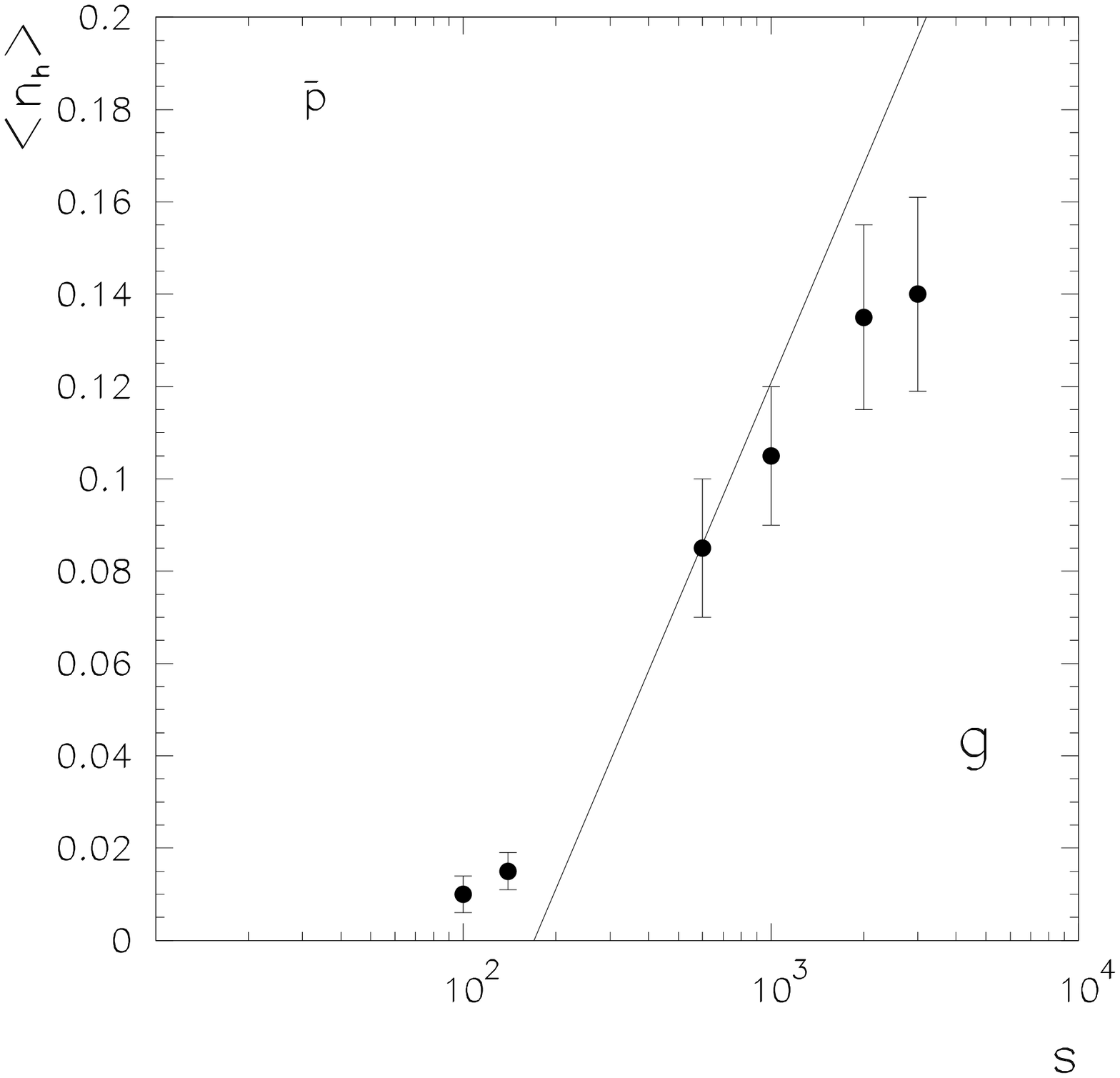,width=9.0cm}}
\centerline{Fig.10}
\end{figure}

Consider first the meson production processes. In Figs.10-12 the data
on average multiplicities of secondary mesons in $pp$ (Fig.10),
$\pi^{\pm}p$ (Fig.11) and $K^-p$ (Fig.12) collisions are presented. The
straight lines correspond to the predictions of the quark combinatorial
calculus. In each case there is a satisfactory agreement between theory
and experiment.

Concerning baryons and baryon resonances, the experimental data and
the corresponding predictions of the quark model agree only roughly.
E.g., the ratios $\Lambda/\Sigma^0$, $\Sigma^+(1385)/\Sigma^0$ and
$\Sigma^-(1385)/\Sigma^0$ satisfy the prediction quite well, what
corresponds to the idea of baryons produced in $SU(6)$ multiplets. The
same ratios indicate that there might be a significant contribution of
higher resonances. (For details, see [27]).

\newpage
\begin{figure}
\centerline{\epsfig{file=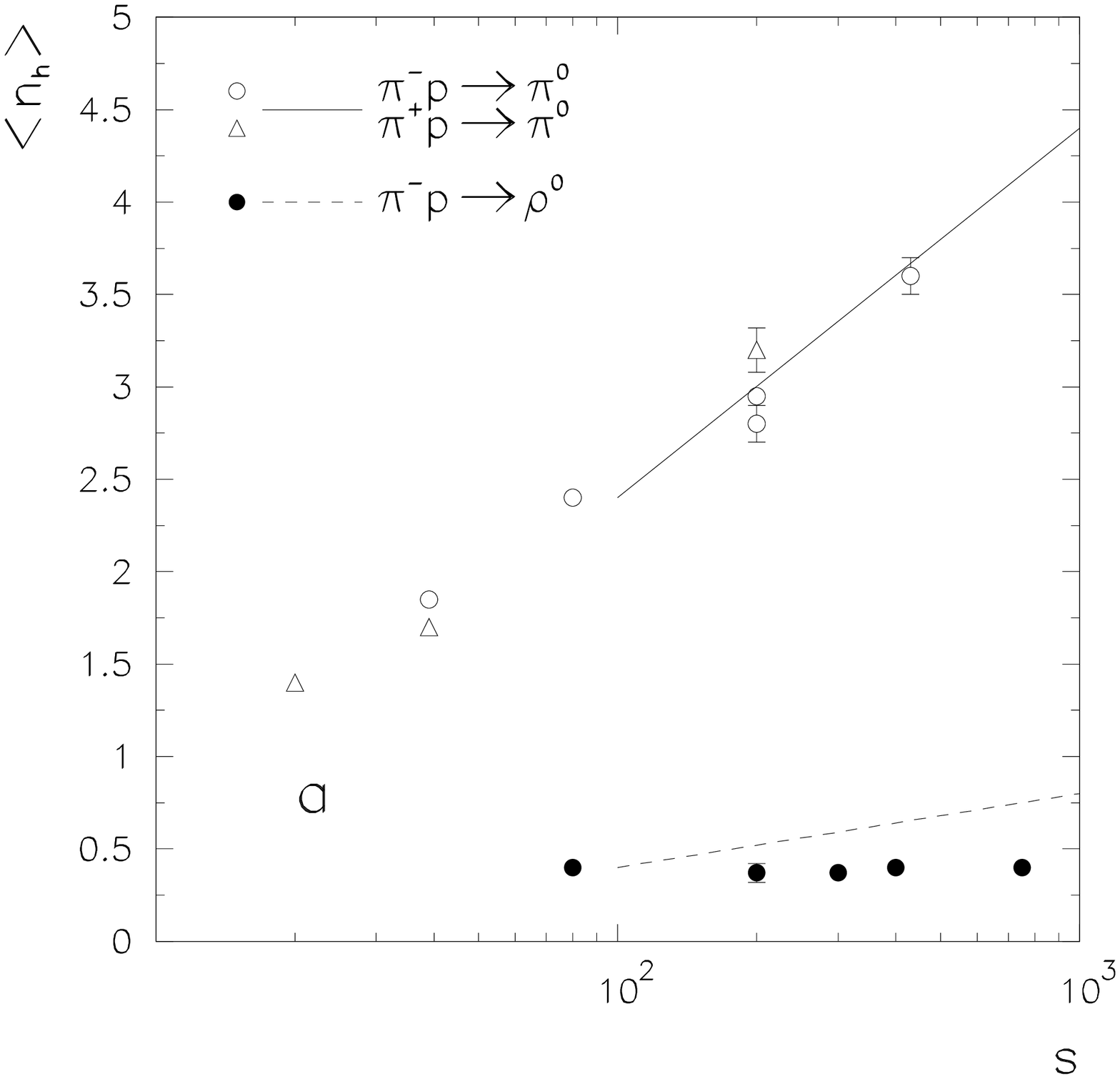,width=9.0cm}
            \epsfig{file=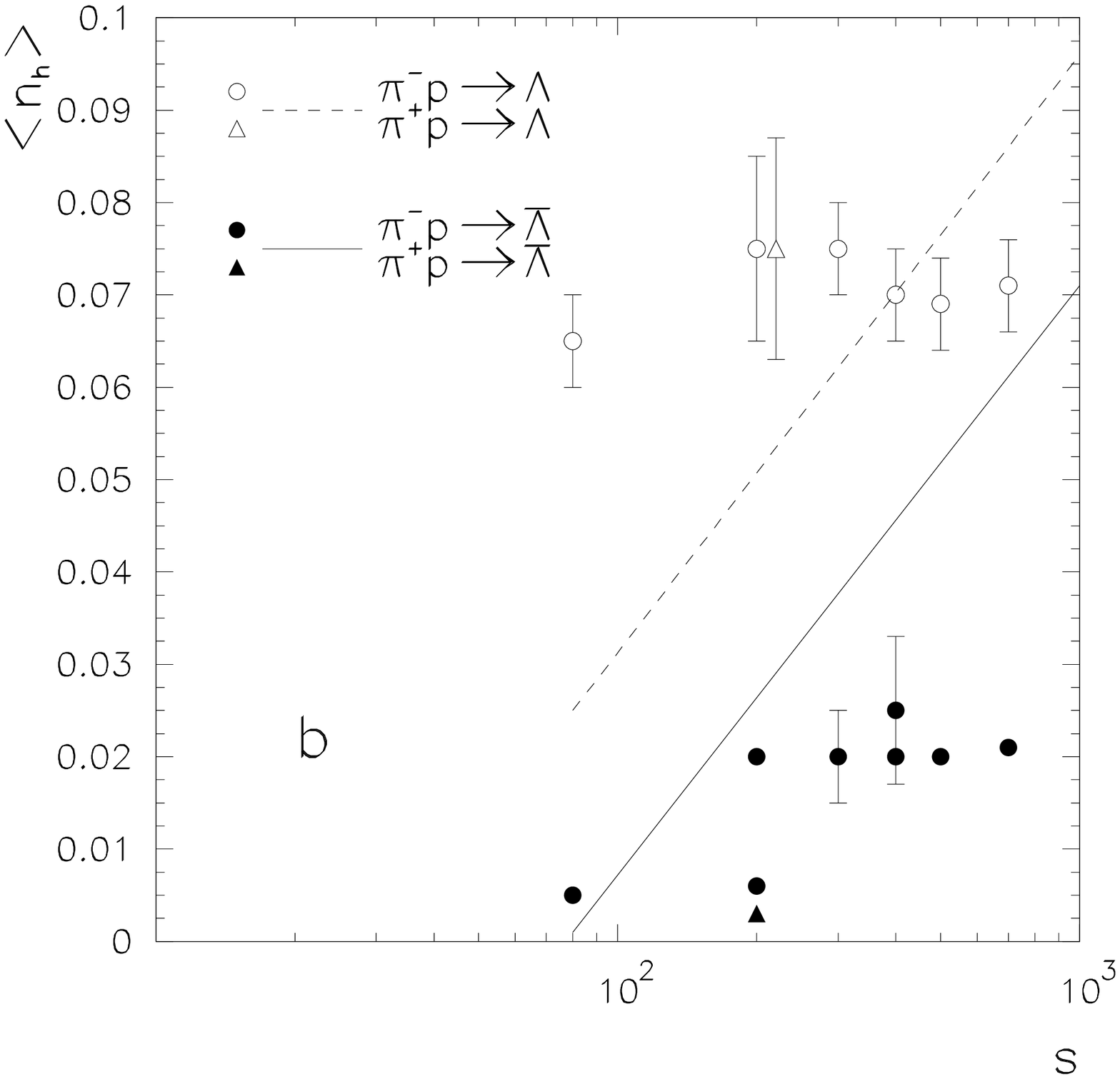,width=9.0cm}}
\centerline{\epsfig{file=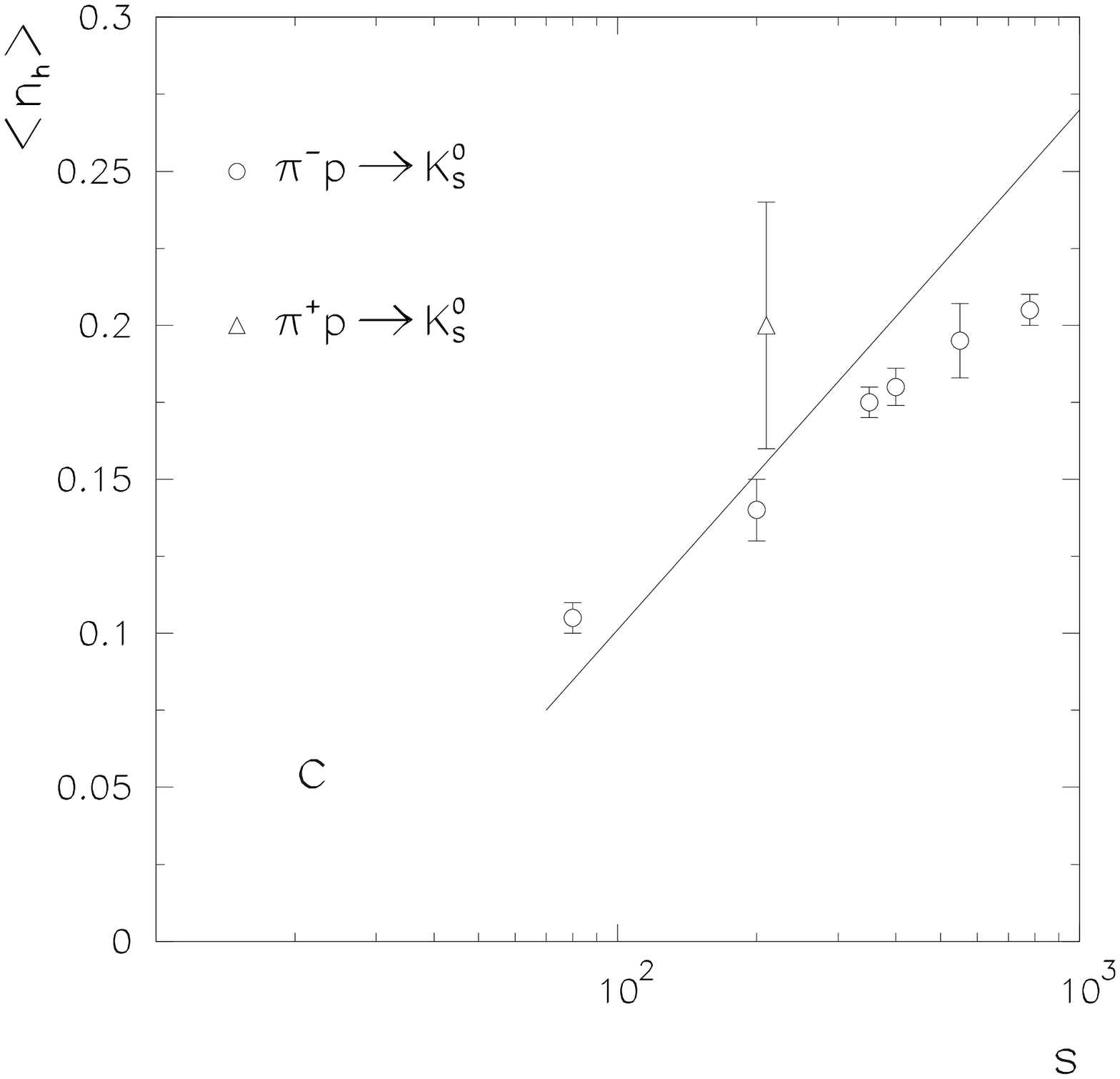,width=9.0cm}}
\centerline{Fig.11}
\end{figure}

\newpage
\begin{figure}
\centerline{\epsfig{file=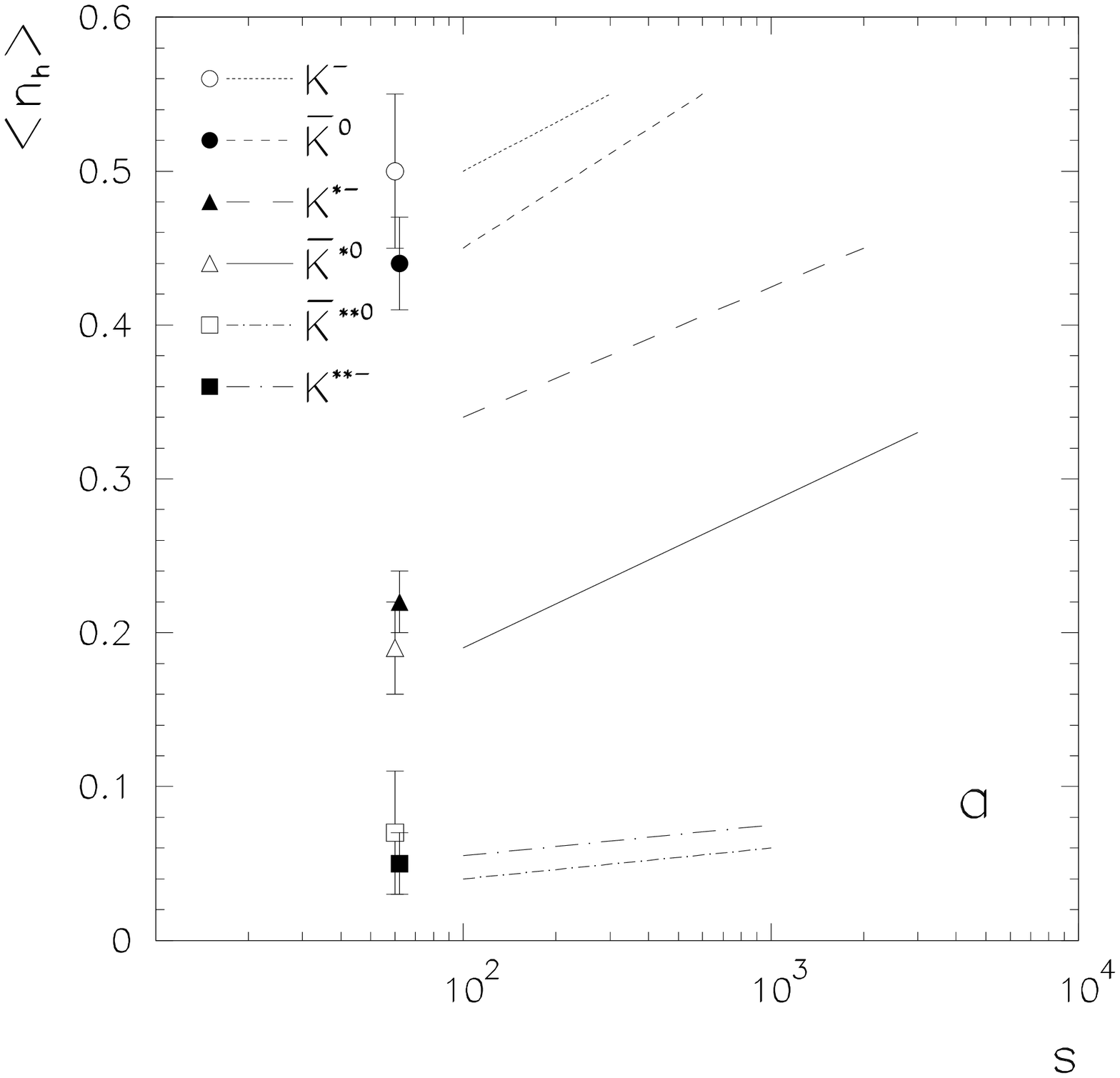,width=7.0cm}
            \epsfig{file=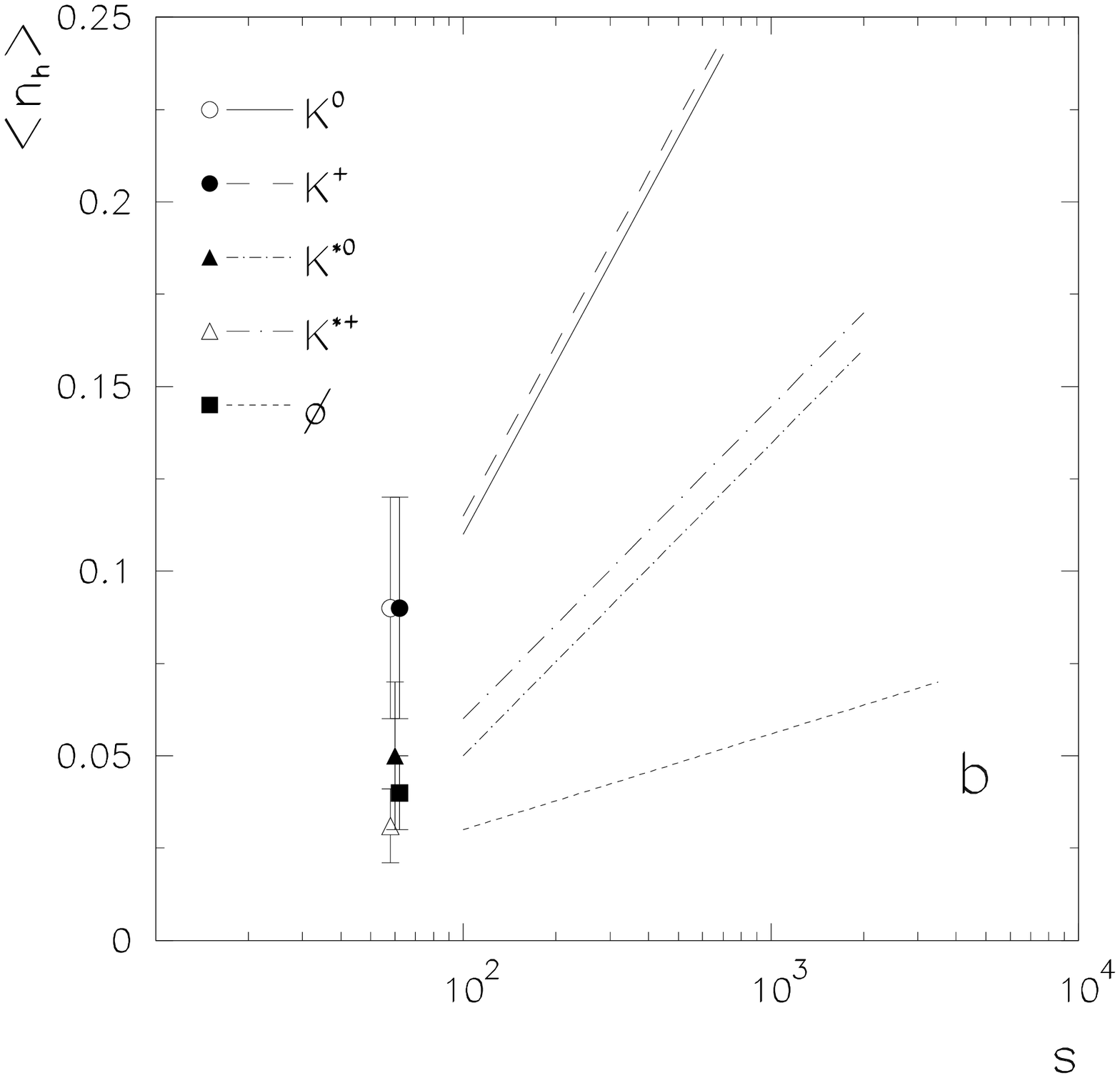,width=7.0cm}}
\centerline{\epsfig{file=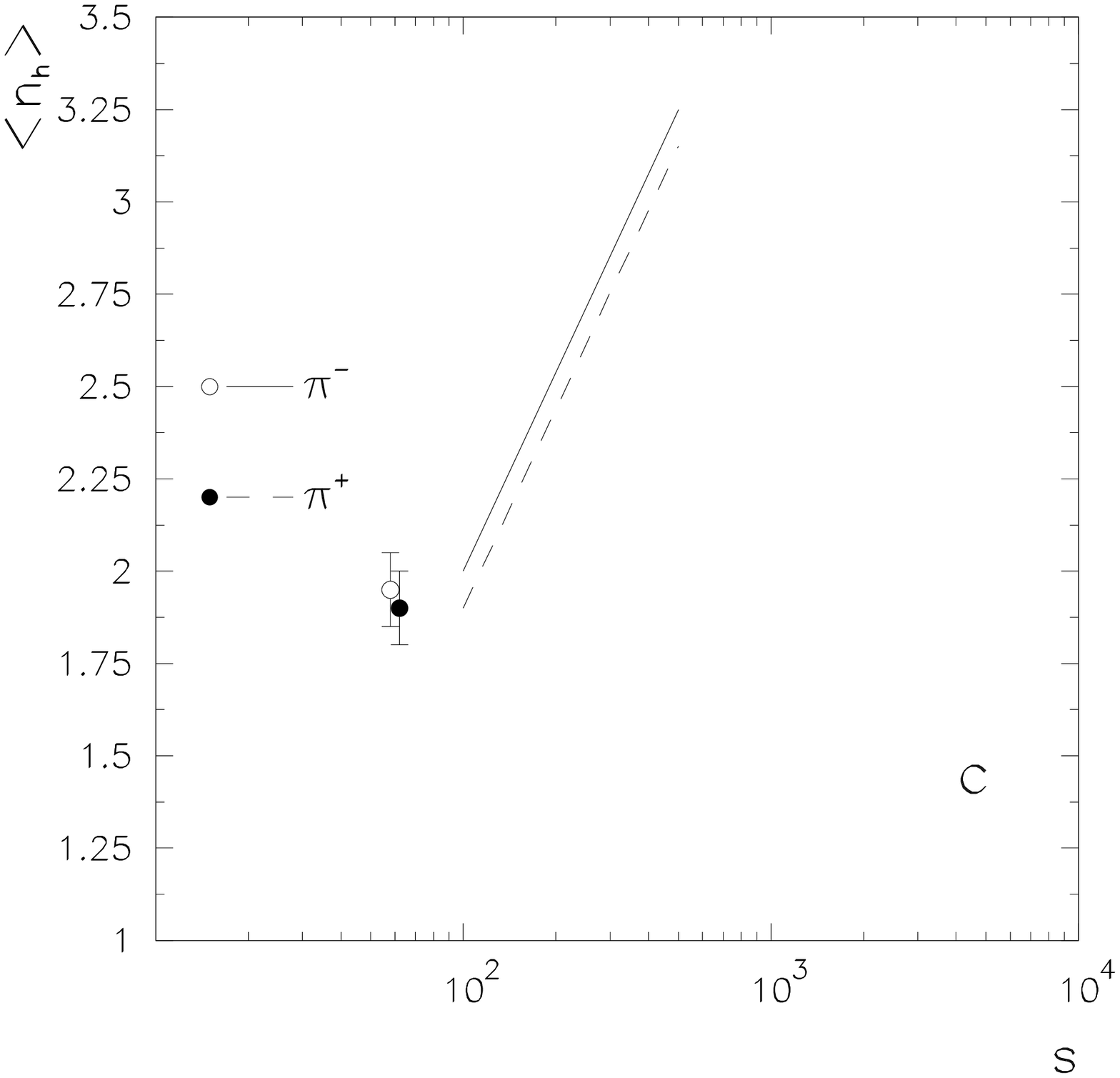,width=7.0cm}
            \epsfig{file=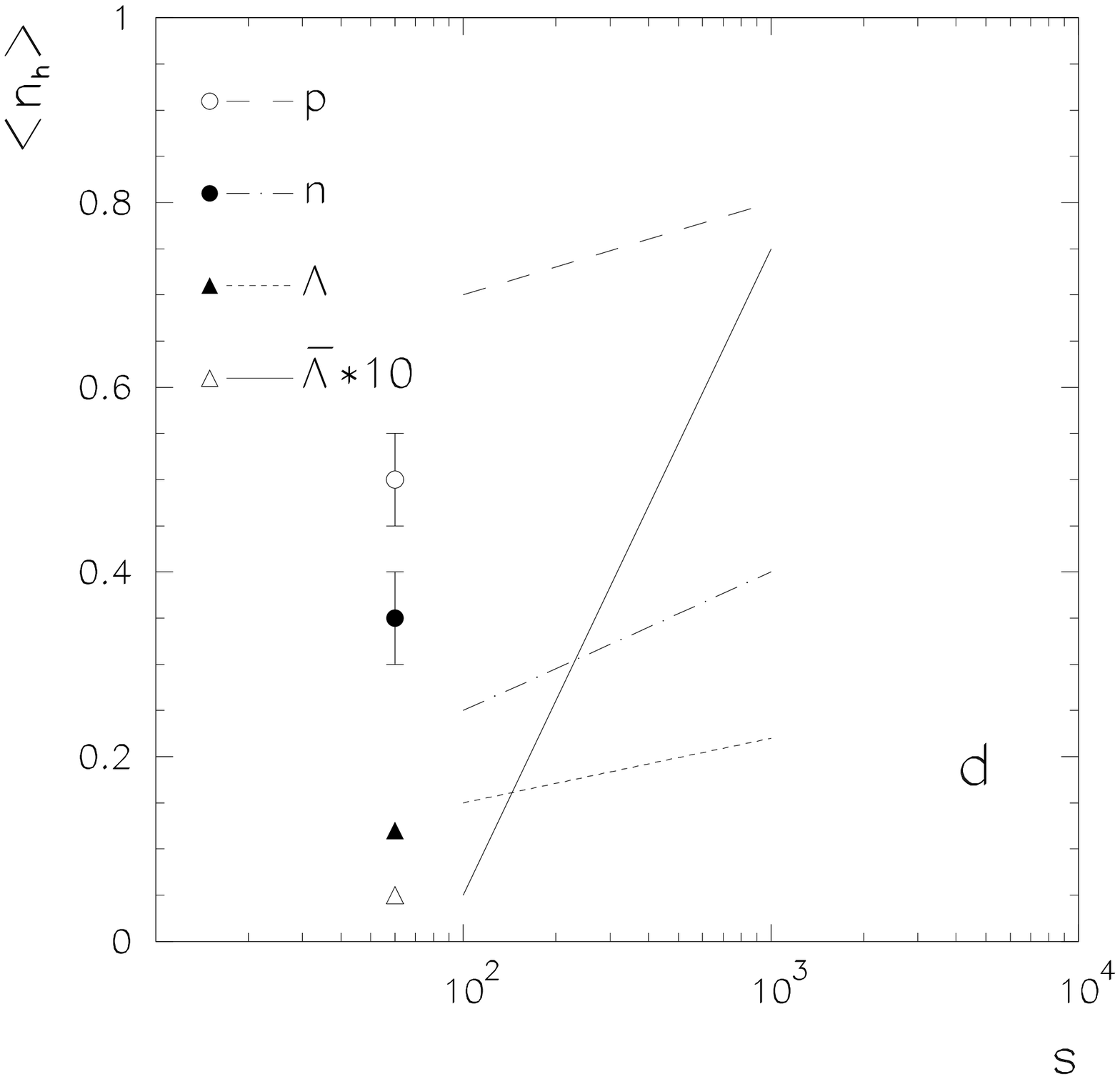,width=7.0cm}}
\centerline{\epsfig{file=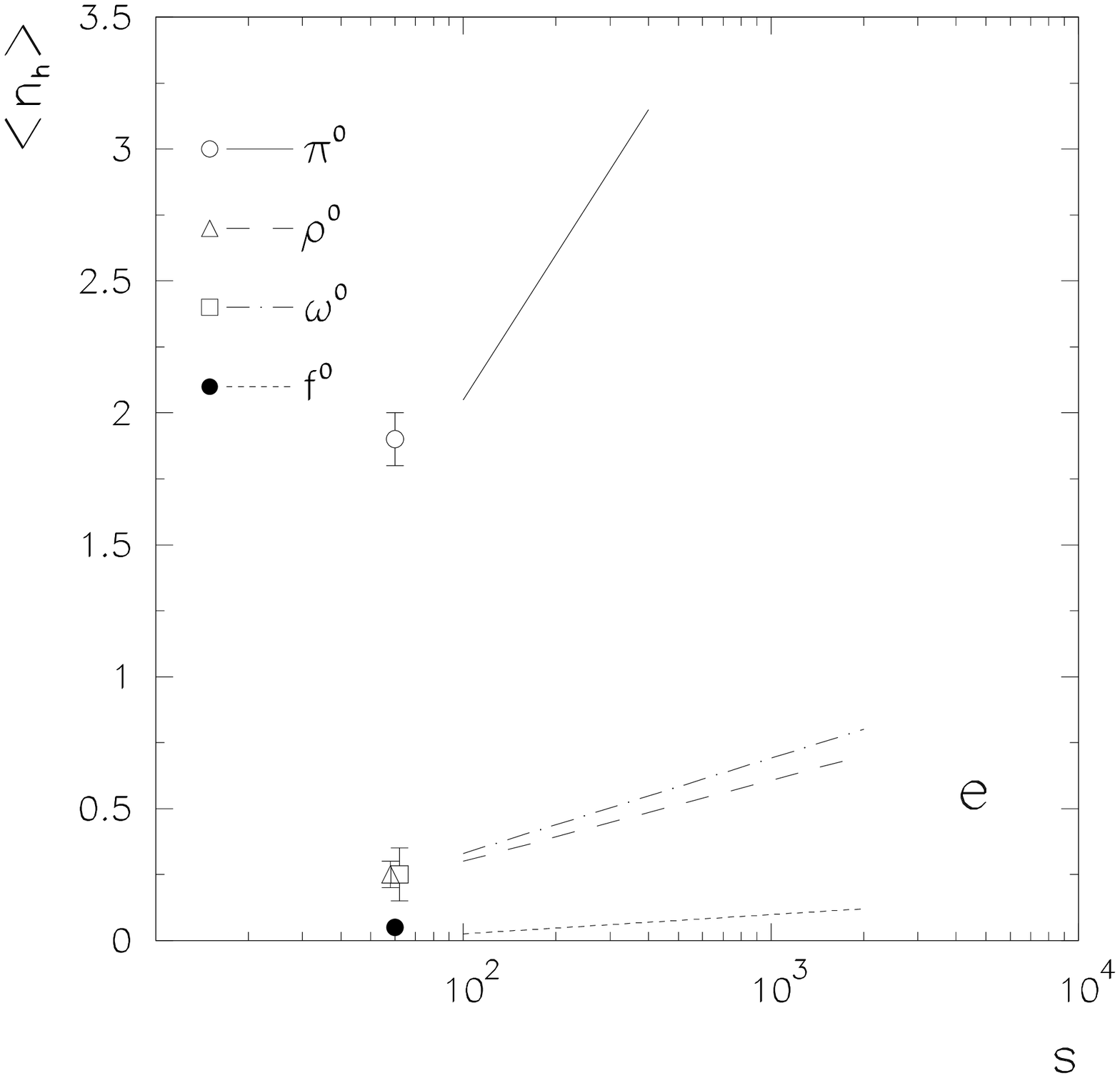,width=7.0cm}}
\centerline{Fig.12}
\end{figure}

As an example for a quite good agreement of the predictions of quark
combinatorics with the experimental data, let us consider the ratio
of secondaries with total quark spins $0$ and $1$. In the framework
of quark combinatorial calculus one assumes that in the multiparticle
production process a cloud of quarks and antiquarks with non-correlated
spin is formed. In such a cloud the ratio of the number of $q\bar{q}$
pairs with total quark spin $s_{q\bar{q}}=1$ and of those with
$s_{q\bar{q}}=0$ is $3:1$. Supposing that the mesons are formed by
quarks and antiquarks independently of their spin projections, this
ratio has to be also true for the produced mesons. The multiplicity of
meson states with $s_{q\bar{q}}=1$ is proportional to the multiplicity
of $s_{q\bar{q}}=0$ states as $3:1$. In hadron-hadron collisions this
relation is valid for both the fragmentational and the central regions.

The condition $3:1$ has to be fulfilled for mesons belonging to the same
$SU(6)$ multiplet. Examples for that can be the well-known relations
$\rho:\pi = K^*:K = 3:1$ for the directly produced mesons of the lowest
36-plet. Summing over all multiplets, we obtain
\begin{equation}
\label{23}
\frac{\sum_L<n_{M(L;s=1)>}}{\sum_L<n_{M(L;s=0)>}} \equiv \frac{V}{P} = 3.
\end{equation}
It is convenient to verify this relation on secondary $K$-mesons $K$,
$K^*(890)$, $K^*(1420)$, since strange particles appear as decay products
to a less extent than $\pi$-mesons.

Experimental data on $pp\rightarrow$ kaons (405 GeV/c, [41]) and
$K^-p \rightarrow$ kaons (32 GeV/c, [42]) provide a possibility to test
the condition (\ref{23}). (The results of the measurements are presented
in [9], Table 7.1.) The main contribution to
the cross section is given by the production of the $L=0$ multiplet.
Indeed, due to the combinatorial calculus the direct production of the
vector mesons is three times as large as that of the pseudoscalar mesons.
Thus, the total amount of the secondary mesons with $L=0$ is $4/3 V$.
The weight of tensor mesons in the $L=1$ multiplet is $5/12$, hence
$12/5 T$ mesons with $L=1$ are produced. The production of mesons with
$s_{q\bar{q}}=1$ in the multiplets with $L=0$ and $L=1$ is $V$ and
$9/5T$, respectively. (The value $V+\frac{9}{5}T$ which is the
contribution of mesons with $s_{q\bar{q}}=1$ in the $S$-wave and $P$-wave
multiplets is also given in [9],Table 7.1.) As it is seen from the data,
the experimental value is in each case around 75\% of the total cross
sections of kaons, in accordance with the predictions of quark
combinatorics.

\section{Quark combinatorics in hadronic $Z^0$ decays}\label{III}

Precision measurements [29] of hadron production in the $Z^0 \rightarrow
hadron $ decay allow us to clarify some key problems of multiparticle
production processes in quark induced jets. Indeed, in jets the
mechanism of hadronization manifests itself explicitly, and, hence, it
may be especially important to investigate multiparticle production in
such processes.

Having in mind the intense experimental search for exotic meson states,
it is desirable to obtain an independent verification of the basic
statements of quark combinatorics. The quark combinatorial calculus
is now widely used as a tool to investigate meson resonances of
masses 1.0-2.5 GeV in order to determine their quark-gluon content
(see, for instance, [33]-[35]). While in these investigations the
qualitative picture of the cloud structure of sea quarks is not
necessarily used, the notion of the "suppression parameter" for the
production of strange quarks continues to play an important r\^{o}le,
and it is identical to that used in multiparticle production processes.

We consider yields of vector ($\rho, \omega, \bar{K}^*$) and
pseudoscalar ($\pi, K$) mesons in hadronic $Z^0$ decays. There exists
rich experimental information about these processes, determined by
transitions $Z^0 \rightarrow q\bar{q} \rightarrow hadrons$. Light-flavour
mesons, produced in quark jets $Z^0 \rightarrow u\bar{u}$,
$Z^0 \rightarrow d\bar{d}$ and $Z^0 \rightarrow s\bar{s}$, are created
with probabilities in the proportion
\begin{equation}
\label{23a}
u\bar{u} : d\bar{d} : s\bar{s} = 0.26 : 0.37 : 0:37 .
\end{equation}
The large mass of the $Z^0$ boson enables us to observe in the
hadronic decays $Z^0 \rightarrow q\bar{q} \rightarrow hadrons$
the characteristic features of both multiparticle production (central
region of quark jets) and meson decay (fragmentation production)
processes.

Data given in [29] and [53]-[55] provide spectra of vector and
pseudoscalar mesons, $d\sigma/dx(Z^0 \rightarrow V+X)$ and
$d\sigma/dx(Z^0 \rightarrow P+X)$, in a broad interval of $x$, hence,
it becomes possible to compare them with the predictions of quark
combinatorics [32].

As we have seen, there are two types of relations which reflect different
aspects of the mechanism of multiparticle production. The first type
connects secondary particles belonging to the same $SU(6)$ multiplet.
These are, for example, relations between vector mesons and pseudoscalar
mesons. Assuming that quarks are created arbitrarily, without colour
correlation, quark combinatorics predict a production probability
proportional to the spin states.In the case of vector mesons this is
$V/P = 3$. These relations are valid, however, only for prompt particle
production and not for particles which are decay products of higher
resonances.

The second type of relations in quark combinatorics which can be
investigated in the decays of $Z^0$ bosons into hadrons is that between
secondary mesons and baryons. The hadronic decays of $Z^0$ bosons are
determined by processes like $Z^0 \rightarrow q\bar{q} \rightarrow
hadrons$ . We consider decays into light hadrons; as we have
seen, light quarks $u\bar{u}$, $d\bar{d}$ and $s\bar{s}$ are produced
with nearly equal probabilities, there are no distinguished flavours.

\subsection{Prompt production in the central region of the quark jet
and the vector-pseudoscalar ratio}\label{3.1}

The prompt verification of (\ref{6}) and (\ref{23}) is rather difficult,
since in multiparticle production processes a large number of resonances
is produced; we took this into account by (\ref{10}), (\ref{11}). One
can try to overcome the ambiguities related to the resonance production
by considering all existing resonances and their decays into all possible
channels. This is the scenario suggested in [46],[47]. However, there
are some problems. Indeed, the number of resonances observed and cited
in the compilation [48] is a comparatively small fraction of the whole
set of existing states. This can be seen, e.g., from recent investigations
of meson production data [49]-[51], where a large number of new meson
states with masses in the region 1950-2350 MeV is reported. It is obvious
that those resonances are first discovered which can be easily detected;
one should also have in mind that all observed resonances have multiplet
partners which are produced with approximately the same probabilities.
Their decays form a background which prevents us to check directly
(\ref{6}) and (\ref{23}). Another obstacle is the effect of accumulation
of widths of overlapping resonances, which has been seen for
scalar-isoscalar mesons in the region 1200-1600 MeV [51],[53]. As a
result of width accumulation, a broad state ($\Gamma/2 \sim 400 MeV$)
was formed; as it was said in [52], similar states can exist in other
waves, other mass regions. At the time being, it does not seem to be
possible to take into account the productions and decays of such broad
states. Still, the investigation of jet processes in the large $x$ region
opens a way to test the relations (\ref{6}) and (\ref{23}).

The decay processes increase the contribution of lighter particles, such
as pions and kaons in the case of mesons, and nucleons in the case of
baryons. However, in jet processes $Z^0 \rightarrow q\bar{q} \rightarrow
hadrons$ considered in [29], [53]-[55], the spectra reach the maximum
when the momentum $x$ carried by the hadron is small, and they decrease
rapidly when $x\rightarrow 1$. This leads to a rapid increase in the
contribution of the promptly produced particles with the increase of
$x$, since the decay products carry only a fraction of $x$ of the
initial resonance. It is just this feature which enables us to estimate
the probability ratios for promptly produced hadrons.

Our analysis shows that the ratio $V_{prompt}/P_{prompt}$ is the
same for the Pomeron ladder in hadron--hadron collisions and
for quark jets, despite of the different structure of the colour
exchanges in these processes.

Investigations of the QCD-Pomeron [56],[57] shed light on the
quark--gluon structure of the multiperipheral ladder in hadron
collisions and allowed us to deal with meson yields in the
central region on a new level.

When considering central production in the $Z^0 \to q\bar q \to
hadrons$ decay, we start with the standard mechanism of soft colour
neutralisation of the outgoing quarks: newly-born quark--antiquark
pairs are produced in multiperipheral ladder (see Fig.13a), which
provides the transfer of the colour from antiquark to quark. The
discontinuity of the self-energy diagram of Fig.13b (determined by
cutting through hadronic states, dashed line in Fig.13b) defines
the transition cross section $Z^0 \to hadrons$, while the
quark-gluon block inside the big quark loop determines the
confinement forces.
Similarly, the inclusive production cross section of the meson in
the central region is provided by the discontinuity of the diagram
of Fig.13c. The quark loop $meson\to q\bar q\to meson$ shown in
Fig.13c with the production of vector or pseudovector mesons
determines the relative probabilities of these particles.  The chain
of the quark loops shown in Figs.13b, 13c and 13d (below we denote
this chain as $A$) contains both colour singlet $(c=1)$ and colour
octet $(c=8)$ components: $A=A_1+A_8$. According to the rules of
$1/N$-expansion [19], the main contribution is due to the
octet component. The idea that the quark leaves the confinement
trap by the production of new quark--antiquark pairs is rather old;
(see, for example, [9], Sections 7 and 9, as well as [10].
Following Gribov's ideas in understanding the confinement mechanism
[8], we use the jet structure shown in Fig.13a assuming that the
$t$-channel exchange by quark is a constructive element of the jet.
\newpage
\begin{figure}
\centerline{\epsfig{file=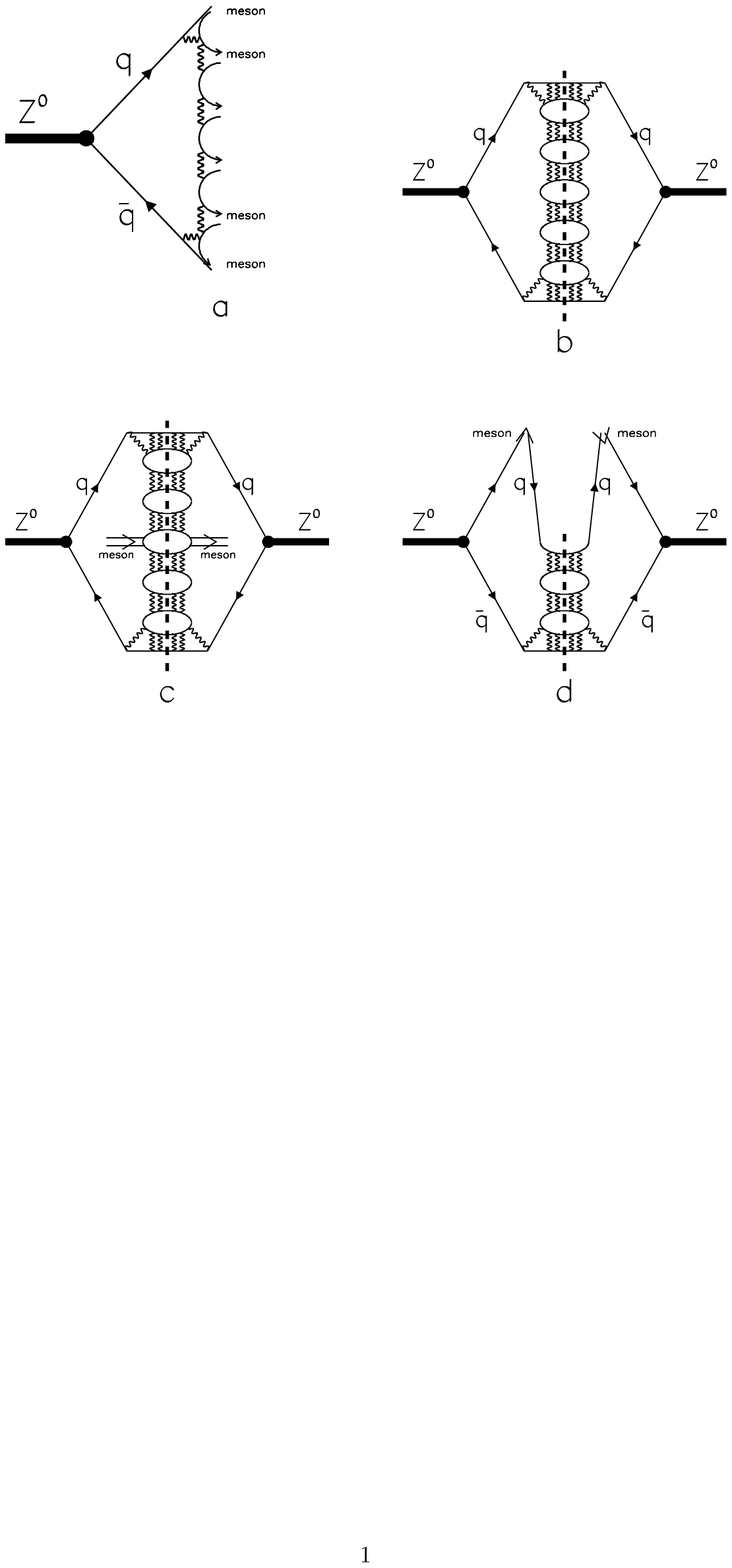,width=13.0cm}}
\centerline{Fig.13}
\end{figure}

Spectroscopic calculations (see, for example, [60]) support the
hypothesis about the scalar type of confinement forces; accordingly, we
assume that the chain $A$ realizes the $t$-channel exchange with
$J^P=0^+$.

The calculation of the block for central meson production proves that
Eq.(\ref{23}) is satisfied, if the wave functions of mesons ($V$ and $P$)
belonging to the same multiplet are equal.

Let us discuss the prompt production of vector and
pseudoscalar mesons (for definiteness, $\rho$ and $\pi$) in the
central region of a quark jet. The production cross section is
determined by the discontinuity of the diagram shown in Fig.13c;
it is redrawn in Fig.14a. A specific feature of the production
of $\rho$ and $\pi$ is the presence of a loop diagram, which is
shown separately by Fig.14b. Below, we calculate these loop
diagrams for $\rho$ and $\pi$ using the spectral integration
technique which is discussed in detail in [58],[59]. Within
this technique the loop diagrams are expressed in terms of the
$\rho$ and $\pi$ light--cone wave functions.

But first, let us present the result of our calculations.

Direct calculations lead to the following formulae for the
inclusive cross section of the  $\rho$ and $\pi$ mesons at $x\sim
0$:
\begin{eqnarray}
\label{24a}
&& \frac{d\sigma}{dx}(Z^0\to\rho
+X)=\frac1{16\pi^3}\int\limits^1_0 \frac{d\xi}{\xi(1-\xi)}\int
d^2{\bf k}_\perp \psi^2_\rho(\xi,{\bf
k}_\perp)\cdot 3\Pi_Z(W_1^2,W_2^2) \nonumber\\
&& \frac{d\sigma}{dx}(Z^0\to\pi +X)=\frac1{16\pi^3}\int\limits^1_0
\frac{d\xi}{\xi(1-\xi)}\int d^2{\bf k}_\perp \psi^2_\pi(\xi,{\bf
k}_\perp)\cdot \Pi_Z(W_1^2,W_2^2)
\end{eqnarray}
Here $\psi_{\rho}$ and $\psi_{\pi}$ are quark wave functions of
$\rho$ and $\pi$ mesons, $\xi$ and ${\bf k}_{\perp}$ are quark
light--cone variables (the fraction of momentum carried by a quark along
the $z$-axis and its momentum in the $(x,y)$-plane, respectively). In
(\ref{24a}) one can see explicit expressions related to the production of
$\rho$ and $\pi$ mesons. The rest (contributions from the large
quark loop as well as from ladder diagrams) is denoted in (\ref{24a})
as $\Pi_Z(W_1^2,W_2^2) $, which depends on the invariant energies
squared for the quark chains, $W_1^2$ and $W_2^2$. Multiperipheral
kinematics gives
\begin{equation}
\label{25a}
W_1^2 \; W_2^2 \simeq \xi(1-\xi)(m^2+k^2_{\perp})M^2_Z \; .
\end{equation}
Here $M_Z$ is the mass of the $Z^0$ boson.

The factor 3 in the $\rho$ production cross section is the result
of summation over polarizations of the vector particle.

Equation (\ref{24a}) demonstrates directly that, if quark wave functions
of $\rho$ and $\pi$ are identical (what is assumed by the quark
multiplet classification of these mesons), then at $x\sim 0$ we
have $d\sigma(Z^0\to\rho +X)/dx : d\sigma(Z^0\to\pi +X)/dx\, =\, 3:1 $.
Let us repeat that both (\ref{24a}) and the diagrams of
Figs.13c, 14a, 14b stand for promptly produced mesons.

\newpage
\begin{figure}
\centerline{\epsfig{file=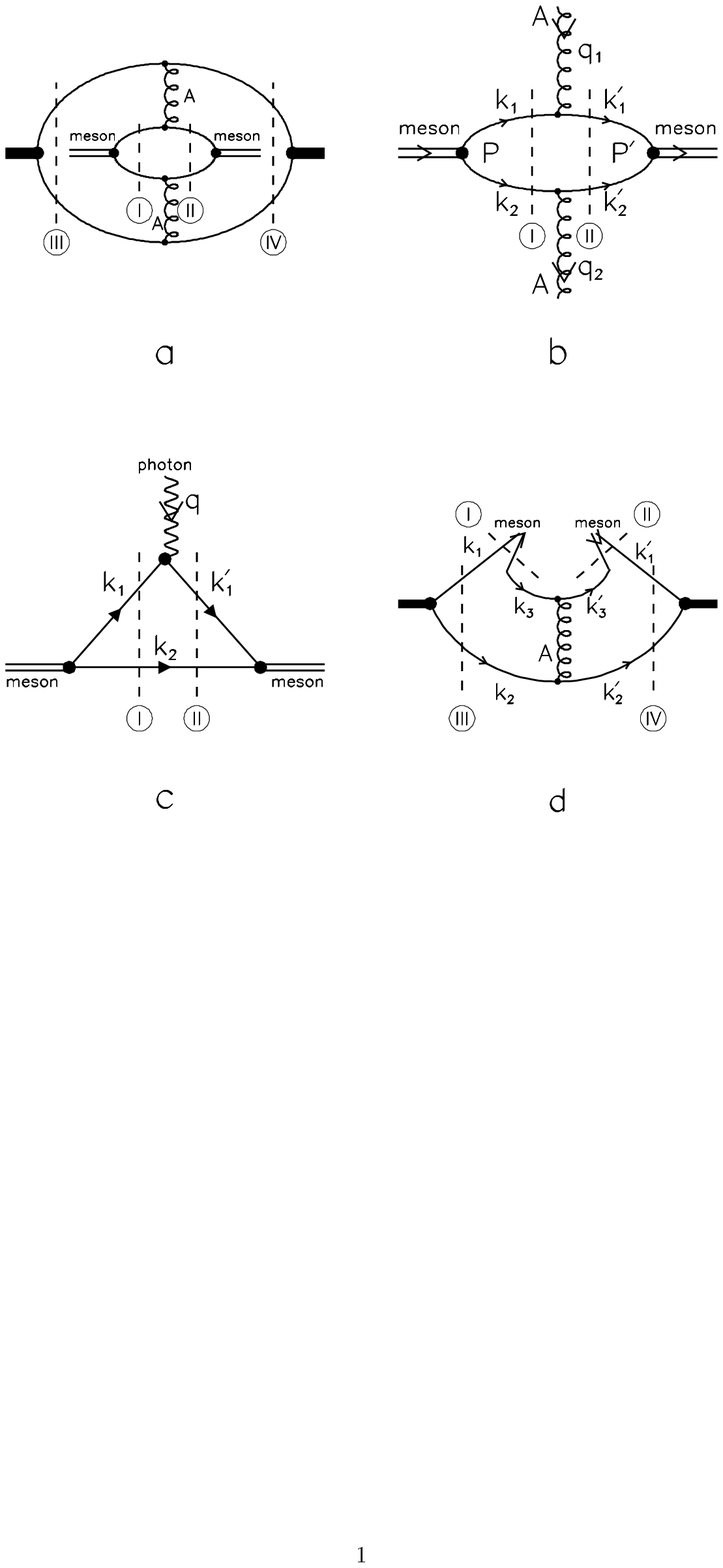,width=13.0cm}}
\centerline{Fig.14}
\end{figure}

Let us calculate the diagram of Fig.14a in terms of a spectral
integration (see also [61],[62]):
\\ (i)  quark loops in Fig.14a are taken
energy-off-shell; \\
(ii) for the energy-off-shell quark loops the discontinuities are
calculated (corresponding
cuts are shown by the dotted lines I, II, III and IV); \\
(iii) the spectral integrals are determined by the discontinuities
being the integrands.

First, consider the double spectral integral which corresponds to
the cuts III and IV, which are  the spectral integrals over
effective masses squared,  $M^2$ and $M'^2$, in the transitions
$Z^0 \to q\bar q$ and $q\bar q \to Z^0$:
\begin{eqnarray}
\label{26a}
\lefteqn{\int \limits_{4m^2}^{\infty}\frac {dM^2dM'^2}{\pi^2} \,
\frac{g_Z(M^2)g_Z(M'^2)}{(M^2-M_Z^2)(M'^2-M_Z^2)} } \nonumber\\
& \times & \int d\Phi(P_Z;p_1,p_2) d\Phi(P'_Z;p_3,p_4)
S_{Z}\; T(p;q_1,q_2) A(W^2_1,q^2_1)A(W^2_2,q^2_2) .
\end{eqnarray}
Here $g_Z(M^2)$ is the vertex function for the
$Z^0\to q\bar q$ transition and $M_Z$ is the $Z^0$ boson mass. The
factors $d\Phi(P_Z;p_1,p_2)$ and $ d\Phi(P'_Z;p_3,p_4)$ are  phase
spaces related to the cuts III and IV. $S_{Z}$ is the spin factor
for the big quark loop in Fig.14a. The amplitudes
$A(W^2_i,q^2_i)$ ($i=1,2$) refer to the quark-gluon chains with
the $t$-channel scalar quantum numbers, and $T(p;q_1,q_2)$ is the
block which corresponds to the small quark loop (the $q\bar q$ loop
for production of $\rho$ and $\pi$ mesons).

The characteristic feature of the spectral integral (\ref{26a}) is the
large value of $M_Z$. Because of that one can replace, with a rather
good accuracy, the poles of the spectral integrand by
half-residues:
\begin{equation}
\label{27a}
\frac{1}{M^2-M_Z^2}  -i\pi \delta (M^2-M_Z^2),
\frac{1}{M'^2-M_Z^2} \to i\pi \delta (M'^2-M_Z^2) \; .
\end{equation}
Equation (\ref{26a}) stands for the discontinuity of the
amplitude which results in different signs for the half-residues in
(\ref{27a}). Equation (\ref{27a}) means that the block inside of the
big quark loop can be considered, with a good accuracy, as a block of
the real process. This is a well-known feature of the high-energy
jets; below we use it for estimating meson production amplitudes.
The spectral integral (\ref{26a}), determined by the big quark loop, is
quite common for the $\pi$ and $\rho$ production. More interesting is the
spectral integral which corresponds to the amplitude $T(p;q_1,q_2)$
that is the $\rho$ or $\pi$ meson production block. This block is shown
separately in Fig.14b.

The amplitude of the loop diagram of Fig.14b represented as a
double dispersion integral is:
\begin{equation}
\label{28a}
{\mbox M} \ =\
T(p;q_1,q_2)2p_0(2\pi)^3\delta^{(3)}({\bf p}+{\bf q}_1-{\bf q}_2
-{\bf p}')\ ,
\end{equation}
\[T(p;q_1,q_2)\ =\int\limits^\infty_{4m^2}\frac{dsds'}{\pi^2}\int
d\phi \; \frac{G_{\rm meson}(s)}{s-\mu^2} \frac{G_{\rm
meson}(s')}{s'-\mu^2}\; S_{\rm meson}\ . \]

Here $\mu$ is the mass of produced meson, $s$ and $s'$ are
invariant masses squared in the intermediate $q\bar q$-states;
$G_{\rm meson}$ is the vertex for the $meson \to q\bar q$
transition. The ratio $G_{\rm meson}(s)/(s-\mu^2)$ determines the
wave function of the produced meson up to the spin factor, and
$S_{\rm meson}$ is the spin factor of the loop diagram Fig.14b.

In the approximation given by (\ref{27a}), when the jet block
inside of the big quark loop is considered as a real process, one
may fix $q_1 =q_2 =0$; then the inclusive cross section is
proportional  to $T(p;0,0)$. After some calculations,
the formula for $T(p;0,0)$ acquires a rather simple form:
\begin{equation}
\label{29a}
 T(p;0,0)\ =\ \frac1{16\pi^3}\int\limits^1_0 \frac{d\xi}{\xi(1-\xi)}
\int d^2k_\perp\ \left(\frac{G_{\rm meson}(s)}{s-\mu^2} \right)^2
S_{\rm meson}\ ,
\end{equation}
where $s=m^2_\perp/\left(\xi(1-\xi)\right )$.

The amplitude $T(p;0,0)$ alone does not determine the inclusive
cross section $d\sigma /dx (Z^0$ $\to {\rm meson} +X)$ because the
amplitudes $A(W^2_1,0)$ and $A(W^2_2,0)$ depend on $\xi$ and
$k^2_\perp $, see (\ref{25a}). Taking into account this
dependence, one has at $x\sim 0$:
\begin{eqnarray}
\label{30a}
\lefteqn{\frac{d\sigma}{dx}(Z^0\to {\rm meson} +X) \sim} \nonumber\\
& & \frac1{16\pi^3}\int\limits^1_0
\frac{d\xi}{\xi(1-\xi)} \int d^2k_\perp\ \left(\frac{G_{\rm
meson}(s)}{s-\mu^2} \right)^2 S_{\rm meson}\ \Pi (W^2_1, W^2_2) \;.
\end{eqnarray}
The spin factor $S_{\rm meson}$ is closely related to the

The spin factors $S_\rho$ and $S_\pi$ at $q_1=q_2=0$ are:
\begin{eqnarray}
\label{31a}
&&S_\pi =\ -{\rm Sp}\left(i\gamma_5(\hat k_1+m)(\hat
k'_1+m)i\gamma_5(-\hat k'_2+m)(-\hat k_2+m)\right) \nonumber\\
&& S_\rho=\ -{\rm Sp}\left(\gamma^\perp_\alpha(\hat k_1+m)(\hat
k'_1+m) \gamma^\perp_\alpha(-\hat k'_2+m)(-\hat k_2+m) \right)\ .
\end{eqnarray}
We have taken into account here that the quark--gluon ladder
carries quantum numbers of the scalar state, $J^P=0^+$; hence the
quark-ladder vertex is unity. The $\rho$-meson vertex is:
\begin{equation}
\label{32a}
\gamma^\perp_\alpha=g^\perp_{\alpha\alpha'}\gamma_{\alpha'}\ ,
\quad g^\perp_{\alpha\alpha'}=g_{\alpha\alpha'} -\frac{P_\alpha
P_{\alpha'}}{P^2}\ .
\end{equation}
In the spin factor $S_\rho$ the summation is performed over polarizations
of the meson. For the spin factors we have:
\begin{equation}
\label{33a}
S_\pi=8m^2s\ , \quad  S_\rho=16\,m^2(s+2m^2)\ .
\end{equation}
Let us now demonstrate that similar spin factors determine the
normalization of wave functions of the $\rho$-meson and the pion.

In the framework of the light-cone technique it is reasonable to
introduce the wave function of a particle and its normalization
using the form factor of the particle. The procedure of definition
of the wave function is discussed in detail in [58],[59].
Schematically, for the $q\bar q$ state this procedure looks as
follows.

The form factor of a composite system (for definiteness, we
consider the pion form factor) is determined by the triangle
diagram  Fig.14c, where the photon interacts with the composite
system. The form factor is represented as a double spectral
integral over masses of the incoming and outgoing pion;
corresponding cuttings are shown by the dashed lines I and II in
Fig.14c.
The structure of the amplitude of the triangle diagram for the
pion has the following form:
\begin{equation}
\label{34a}
A^{(tr)}_\nu\ =\ (p_\nu+p'_\nu)\,F_\pi(q^2)\ ,
\end{equation}
where $p$ and $p'$ are momenta of the incoming and outgoing pions, the
index $\nu$ refers to the photon polarization and $F_\pi(q^2)$ is the
pion form factor which, in terms of the double spectral representation,
can be written as
\begin{equation}
\label{35a}
F_\pi(q^2)=\int\limits^\infty_{4m^2}\frac{dsds'}{\pi ^2} \int
d\Phi^{(\rm tr)}(k_1,k'_1,k_2)\frac{G_\pi(s)}{s-\mu^2}\,T_\pi(s,s',q^2)\ .
\end{equation}
At $q^2=0$ one has $F_\pi(0)=1$. Direct calculations of equation
(\ref{35a}) in the limit $q^2\to 0$ give:
\begin{equation}
\label{38a}
1 = \int_{4m^2}^{\infty} \frac{ds}{\pi}\left(\frac{G_\pi(s)}{
s-\mu^2}\right)^2 \rho(s)\, S^{(wf)}_\pi(s)\ .
\end{equation}
Here $\rho(s)$ is the phase volume of the $q\bar q$ system
and $S_\pi^{(wf)}(s)$ is the trace of the quark loop diagram
for the pion.
Using light-cone variables we come to the following form of
(\ref{38a}):
\begin{equation}
\label{41a}
1\ =\ \frac1{16\pi^3}\int\limits^1_0\frac{d\xi}{\xi(1-\xi)}\int d^2{\bf
k}_\perp\left(\frac{G_\pi(s)}{s-\mu^2}\right)^2 2s\ ,
\end{equation}
where $s=(m^2+k^2_\perp)/\left (\xi(1-\xi)\right)$. This equation
enables us to introduce the pion wave function in the form
\begin{equation}
\label{42a}
\psi_\pi(\xi,{\bf k}_\perp)\ =\ \frac{G_\pi(s)}{s-\mu^2}\sqrt{2s}\ ,
\end{equation}
which is normalized by the standard requirement.

Likewise, we introduce the $\rho$-meson wave function: it is
defined by the form factor which is the spin matrix $F_{\alpha
\alpha'}(q^2)$. In problems that do not deal with polarization
properties of the vector particle, it is convenient to work with
the trace of the form factor matrix, $\sum_\alpha F_{\alpha
\alpha}(q^2)$, which is normalized by
\begin{equation}
\label{43a}
\sum_{\alpha=1,2,3} F_{\alpha\alpha}(0)\ =\ 3\ .
\end{equation}
The trace $\sum_\alpha F_{\alpha \alpha}(q^2)$ is determined by the
expression analogous to (\ref{35a}), with evident substitutions
$G_\pi \to G_\rho$ and $T_\pi \to T_\rho$. As a result, we obtain
the normalization for the averaged form factor:
\begin{equation}
\label{44a}
1\ =\ \frac13\sum_{\alpha=1,2,3} F_{\alpha\alpha}(0)\ =\
\int\limits^\infty_{4m^2}\frac{ds}\pi
\left(\frac{g_\rho(s)}{s-\mu^2}\right)^2 \rho(s) S^{(wf)}_\rho(s)\,
\end{equation}
where
\begin{eqnarray}
\label{45a}
\lefteqn{ S^{(wf)}_\rho(s)\ =}\nonumber\\
& & -\frac13\ {\rm Sp}\left[\left(\gamma_\alpha -P_\alpha\frac{\hat
P}{P^2}\right)(\hat k_1+m)\left(\gamma_\alpha -P_\alpha\frac{\hat
P}{P^2}\right)(-\hat k_2+m)\right] .
\end{eqnarray}
The $\rho \to q\bar q$ vertex, $ \gamma_\alpha -P_\alpha \hat
P/P^2$, selects three degrees of freedom of the $\rho$-meson. We
have :
\begin{equation}
\label{46a}
S^{(wf)}_\rho(s)\ =\ \frac43 (s+2m^2)\ .
\end{equation}
In the infinite momentum frame, (\ref{44a}) is re-written
as
\begin{equation}
\label{47a}
1\ =\ \frac1{16\pi^3}\int\limits^1_0
\frac{d\xi}{\xi(1-\xi)}\int d^2{\bf k}_\perp\psi^2_\rho(\xi,{\bf
k}_\perp)\ ,
\end{equation}
where
\begin{equation}
\label{48a}
\psi_\rho(\xi,{\bf k}_\perp)\ =\ \frac{G_\rho(s)}{s-\mu^2} \sqrt{\frac43
(s+2m^2)}\ .
\end{equation}

The normalization conditions for pion and $\rho$-meson wave
functions define unambigously the ratio of yields for prompt
production: $\rho/\pi=3$, if the wave functions of these mesons
are similar. Indeed, the expression (\ref{30a}) represented in
terms of wave functions $\psi_{\rho}$ and $\psi_{\pi}$ gives us
(\ref{24a}) immediately.

We have not taken into account explicitly the colour degrees of
freedom in the derivation. This, however, can be easily done. For the
$meson \to q\bar q$ vertex the colour operator is equal to $I/\sqrt{N_c}$,
where $I$ is a unity matrix in colour space. We have two colour
amplitudes, singlet and octet, for the chain of the quark loop
diagrams, $A_1$ and $A_8$. The couplings of the amplitudes $A_1$ and
$A_8$ to quarks ($g(A_1)$ and $g(A_8)$) are proportional to
$I$ and $\lambda$ (Gell--Mann matrices). All colour operators are
the same for both pion and $\rho$-meson production. Because of
that, the colour factors are not important for the ratio
$\rho/\pi$ -- they are identical and cancel in the production
ratio.

As was stated above, the main contribution into inclusive meson
production comes from the ladder diagram $A_8$. This is due to the
fact that the coupling constant for the amplitude with $c=8$ is
larger than for $c=1$. In terms of $1/N_c$ expansion
$g(A_1)/g(A_8) \sim 1/\sqrt{N_c}$.

\subsection{Inclusive production of mesons in the fragmentation region}
\label{3.2}

The equality (\ref{23}) is valid for prompt meson production, while the
decays of highly excited states violate this ratio, as is seen in
the experiment, providing the increase of the rate of light mesons
due to resonance decay. As to $\rho/\pi$ and $K^*/K$, the decays increase
the contribution of pseudoscalar component. According to [29],
$\rho^0/\pi^0=0.15\pm 0.03$, and $K^*/K=0.40\pm 0.06$. This indicates
a large contribution into spectra from the decays of highly excited
states.

One should stress that the ratios $V/P$ for beauty and charmed
mesons are saturated in the fragmentation region due to the
transitions $Z^0\to b\bar b$ and $Z^0\to c\bar c$. Therefore, in
Section 3.2 we re-analyse quark combinatorics for the fragmentation
region of the hadronic $Z^0$ decay.

The problem of the production and decay of highly excited states
in hadron--hadron collisions had been discussed in [61]. The
conclusion was similar: average multiplicities of the produced
light mesons and baryons result mainly from the cascade decays of
the highly excited resonances.

The existence of the  decay of highly excited resonances is a
reality that one should take into account in the verification of
quark combinatorial rules. We discuss several ways of solving this
problem.

One way is to check quark combinatorics for heavy particle yields,
where the cascade multiplication is suppressed. An ideal example
could be the production of mesons containing a $b$-quark. In fact,
for the beauty mesons the ratio $B^*/B$ observed in the experiment
agrees with (\ref{23}). When the lowest $S$-wave multiplet dominates in
the production of heavy mesons, then one has (provided the equation
(\ref{23}) is fulfilled): $B\simeq B_{prompt}+B^*_{prompt}=4B_{prompt}$,
and the ratio of vector and pseudoscalar mesons is $B^*/B\simeq 0.75$.

In experiments on $Z^0$ decay it was observed:  $B^*/B=0.771\pm
0.075$ [62], $0.72\pm 0.06$ [62], $0.76\pm 0.10$ [64],
$0.76\pm 0.09$ [66]; the mean value is $0.75\pm
0.04$.

For charmed mesons $D^*/D=0.60\pm 0.05$ [66], $0.62\pm 0.03$ [67],
$0.57\pm 0.05$ [68]. The mean value is $0.61\pm
0.03$, which means a rise of the contribution from the decay of the
non-$S$-wave multiplets.

The production cross section of mesons in the fragmentation region
is determined by the discontinuity of the diagram of Fig.13c (the
cutting of ladder diagram is shown by dashed line). Direct
calculations demonstate that (\ref{23}) is satisfied with a rather good
accuracy for the fragmentation region as well, provided the wave
functions of vector and pseudoscalar mesons are equal.

Investigations of meson production in the fragmentation region
open the way to test the rules of quark combinatorics for
light--flavour hadrons, and to verify (\ref{23}) in particular. As is
said above, in the spectra of light--flavour hadrons the
contribution of the component related to the decay of highly
excited states dominates. Still, in the case of jet processes this
component dominates in the central region, at $x\sim 0$, but not
in the fragmentation one. The hadronic spectra for jets are
maximal at $x\sim 0$, and they decrease rapidly with the growth of
$x$. As a result, the component which comes from a resonance decay
decreases quickly, because the decay products share the value
$x_{resonance}$, thus entering the region of smaller $x$. In due
course this effects a fast growth of relative contributions from
prompt particle production. Therefore, the measurements of
particle yields at $x\sim 0.5-1.0$ provide an opportunity for
model--independent testing of quark combinatorics.

The inclusive production cross section of mesons in the
fragmentation region is determined by the discontinuity of the
diagram of Fig.14d. The spectral representation for this diagram is
written as an integral over the masses of initial and final $q\bar
q$ states in the transitions $Z^0\to q\bar q$ and $q\bar q\to Z^0$
and over $q\bar q$ masses in the transitions $q\bar q\to meson$
and $meson\to q\bar q$. The amplitude of the diagram of Fig.14d
reads:
\begin{eqnarray}
\label{49a}
&& \int \limits_{4m^2}^{\infty} \frac{dM^2dM'^2}{\pi^2}\
\frac{g_Z(M^2)g_Z(M'^2)}{(M^2-M^2_Z) (M'^2-M^2_Z)}\int
\limits_{4m^2}^{\infty} \frac{dsds'}{\pi^2}\psi_{\rm
meson}(s)\psi_{\rm
meson}(s')  \nonumber\\
&\times&\;  d\phi_3(k_1,k_2,k_3)d\phi_3(k'_1,k'_2,k'_3)
A(W^2,(k_2-k'_2)^2) \frac{S^{(fr)}_{\rm meson}}
{\sqrt{S^{(wf)}_{\rm meson}(s)S^{(wf)}_{\rm meson}(s')}}\ .
\end{eqnarray}
The vertices $g_Z(M^2)$ and $g_Z(M'^2)$ are written for the
$Z^0\to q\bar q$ and $q\bar q\to Z^0$ transitions. Spectral
integrals over $s$ and $s'$ stand for $q\bar q\to meson$ and
$meson\to q\bar q$ (where $meson$ means $\pi,\rho$). The wave
function $\psi_{\rm meson}$ of the produced meson was introduced
in an explicit form in Section 3.1 for the pion and the
$\rho$-meson, and the factors $d\phi_3(k_1,k_2,k_3)$ and
$d\phi_3(k'_1,k'_2,k'_3)$ define the integration over phase spaces
in the left- and right-hand parts of the diagram of Fig.14d:
\begin{eqnarray}
\label{50a}
&& d\phi_3(k_1,k_2,k_3)= \frac12\, \frac{d^3k_1}{(2\pi)^3 2k_{10}}
\frac{d^3k_2}{(2\pi)^32k_{20}}\,  \nonumber \\
&&\times (2\pi)^4\delta^{(4)}\left(\tilde P-k_1-k_2\right)\  \frac12\,
\frac{d^3k_3}{(2\pi)^32k_{30}}\, (2\pi)^4\delta^{(4)}(P-k_1-k_3) .
\end{eqnarray}
Here $\tilde P^2=M^2$ and $ P^2=s$.

The block $A\left (W^2, (k_2-k_2')^2\right ) $ defines the
multiperipheral ladder (wavy line in Fig.14d). This block depends
on the momentum transfer squared $(k_2-k'_2)^2$ and the total
energy squared $W^2$:
\begin{equation}
\label{51a}
W^2\ \simeq\ M^2_Z(1-x)\ ;
\end{equation}
$x$ is the momentum fraction carried by the produced meson:
$x=2p/M_z$, where $p$ is the longitudinal component of meson
momentum, $p_{\rm meson}=(p+\mu^2_\perp/2p, 0,p)$.

The spectra $d\sigma(Z^0\to meson+X)/dx$ fall rapidly with
increasing $x$: this decrease is governed by $A\left
(W^2,(k_2-k'_2)\right )$.

All the characteristics of (\ref{49a}) listed above are the same
for both pion and $\rho$-meson production, the wave functions
$\psi_{\pi}$ and $\psi_{\rho}$ are also supposed to be the same.
The difference may be contained in the spin factors $S^{(fr)}_\pi$
and $S^{(fr)}_\rho$ which are
\begin{eqnarray}
\label{52a}
S^{(fr)}_\pi &=&(-){\rm
 Sp}\bigg[\gamma'^\perp_\nu(1+R\gamma_5)(\hat k'_1+m)i\gamma_5
(\hat k'_3+m) (\hat k_3+m) \nonumber\\
&\times &\;  i\gamma_5(\hat k_1+m)\gamma^\perp_\nu
(1+R\gamma_5)(-\hat k_2+m)(-\hat k'_2+m)\bigg];
\nonumber\\
S^{(fr)}_\rho&=&(-){\rm
 Sp}\bigg[\gamma'^\perp_\nu(1+R\gamma_5)( \hat
k'_1+m)\gamma'^\perp_\alpha(\hat k'_3+m)(\hat k_3+m)
\nonumber\\
&\times &\; \gamma^\perp_\alpha (\hat
k_1+m)\gamma^\perp_\nu(1+R\gamma_5)(-\hat k_2+m)(-\hat
k'_2+m)\bigg].
\end{eqnarray}
Here $R$ is determined by the ratio $g_A/g_V$
and the factor $\gamma_\nu^\perp (1+R\gamma_5)$ is related to the
vertex $Z^0 \to q\bar q$ which is determined by the vector and
axial--vector interactions (the ratio of coupling constants is
$\sim 2.63$ for the $u$ quark
and $\sim 1.43$ for the $d$ quark).

After some calculations (for details see [32]) we get for the light
mesons
\begin{equation}
\label{53a}
\frac{S^{(fr)}_\pi}{S^{(wf)}_\pi}\simeq2M^2_Z(1+R^2)\ , \quad
\frac{S^{(fr)}_\rho}{S^{(wf)}_\rho}\simeq6M^2_Z(1+R^2)\ ,
\end{equation}
which gives the ratio of the prompt yields
$\rho:\pi=3:1$ in the fragmentation region $x\sim 0.5-1$ (more
generally, $V:P=3:1$ for hadronic decays $Z^0\to q\bar q$ with
$q=u,d,s$). The same ratio appears for the production of heavy quarks
$Z^0=Q\bar Q$, where $Q=c,b$. For example, in the case of $b$
quark the spin factors are:
\begin{eqnarray}
\label{54a}
&& \frac{S^{(fr)}_B}{S^{(wf)}_B}\ \simeq\
2\left[M^2_Z+2m^2_b+R^2(M^2_Z-4m^2_b)\right]\ , \nonumber\\
&& \frac{S^{(fr)}_{B^*}}{S^{(wf)}_{B^*}}\ =\
6\left[M_Z^2+2m^2_b+R^2( M^2_Z-4m^2_b)\right]\ ,
\end{eqnarray}
and thus the ratio $B^*_{prompt}:B_{prompt}$ also equals 3.

We have seen that $V/P=3$ in the fragmentation region as well as
in the central region. However, in the central region the
comparison of quark combinatorics with experiment is hampered by
the presence of a number of the decay products of highly excited
resonances, while in the fragmentation region this contribution is
suppressed by rapidly decreasing spectra. This means that the
fragmentation region allows us to perform a model--independent
verification of quark combinatorics. We compare the calculated ratio with
ALEPH data [29]. For meson spectra at $x\sim 0.2-0.8 $ we have fitted
the spectra $(1/\sigma_{tot})d\sigma/dx$ to the sum of exponents
$\sum C_ie^{-b_ix}$; the calculation results are presented in Fig.15
for $\pi^\pm, \pi^0, \rho^0$ and $(p,\bar p)$.
The ratio of fitting curves drawn with calculation errors (shaded area)
is shown in Fig.16a for $\rho/\pi$. We see that for $0.6 < x <0.8$
the data are in reasonable agreement with the prediction $\rho/\pi=3$.
Figures 16b,c demonstrate the ratios $K^{*0}/K^0$ and $K^{*\pm}/K^\pm$:
the data do not contradict the prediction, though the errors are too
large to conclude anything more definite.

\newpage
\begin{figure}
\centerline{\epsfig{file=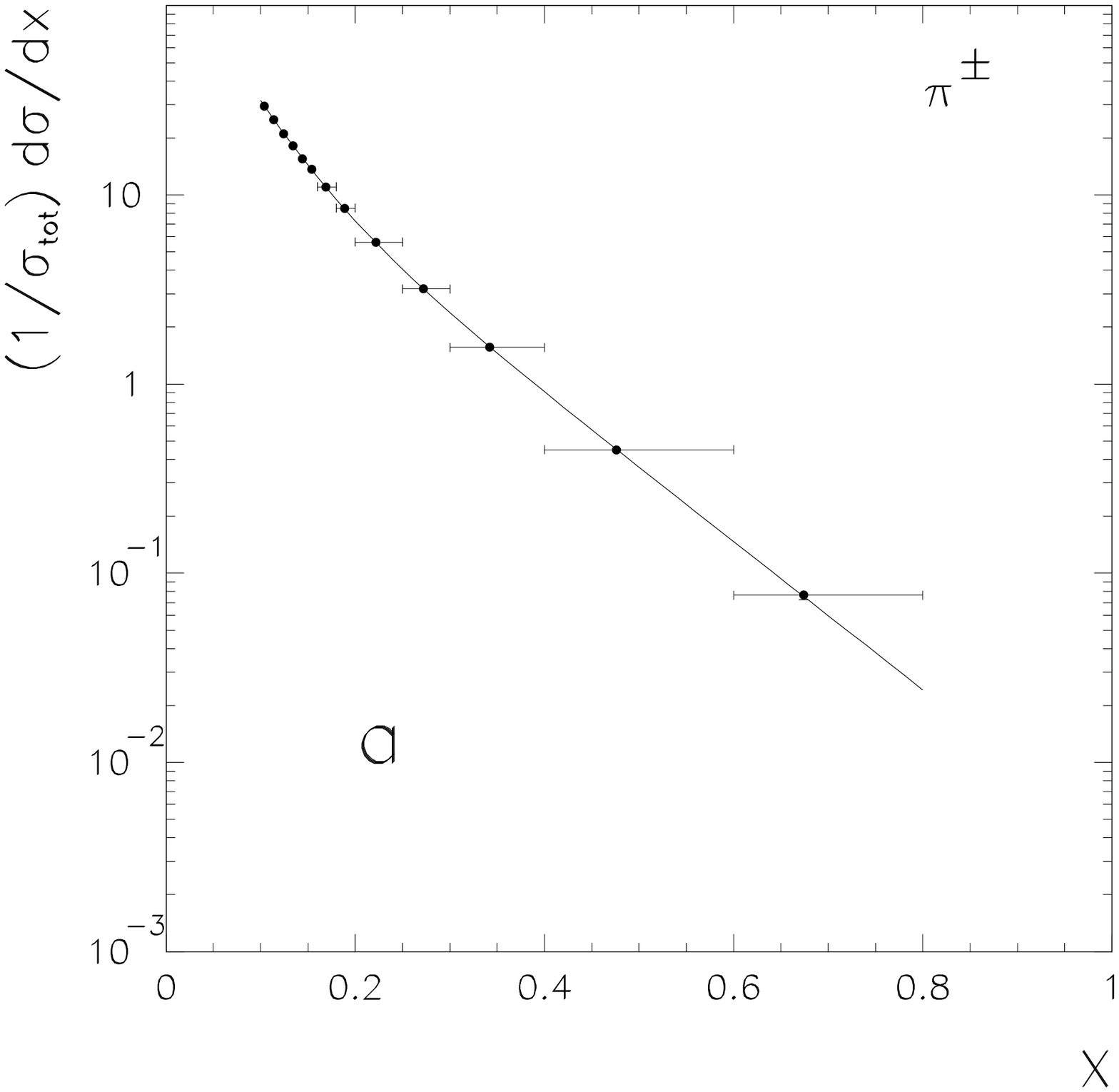,width=9cm}
            \epsfig{file=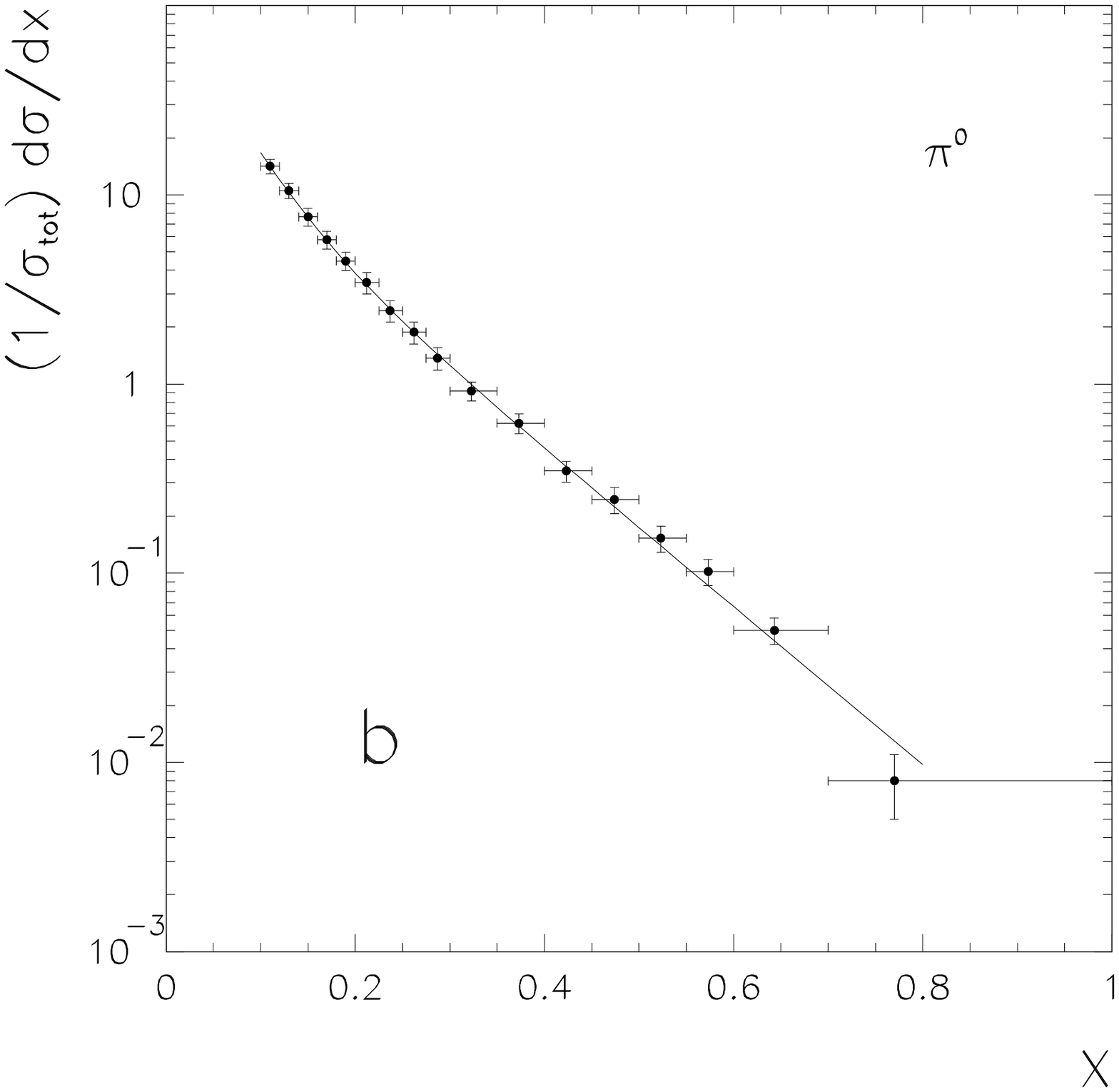,width=9cm}}
\centerline{\epsfig{file=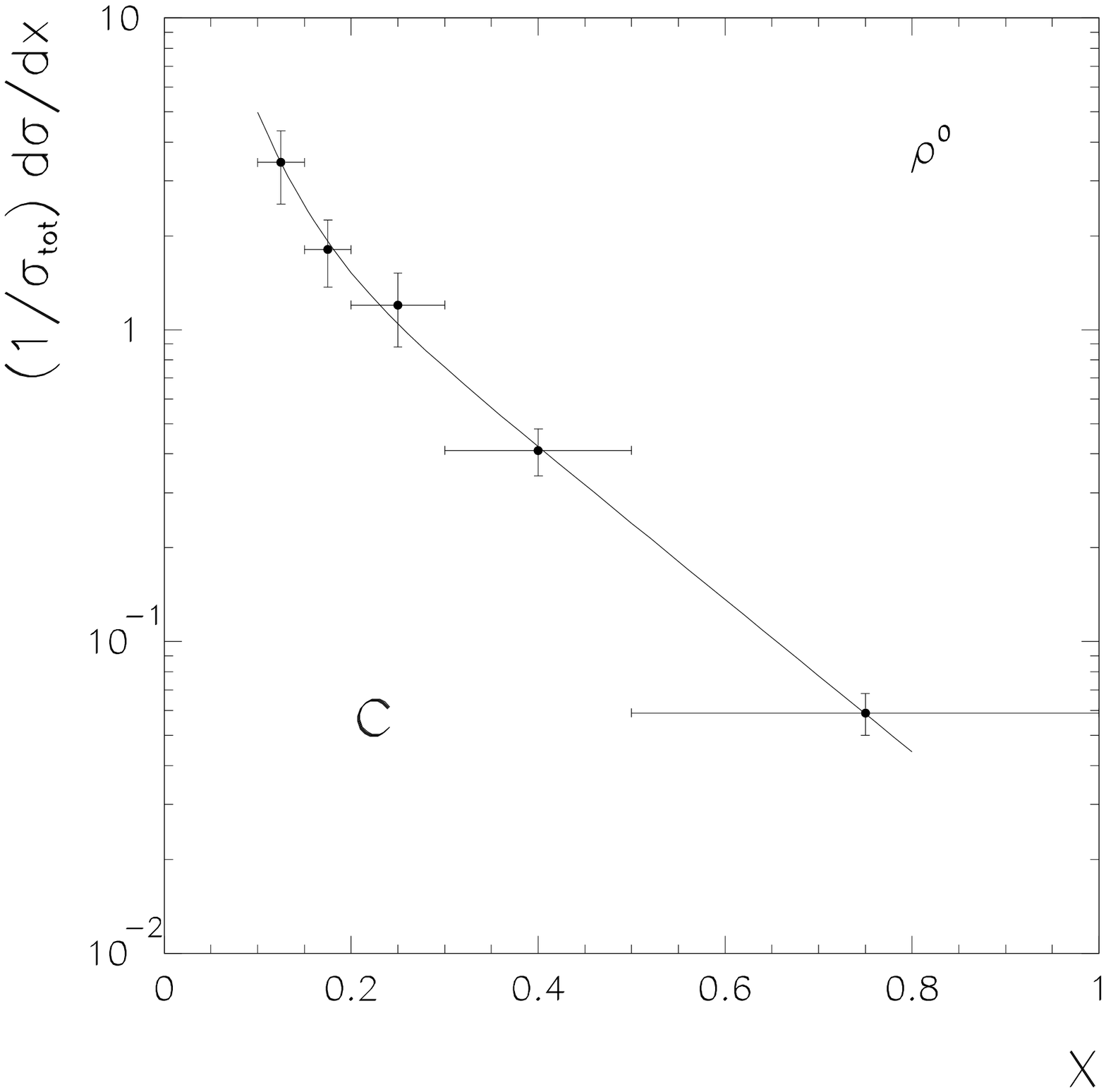,width=9cm}
            \epsfig{file=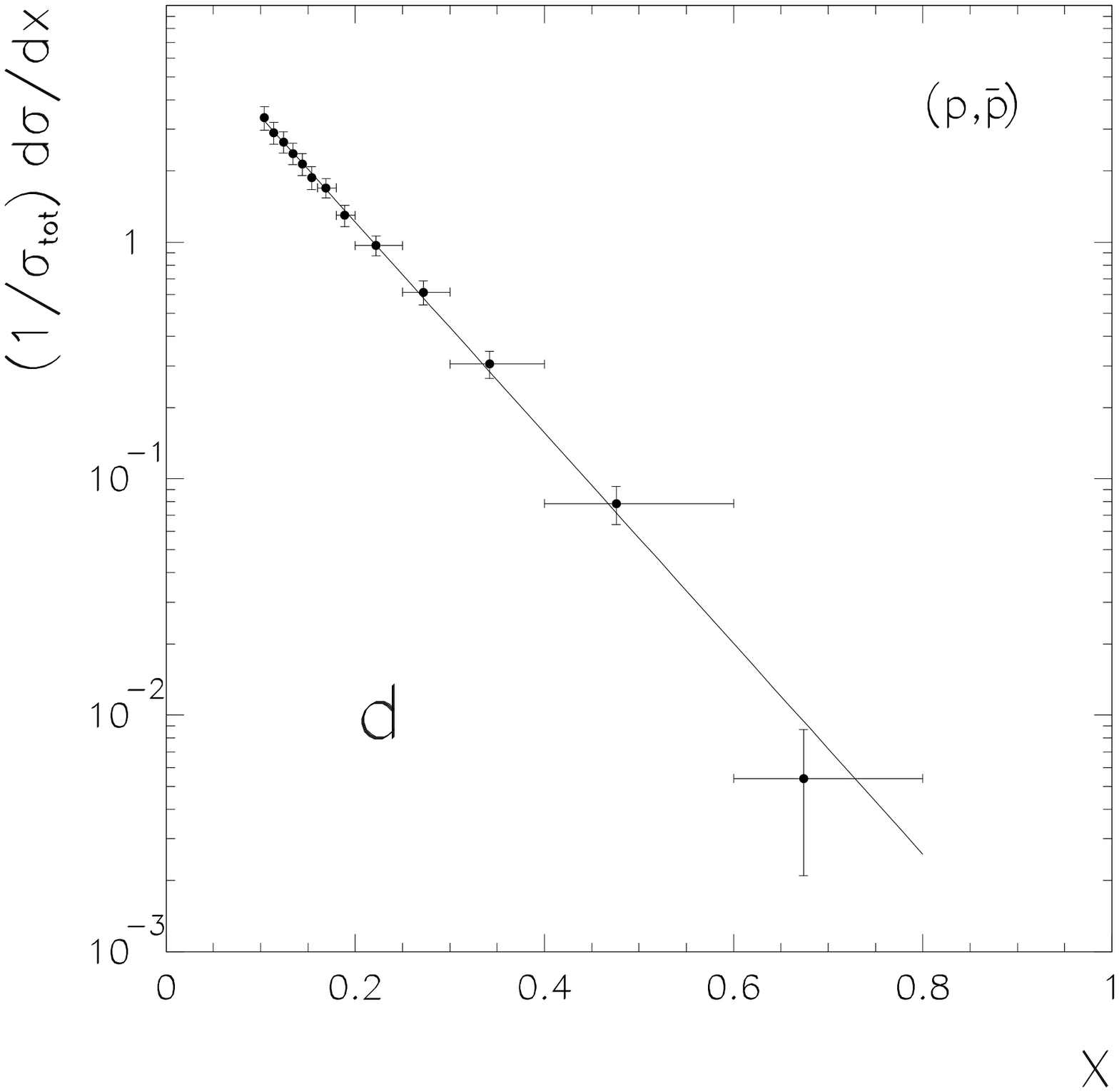,width=9cm}}
\centerline{Fig.15}
\end{figure}

\newpage
\begin{figure}
\centerline{\epsfig{file=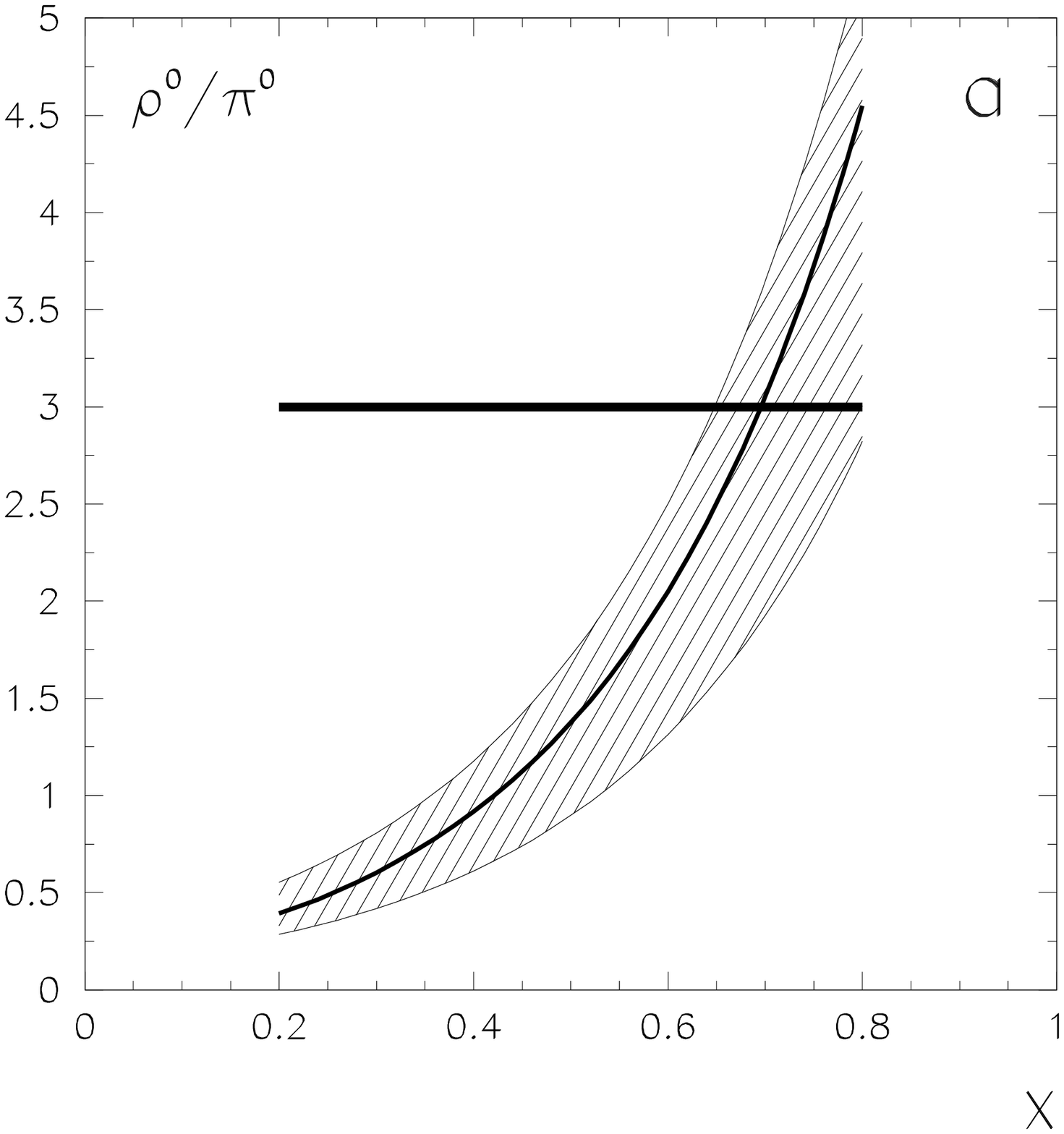,width=7.5cm}
            \epsfig{file=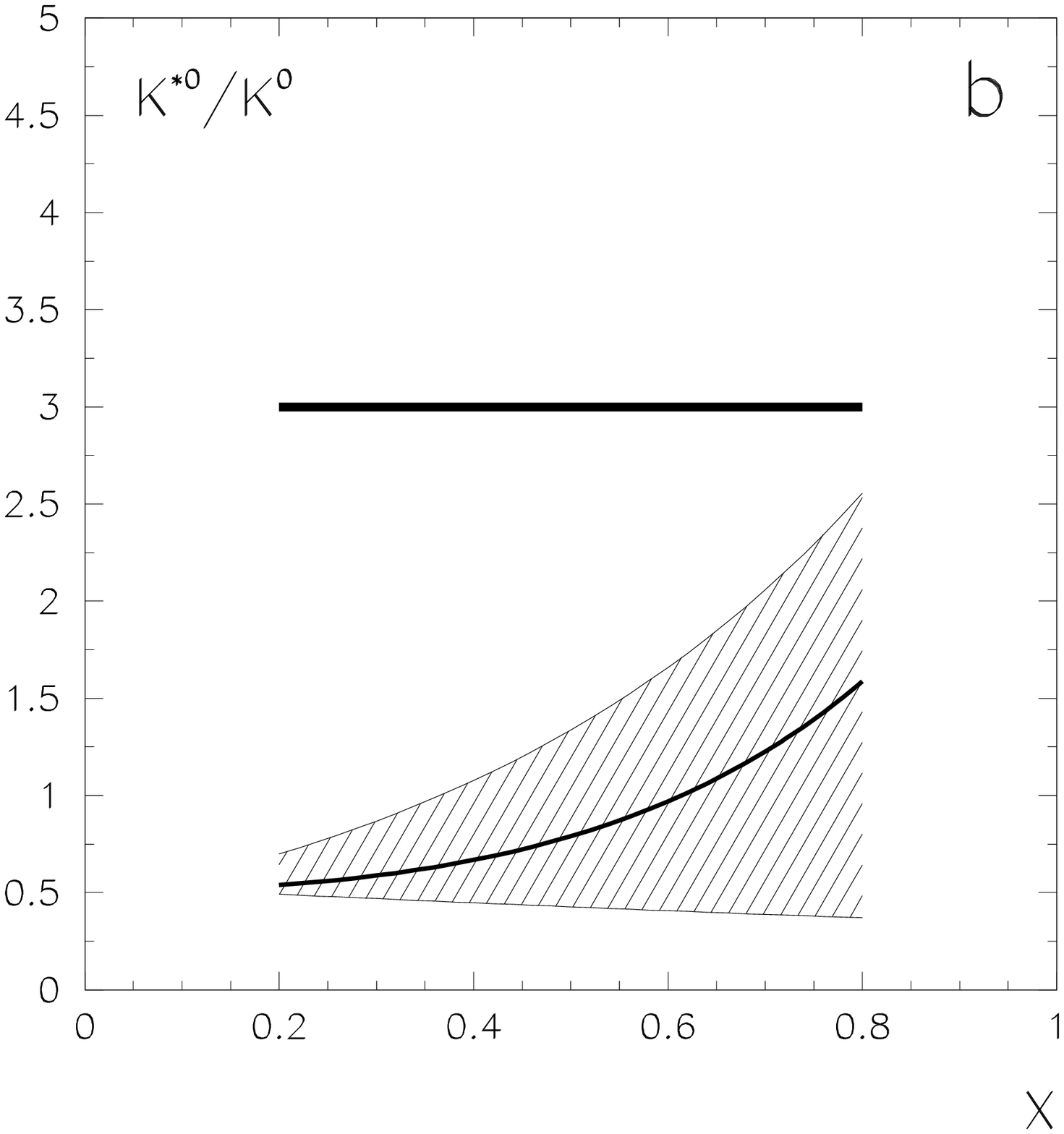,width=7.5cm}}
\vspace{-0.5cm}
\centerline{\epsfig{file=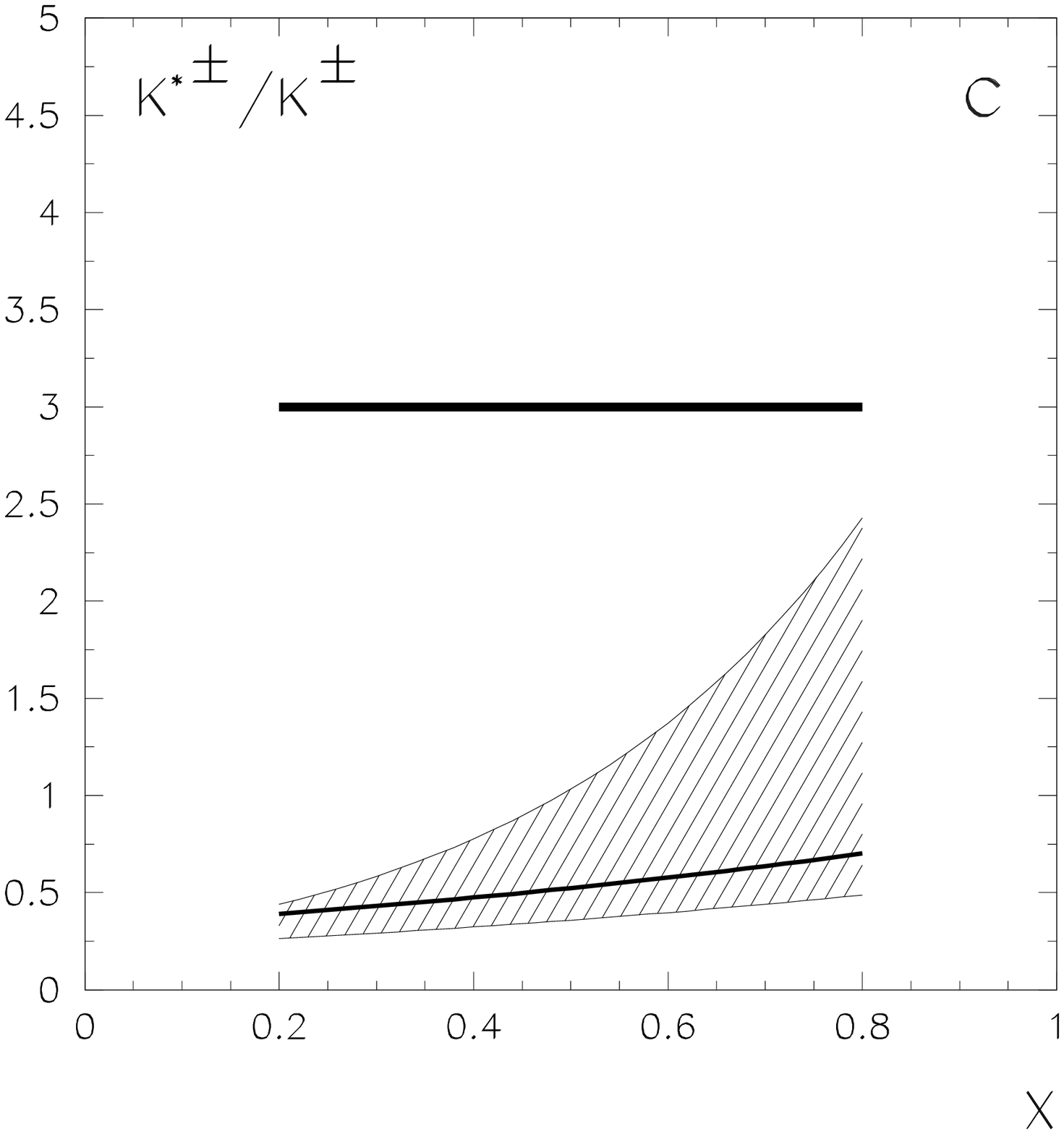,width=7.5cm}
            \epsfig{file=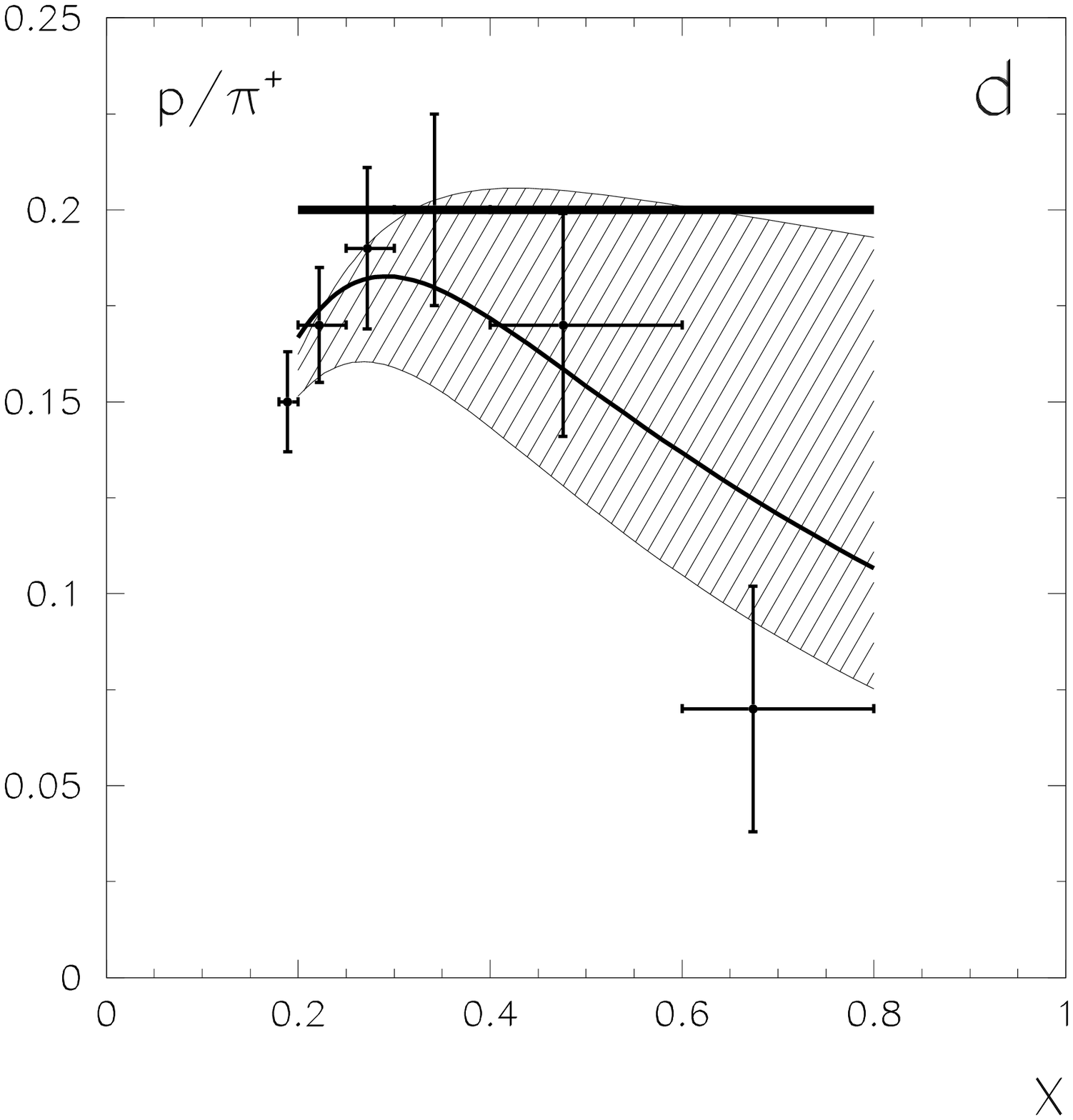,width=7.5cm}}
\vspace{-0.5cm}
\centerline{\epsfig{file=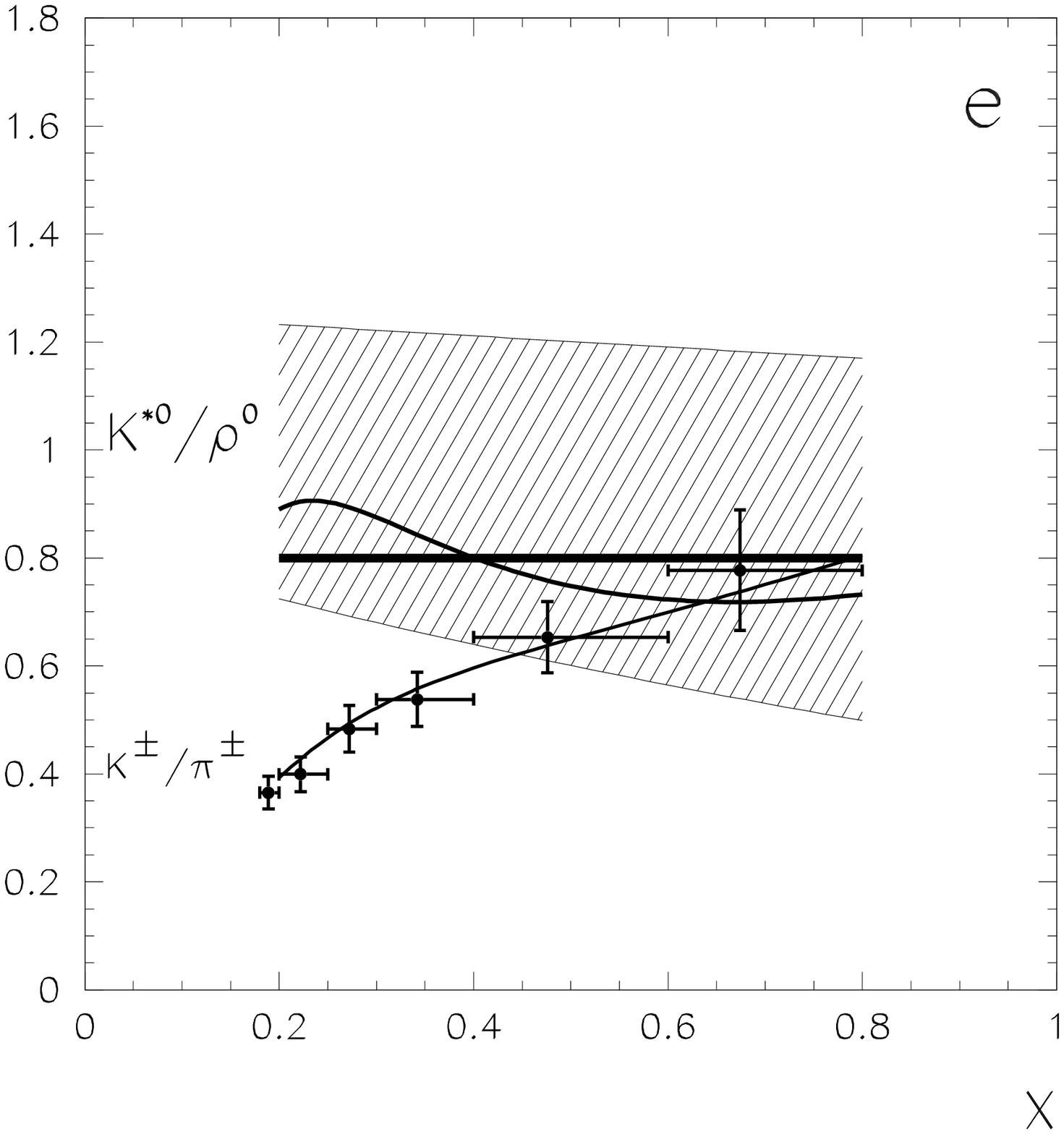,width=7.5cm}
            \epsfig{file=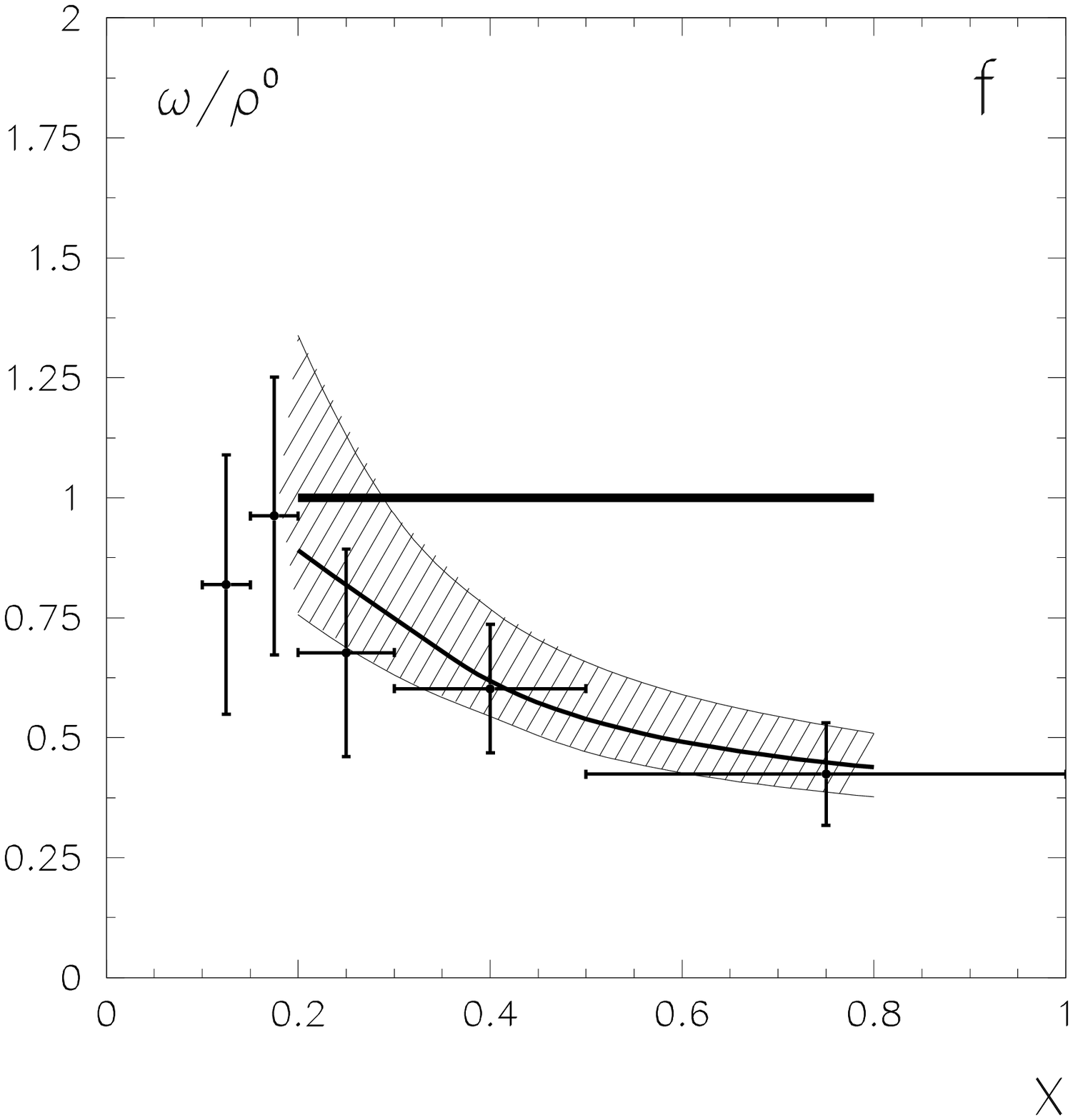,width=7.5cm}}
\centerline{Fig.16}
\end{figure}

\subsection{Suppression of the strange and heavy quark productions}
\label{3.3}

In hadronic multiparticle production processes (in jet processes
of the type of $Z^0\to hadrons$ or in hadron--hadron collisions at
high energies) the production of strange quarks is suppressed.
Strong suppression is observed for the production of heavy quarks
$Q=c,b$. One can assume that this suppression, being of the same
nature for different reactions, is related to the mechanism of the
production of new quarks at large separations of colour objects.
This mechanism is seen in its pure form in the two--particle
decays (the corresponding diagram is shown in Fig.17a).  The
block of the production of a new $q\bar q$ pair in the
two--particle  decay is the same as that of meson production in
jet processes (Fig.17b), so it is reasonable to suppose that the
suppression mechanism of the production of new quarks is similar
for these processes.

The decay of the $q\bar q$ state takes place as follows: the
quarks of the excited state leave the region where they were kept
by the confinement barrier, and at a sufficiently large separation
a new quark--antiquark pair will be produced inevitably: together
with the incident quarks, these new quarks then form mesons (i.e.
free particles). Schematically, this process (which is the leading
one in terms of the $1/N$ expansion) is shown on the diagram of
Fig.17a: two quarks fly away (with the momenta $p_1$ and $p_2$),
and at large quark separations the gluonic field produces a new
$q\bar q$ pair (the quarks with momenta $k_2$, $k_3$); then the
primary quark (now with momentum $k_1$) joins the newly-born one
($k_2$) creating a meson. Similarly,  another newly-born quark
($k_3$) joins the other primary quark (now with momentum $k_4$)
producing the second meson.

The block with the quark-antiquark production, that is the
transition
\begin{equation}
\label{55a}
q(p_1)+\bar{q}(p_2) \to q(k_1) + \bar{q}(k_2)+q(k_3)+\bar{q}(k_4)
\end{equation}
is the key process that
determines the decay physics; it is shown separately in Fig.17c.
The process (\ref{55a}) is responsible for the fact that quarks leave
the confinement trap. For modeling this, quark combinatorics uses the
hypothesis of soft hadronization.  It suggests that in the ladder of
produced quarks (Figs.13a, 17c) the contribution comes from small
momentum transfers (of the order of $R^{-2}_{confinement}$). In
the framework of the space--time picture this means that new
$q\bar q$ pairs are produced at large separations, when colour
objects leave the confinement well.
\newpage
\begin{figure}
\centerline{\epsfig{file=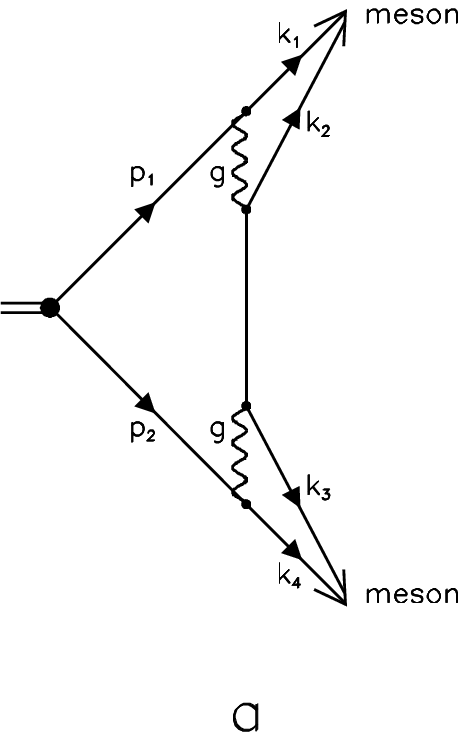,width=5.0cm}\hspace{1cm}
            \epsfig{file=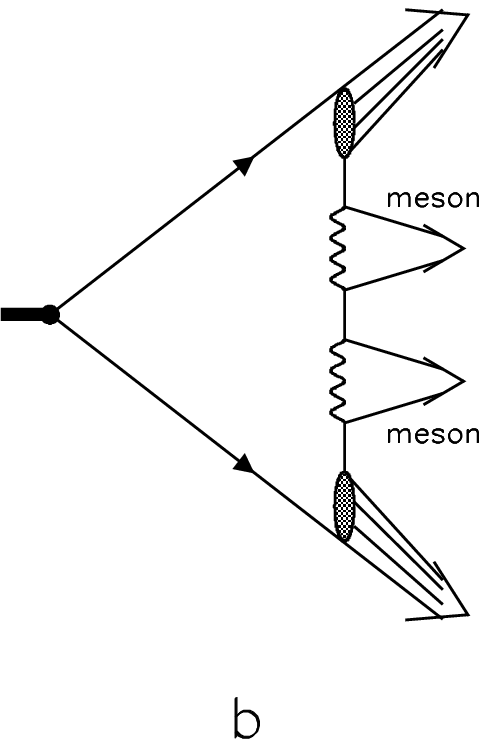,width=5.0cm}\hspace{1cm}
            \epsfig{file=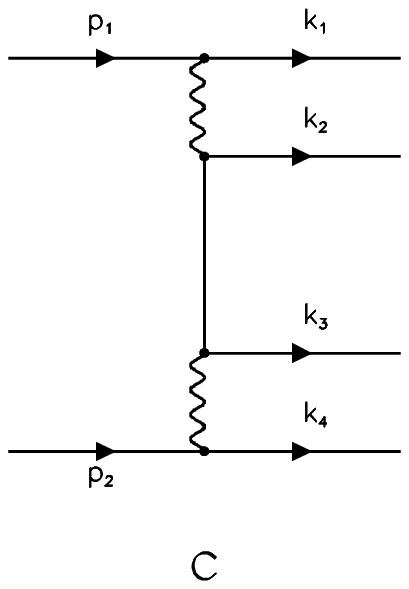,width=5.0cm}}
%\centerline{\epsfig{file=f5c_ann.eps,width=5.0cm}}
\centerline{Fig.17}
\end{figure}

The soft hadronization hypothesis applied to the decay processes
treats the ladder diagram of Fig.17c for the decay amplitude of
Fig.17a in the same way as for jet production, Fig.17b: process
(\ref{55a}) is an elementary subprocess both for the high--energy
ladder and the two--particle decay amplitude, and the momentum
transfers which enter the amplitude (\ref{55a}) appear to be
small in the hadronic scale, $\sim R^{-2}_{confinement}$.

Let us consider in detail the decay amplitude of Fig.17a,
performing calculations, as before, in the framework of
the spectral representation with the light-cone wave functions for
$q\bar q$ states.

By using the notations
\begin{eqnarray}
\label{56a}
&& P\ =p_1+p_2, \quad k_{12}\ =k_1+k_2, \quad k_{34}\ =k_3+k_4,
 \nonumber\\
&& M^2\ =(k_1+k_2)^2, \quad s_{12}\ =(k_1+k_2)^2, \quad s_{34}\
=(k_3+k_4)^2
\end{eqnarray}
we have the following spectral representation for the amplitude:
\begin{eqnarray}
\label{57a}
&&A(q\bar q \,\; {\rm state} \to {\rm two} \, {\rm mesons})= \int
\limits_{4m^2}^{\infty} \frac {dM^2}{\pi} \Psi_{in}(M^2)
d\Phi(P;p_1,p_2) \int \limits_{(m+m_s)^2}^\infty
\frac {ds_{12}ds_{34}}{\pi^2}     \nonumber \\
&&\times \; t(p_1,p_2;k_1,k_2,k_3,k_4)d\Phi(k_{12};k_1,k_2)\;
d\Phi(k_{34};k_3,k_4) \Psi_1(s_{12}) \Psi_2(s_{34})\ .
\end{eqnarray}
Here the masses of newly--born quarks $i=2,3$ are denoted as
$m_s$, thus opening the way to consider the decay with strange
quark production. The transition  amplitude (\ref {55a}) of Fig.
17c is denoted as  $t(p_1,p_2;k_1,k_2,k_3,k_4)$. The decay
amplitude (\ref {56a}) is written in terms of meson wave
functions: for the initial state it is $\Psi_{in}(M^2)$, and for
outgoing mesons they are $\Psi_1(s_{12})$ and $\Psi_2(s_{34})$.

Thus, the decay amplitude $A$ is a convolution of the transition
amplitude (\ref{55a}) with wave functions of initial and outgoing
mesons:
\begin{equation}
\label{58a}
A(q\bar q \,\; {\rm state} \to {\rm two} \, {\rm
mesons})= \Psi_{in}\otimes t \otimes \Psi_1 \Psi_2\ .
\end{equation}

Let us turn to the principal point, namely: the evaluation of the
region in momentum space selected by the transition amplitude of
Fig.17c within the assumption of soft hadronization.

The hypothesis of soft hadronization means that the ladder diagram
of Fig.17c has a peripheral structure: it requires the momentum
transfers to the $q\bar q$ block to be small, of the order
of $1/R^2_{confinement}\;$. So
\begin{eqnarray}
\label{59a}
&&-(p_1-k_1)^2 \simeq ({\bf p}_{1\perp}-{\bf k}_{1\perp})^2 \sim
R^{-2}_{confinement}\; , \nonumber\\
&&-(p_2-k_4)^2 \simeq ({\bf p}_{2\perp}-{\bf k}_{4\perp})^2 \sim
R^{-2}_{confinement}\; .
\end{eqnarray}
Likewise, the momentum transfer squared in the quark propagator
(see Fig.17c) is of the order
\begin{equation}
\label{60a}
(p_1-k_1-k_2)^2 \simeq -({\bf p}_{1\perp}-{\bf k}_{1\perp}-{\bf k}_{2\perp})^2.
\end{equation}

In the simplest approximation, taking into account only the $t$-channel
propagators, we can write the formula for $t(p_1,p_2;k_1,k_2,k_3,k_4) $
in the form
\begin{eqnarray}
\label{61a} &&t(p_1,p_2;k_1,k_2,k_3,k_4)=\frac{g}{m^2_g+({\bf p}_{1\perp}
-{\bf k}_{1\perp})^2}
\nonumber \\
&&\times  \frac{g^2 \left
(m_s-\gamma({\bf p}_{1\perp}-{\bf k}_{1\perp}-{\bf k}_{2\perp})\right  )}
{m^2_s+({\bf p}_{1\perp}-{\bf k}_{1\perp}- {\bf k}_{2\perp})^2} \cdot
\frac{g}{m^2_g+( {\bf p}_{2\perp}-{\bf k}_{4\perp})^2}\ ,
\end{eqnarray}
where $\gamma$ is the Dirac matrix. To avoid ultrared
divergence, the effective mass of the soft gluon is introduced
into the gluon propagator  (see, for example, [69]).

The equation (\ref{61a}) does not state that the transition
amplitude $t(p_1,p_2;k_1,k_2,$ $k_3,k_4)$ selects large distances. At
this point the amplitude (\ref{61a}) may be, however, improved by
incorporating form factors into the gluon emission vertex:
\begin{equation}
\label{62a}
g\to g(({\bf p}_{1\perp}-{\bf k}_{1\perp})^2), \quad g\to
g(({\bf p}_{2\perp}-{\bf k}_{4\perp})^2)\ .
\end{equation}
With equations (\ref{61a}) and (\ref{62a}) the amplitude $A$ reads:
\begin{eqnarray}
\label{63a}
&&A(q\bar q \to {\rm two} \, {\rm mesons})
=\int \limits_0^1 \frac{dx_1dx_2\delta(1-x_1-x_2)}{16\pi^2 x_1x_2}
\int d{\bf p}_{1\perp}d{\bf p}_{2\perp}
\delta({\bf p}_{1\perp}+{\bf p}_{1\perp})
\nonumber \\
&&\times \int \limits_{y_1\gg y_2}
\frac{dy_1dy_2\delta(1-y_1-y_2)}{16\pi^2 y_1y_2}\cdot \int
d{\bf k}_{1\perp}d{\bf k}_{2\perp} \delta({\bf k}_{1\perp}+{\bf k}_{2\perp})
\nonumber \\
&&\times\int \limits_{y_4\gg y_3}
\frac{dy_3dy_4\delta(1-y_3-y_4)}{16\pi^2 y_3y_4} \int
d{\bf k}_{3\perp}d{\bf k}_{4\perp} \delta({\bf k}_{3\perp}+{\bf k}_{4\perp})
\Psi_{in}(x_1,x_2;{\bf p}_{1\perp},{\bf p}_{2\perp})
\nonumber \\
&&\times \frac{g (({\bf p}_{1\perp}-{\bf k}_{1\perp})^2 )}
{m^2_g-({\bf p}_{1\perp}-{\bf k}_{1\perp})^2} \frac{g^2 (m_s- \gamma
({\bf p}_{1\perp}-{\bf k}_{1\perp}-{\bf k}_{2\perp})) } {m^2_s
+({\bf p}_{1\perp}-{\bf k}_{1\perp}-{\bf k}_{2\perp})^2 } \frac{g
(({\bf p}_{2\perp }-{\bf k}_{4\perp })^2 ) } {m^2_g+({\bf p}_{2\perp}
-{\bf k}_{4\perp })^2 }
\nonumber \\
&&\times \Psi_1(y_1,y_2;{\bf k}_{1\perp},{\bf k}_{2\perp})
\Psi_2(y_3,y_4;{\bf k}_{3\perp}, {\bf k}_{4\perp}) \ .
\end{eqnarray}
This is the expression for the decay amplitude which makes it
possible to discuss the rules of quark combinatorics.

In a rough approximation that still gives a qualitatively correct
answer, we  neglect the momenta in the propagator of newly--born
quarks:
\begin{equation}
\label{64a}
\frac{g^2\left (m_s-\gamma
({\bf p}_{1\perp}-{\bf k}_{1\perp}-{\bf k}_{2\perp})\right )}{m_s^2+
({\bf p}_{1\perp}-{\bf k}_{1\perp}-{\bf k}_{2\perp})^2} \to \frac{g^2}
{m_s} \,  .
\end{equation}
We have
\begin{equation}
\label{65a}
 A(q\bar q \,\; {\rm state} \to {\rm
two} \, {\rm mesons}) =\frac{\alpha_s}{m_s} \cdot \left (
\Psi_{in}\otimes t \otimes \Psi_1\Psi_2 \right )\ ,
\end{equation}
This equation tells us that the probability to produce
non-strange and strange quarks, $u\bar u:d\bar d:s\bar
s=1:1:\lambda$, is determined by the ratio of masses squared of
non-strange $(u,d)$ to strange $(s)$ quark. Introducing the
constituent quark masses in the soft region, $m_u\simeq m_d\equiv
m = 350$ MeV and $m_s \simeq 500$ MeV, we get:
\begin{equation}
\label{66a}
\lambda \simeq \frac{m^2}{m^2_s}\simeq 0.5 .
\end{equation}
The equations (\ref{64a}) and (\ref{65a}) justify the statements of quark
combinatorics applied to the decay processes [9],[70]-[72]. Of course,
here we suppose that meson wave functions belonging to the same multiplet
are identical.

Equation (\ref{66a}) gives us just a rough evaluation for
$\lambda$, for in (\ref{64a}) we neglected momentum transfers
squared compatible with light quark masses. In more
sophisticated evaluations of $\lambda$, one may take into account
the momentum dependence of the quark propagator:
\begin{equation}
\label{67a}
\frac{1}{m_s^2+ ({\bf p}_{1\perp}-{\bf k}_{1\perp}-{\bf k}_{2\perp})^2}
\to \frac1{m_s^2 +<k^2>} ,
\end{equation}
where $<k^2>$ is a typical momentum
squared inherent to the considered decay process. Therefore,
\begin{equation}
\label{68a}
\lambda =\frac{m^2+<k^2>}{m_s^2+<k^2>} \; .
\end{equation}
For standard decays of light resonances $<k^2>\sim 0.1-0.3$
(GeV/c)$^2$, and this results in the increase of $\lambda$
compared to (\ref{66a}). Indeed, in the analysis [72]
$\lambda \sim 0.7$ was found.  Actually,
(\ref{68a}) demonstrates that $\lambda$ can vary depending on
different types of reactions.

Let us now turn to the processes of the $q\bar q$ pair production
in jet processes $Z^0 \to q\bar q \to hadrons$, which is shown in
Fig.17b. All the above considerations, which have been used for
the decay of a resonance into two mesons, can be applied to this
process. As a result, we obtain the following formula which is a
counterpart of (\ref{65a}):
\begin{equation}
\label{69a}
A(Z^0 \to {\rm two}\, {\rm mesons}+X)=\frac{\alpha_s}{m_s} \cdot \left
(q\bar q \,{\rm from}\,{\rm jet}\,{\rm ladder} \otimes t \otimes
      \Psi_1\Psi_2\right ) .
\end{equation}
This expression differs from (\ref{65a}) by the initial state only, which
is the wave function of $q\bar q$-pair for a jet ladder but not for the
state defined by the wave function $\Psi_{in}$. This means that the
ratio of probabilities of producing a strange and a non-strange
quark are given by the factor $m^2/m^2_s$. So, one has the same
$ \lambda \simeq 0.5$ value as in the decay process, which does not
contradict the experimental data on the ratio of yields $K^{\pm}/\pi^\pm$.
The ratio  $K^{\pm}/\pi^\pm$ as a function of $x$ is shown in Fig.16e.
It is seen that at $x=0.2$ $K^{\pm}/\pi^\pm\simeq 0.35$. With the increase
of $x$ the ratio $K^{\pm}/\pi^\pm$ grows and reaches the value $\sim 0.8$ at
$x=0.7$. Such an increase is rather legible: indeed, $K$
mesons are produced both due to the formation of a new $s\bar s$
pair in the ladder (with the probability $\lambda$) and to the
fragmentation production of an $s\bar s$ pair in the transition $Z^0
\to s\bar s$. Relative probabilities of prompt production of $Z^0
\to u\bar u,d\bar d,s\bar s$  obey the ratio $u\bar u:d\bar
d:s\bar s\simeq 0.26:0.37:0.37$. Because of that the production of
$K$ meson at large $x$ is proportional to
\begin{equation}
\label{70a}
K^+ \sim (0.37 \cdot 1 +0.26\cdot \lambda)\ .
\end{equation}
The same quantity for pions is
\begin{equation}
\label{71a}
\pi^+ \sim (0.37\cdot 1 + 0.26\cdot 1) .
\end{equation}
So, the ratio $K^+/\pi^+$ at large $x$ is equal to
\begin{equation}
\label{72a}
\frac{K^+}{\pi^+}=\frac{0.37+0.26\lambda} {0.37+0.26}\simeq 0.8\
\end{equation}
at $\lambda=0.5$. This value agrees with the
experimental data as it is demonstrated by Fig.16e,
where $\lambda(x)$ determined as the $ K^{\pm}/\pi^{\pm}(x)$-ratio
is shown.

The small value of $K^+/\pi^+$ at $x=0$ is a direct consequence of
large probability to produce highly excited resonances: in the
resonance decay more pions than kaons are produced. For the
problem of breeding of strange and non-strange states in the
decay, it is rather interesting to see the ratio $K^*/\rho$ ---
experimental data for  $K^{0*}/\rho^0$ are shown in Fig.16e too
(shaded area). Remarkably, the ratio  $K^{0*}/\rho^0$  has no
tendency to decrease with decreasing $x$: this means that the rate
of breeding of $K^{0*}$ and $\rho^0$ in decays is approximately
the same. Unfortunately, experimental errors are too large to have
more definite conclusions about the behaviour of $\lambda(x)$.

The supression parameter for the production of a strange quark
cannot be reliably determined. This is because the masses of
strange and non-strange quarks are small compared to the mean
transverse momenta of quarks in the production process, see
(\ref{68a}). We can draw a more definite conclusion about the
suppression parameter $\lambda_Q$ for the production of heavy
quarks $Q=c,b$. This parameter is defined by the same formula
(\ref{63a}) for multiperipheral production, so we have:
\begin{equation}
\label{73a}
\lambda_Q \simeq \frac{m^2}{m^2_Q\ln^2\frac{\Lambda^2}{m^2_Q}}.
\end{equation}
Here we take into account that the gluon--quark coupling
constant decreases with the growth of the quark mass,
$\lambda_Q\sim \alpha^2(m^2_Q)$. The QCD scale constant,
$\Lambda$, is of the order of 200 MeV.

To estimate $\lambda_c$ and $\lambda_b$, let us use $m=0.35$ GeV,
$m_c=M_{J/\Psi}/2=1.55$ GeV, $m_b=M_{\Upsilon}/2=4.73$ GeV and
$\Lambda=0.2$ GeV. Then                                                                                                        \begin{equation}
\label{74a}
\lambda_c \simeq 2.8\times 10^{-3},
\qquad \lambda_b \simeq 1.1\times 10^{-4}.
\end{equation}
The value of $\lambda_c $ should reveal itself in the inclusive
production of $J/\Psi$ and $\chi$ mesons, while $\lambda_b$ is to
be seen in reactions with $\Upsilon$'s: $Z^0 \to (\sum J/\Psi +
\sum \chi)+X$ and $Z^0 \to \sum  \Upsilon +X$. These reactions are
determined by the processes $Z^0 \to c\bar c \to c+(\bar c c +\bar
q q{\rm -sea})+\bar c$  and $Z^0 \to b\bar b \to b+(\bar b b +\bar
q q{\rm -sea})+\bar b $: for the production of a $c\bar c$ or
$b\bar b$ meson a new pair of heavy quarks should be produced,
since the quarks formed at the first stage of the decay, $Z^0 \to
c\bar c $ or $Z^0 \to b\bar b $, have a rather big gap in the
rapidity scale. So, within the definition
\begin{eqnarray}
\label{75a}
&&\lambda_c \simeq \Gamma \bigg (c\bar{c} \to c+(\bar{c} c +\bar{q}q
{\rm -sea} )+\bar{c} \bigg ) /\Gamma (c\bar{c})
\nonumber \\
&&\lambda_b \simeq \Gamma \bigg  (b\bar{b} \to b+(\bar{b} b +\bar{q}
q{\rm -sea} )+ \bar{b} \bigg ) /\Gamma (b\bar{b}),
\end{eqnarray}
we estimate $\Gamma (c\bar{c} \to c+(\bar{c} c +\bar{q} q{\rm -sea} )
+\bar{c})$ and $\Gamma (b\bar{b} \to b+(\bar{b} b +\bar{q} q{\rm -sea})
+\bar{b})$ by the available data from [49].
Roughly, we get
\begin{equation}
\label{76a}
\lambda_c=(2.07\pm 0.23)\times 10^{-3}, \qquad \lambda_b=(0.31
\pm 0.19\times 10^{-4} \, ,
\end{equation}
in a reasonable agreement with (\ref{74a}).

\subsection{Baryon-meson ratio and the Watson-Migdal factor}\label{3.4}

Our present understanding of the multiperipheral ladder is not
sufficient to re-analyse (\ref{7}) on the level carried out in
Sections 3.1 and 3.2 for $V/P$. Nevertheless, the data for decays
$Z^0\to p+X$ and $Z^0\to \pi^++X$ definitely confirm the equation.
Quark combinatorics [9] predicts for $p/\pi^+$
at large $x$:
\begin{equation}
\label{5-hep}
p/\pi^+\simeq 0.20
\end{equation}
In Fig.16d one can see the $p/\pi^+$ ratio given by the fit to the
data [29] (shaded area) and the prediction of quark
combinatorics (\ref{5-hep}): the agreement at $x>0.2$ is quite
good.

Let us comment the result of our calculation $p/\pi^+ \simeq 0.20$
for leading particles in jets. In the jet created by a quark the
leading hadrons are produced in proportions as it is given by
Eq.(\ref{5}): $B_i:2M_i:M$. We consider only the production of
hadrons belonging to the lowest (baryon and meson) multiplets,
and, hence, keep in (\ref{10}), (\ref{11}) only the terms with $L=0$
(hadrons from the quark $S$-wave multiplets). In our estimations
we assume $\beta_0\simeq\mu_0$, and therefore we substitute
$B_i\to B_i(0)$, $M_i\to M_i(0)$ and $M\to M(0)$. The precise
content of $B_i(0)$, $M_i(0)$ and $M(0)$ depends on the
proportions in which the sea quarks are produced. We assume
flavour symmetry breaking for sea quarks (see (\ref{27})).
For the sake of simplicity, we put first $\lambda=0$ (actually
the ratio $p/\pi$ depends weakly on $\lambda$). Then for the
$u$-quark  initiated jet we have:
\begin{eqnarray}
\label{7-hep}
&& B_u(0)\to \frac2{15}p +\frac1{15}n\ +\
(\Delta-\mbox{resonances }), \nonumber\\
&& M_u(0)\to \frac18\pi^+ +\frac1{16}\pi^0
+\frac1{16}(\eta+\eta') +(\mbox{ vector mesons }),  \nonumber\\
&& M(0)\to \frac1{16}\pi^+
+\frac1{16}\pi^0+\frac1{16}\pi^-+\frac1{16} (\eta+\eta')+(\mbox{
vector mesons }).
\end{eqnarray}
The hadron content of the $d$-quark  initiated jet is determined
by isotopic conjugation $p\to n$, $n\to p$, $\pi^+ \to \pi^-$, and
the content of antiquark jets is governed by charge conjugation;
in jets of strange quarks only sea mesons ($M$) contribute to the
$p/\pi^+$ ratio.

Taking into account the ratio $B_i:2M_i:M=1:2:1$ and the
probabilities for the production of quarks of different flavours
$q_i$, given by (1), we obtain $p/\pi^+ \simeq 0.21$ for
$\lambda=0$. We can easily get the $p/\pi^+$ ratio for an
arbitrary $\lambda$: the decomposition of the ensembles $B_i(0)$,
$M_i(0)$, $M(0)$  with respect to hadron states has been performed
in [9] (see Appendix D, Tables D.1 and D.2). But, as
was stressed above, this ratio is a weakly dependent function of
$\lambda$: at $\lambda=1$ we have $p/\pi^+ \simeq 0.20$.

For quark combinatorics the saturation of $q\bar q$ and $qqq$
states by real hadrons is of principal importance. The probability
of saturation is defined by the coefficients $(\mu_L,\mu_L^{(i)})$
and $(\beta_L,\beta_L^{(i)})$ in (\ref{10},\ref{11}). The main
question is what contributions from higher multiplets are not
negligible in the spectra.

Consider in more detail the production of mesons in the central
region: $q\bar q \to M$. The central production of $q\bar q$
states is provided by the diagrams of Fig.18 (loop diagram,
Fig.18a, and interactions of the produced quarks, Fig.18b). The
diagrams of the type of Fig.18b for final state interactions lead
to the relativistic Watson--Migdal factor. To estimate how many
highly excited states are produced, we have to find out which
states are determined by the $q\bar q$ system in the
multiperipheral ladder.

\begin{figure}
\centerline{\epsfig{file=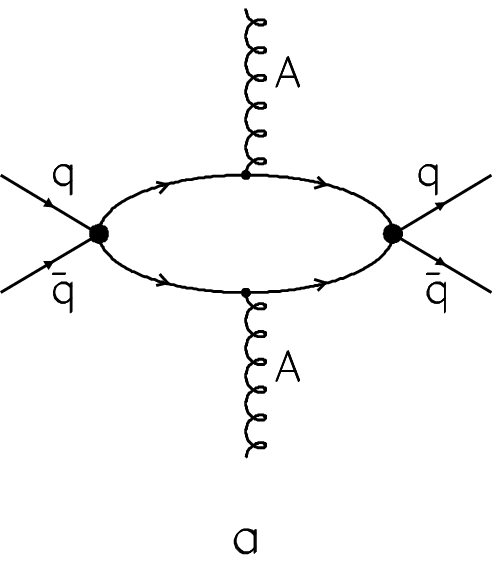,height=5.0cm}\hspace{1cm}
            \epsfig{file=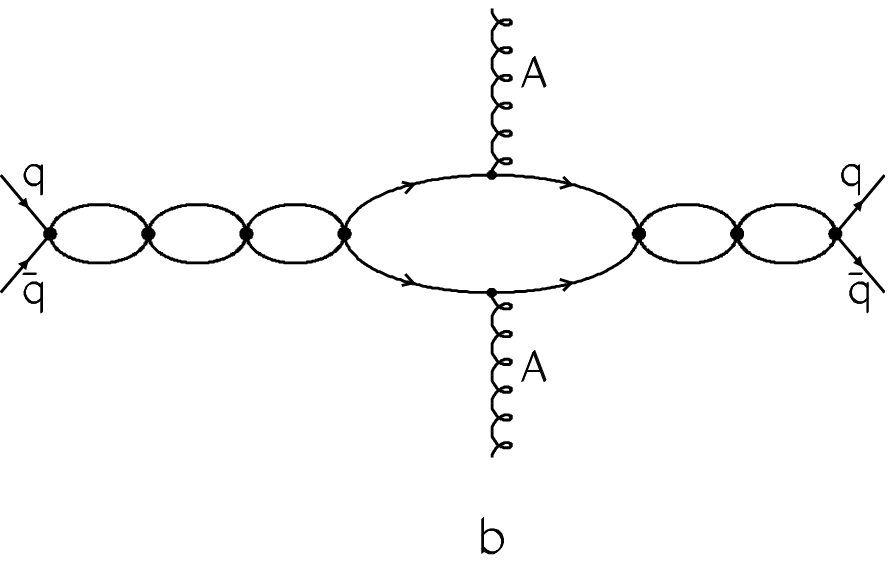,height=5.0cm}}
\vspace{1cm}
\centerline{Fig.18}
\end{figure}
\vspace{1cm}

The constructive element of the ladder is a process shown in Fig.
17b. In this process, as was stressed in Section 3.3, new $q\bar q$
pairs are created at relatively large separations (in hadronic
scale), at $r\sim 1$ fm: these separations are just those in the
$q\bar q\to M$ transitions. The orbital momenta  of the $q\bar q$
system for this transition can be written as $L\sim kr$. For
relative quark momenta $k \le  0.6$ GeV/c, we have
\begin{equation}
\label{75Mig}
L\le  3.
\end{equation}
Relying on the behaviour of the Regge
trajectory, one can understand, to what meson masses $\mu$ this
relation corresponds. The trajectories for $q\bar q$ states are
linear up to $\mu\sim 2.5$ GeV [73]:
\begin{equation}
\label{76Mig}
\alpha(\mu^2) \simeq \alpha(0) +\alpha'(0) \mu^2\ ,
\end{equation}
the slope $\alpha'(0) $ is approximately equal $\alpha'(0) \simeq 0.8$
GeV$^{-2}$ and the intercept belongs to the interval $0.25 \le
\alpha(0) \le 0.5$. Hence, for large $\mu$, the estimation gives
$\mu^2 \sim \alpha(\mu^2)/ \alpha'(0) $; with $\alpha(\mu^2)\sim 3$,
we have $\mu^2\sim 4$ GeV$^2$. So we conclude that in the multiperipheral
ladder it would be natural to expect the production of $q\bar q$
mesons with masses
\begin{equation}
\label{77Mig}
\mu\le 2 \quad {\rm GeV}.
\end{equation}

\section{Hadron - nucleus interaction}\label{IV}

We have demonstrated that the investigation of multiparticle production
processes provides a good possibility to prove the main assumptions of
the presented approach. There exist, however, processes, which allow us
to observe the consequences of the spectator mechanism in a relatively
pure way, i.e. we can prove the hypothesis which is crucial from the
point of view of the hadron structure. These processes are the
hadron - nucleus collisions at high energies. They enable us to test the
hadron structure because of the well-known fact that at sufficiently high
energies the specific picture of the beam fragmentation is not distorted
by possible repeated collisions with the nuclear matter. (For details,
see [9]). As we have discussed it already, due to the parton hypothesis
secondaries need time to be formed ([19], [45]), which is proportional
to their momenta $p$: $\tau \sim \frac{p}{m^2}$. Hence, the constituents
go through the nucleus before forming a secondary hadron, and repeated
interactions with the nucleus become impossible.

In hadron - hadron collisions only one pair of quarks takes part: one of
the incident particle and one of the target. When, however, a hadron
collides with a heavy nucleus, while going through the nuclear matter,
its constituents can interact independently of each other with different
nucleons of the nucleus. In the case of a superheavy nucleus all the
constituents of the projectile would interact so that all three (or two)
quarks of an incident baryon (or meson, respectively) would break up.
(This would mean, e.g., that the multiplicity ratio of the secondaries
for $\pi A$ and $pA$ interactions in the central region were $\sim 2/3$
[37]). For real nuclei (even for $A \sim 200$ ) a part of the constituent
quarks still goes through a nucleus without interacting. The number of
quarks passing the nucleus without interaction determines the multiplicity
of the fragmentational hadrons, i.e. hadrons in the region of large $x$
values.

Hence, in baryon - nucleus collisions three different processes are
possible: one quark is interacting, two go through the nucleus
(Fig.19a); two quarks are interacting and one goes through the nucleus
(Fig.19b) and finally, all three quarks interact (Fig.19c). In
meson - nucleus interactions one (Fig.19d) or two quarks (Fig.19e) of
the incident meson can take part in the interaction.
\begin{figure}
\[   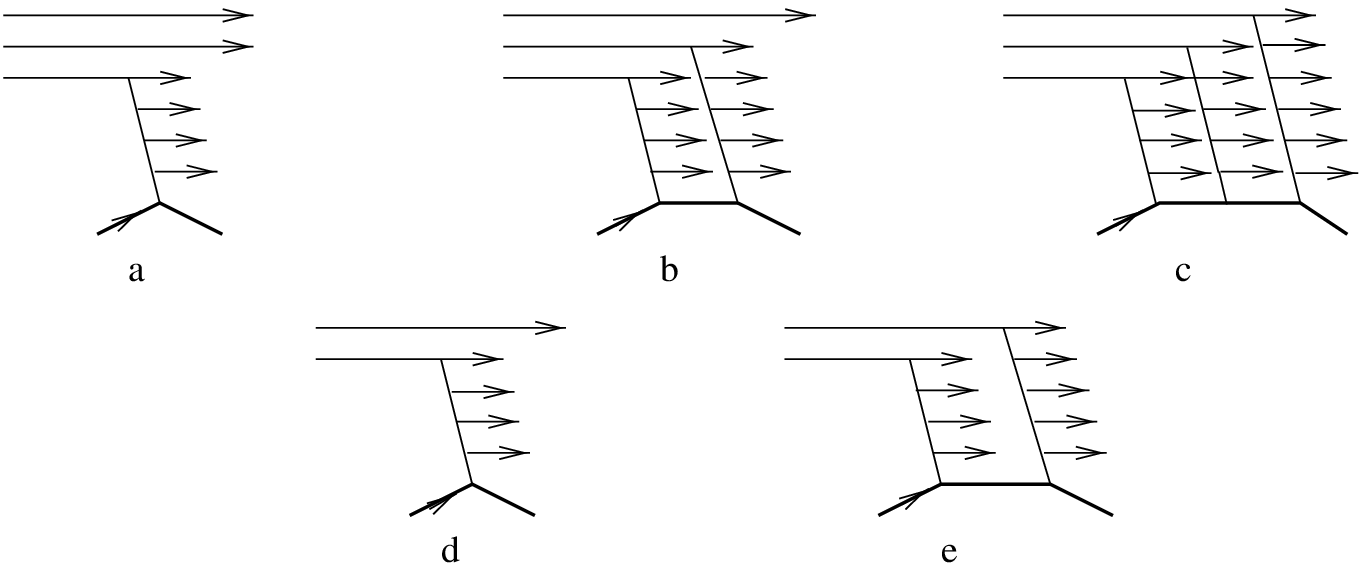  \]
\centerline{Fig.19}
\end{figure}

The probability for a quark to interact can be calculated as a function
of the distribution of the nuclear matter density and the inelastic
quark - nucleon cross section:
\[\sigma_q \equiv \sigma_{inel}^{qN} \approx \frac{1}{3}\sigma_{inel}^{NN}
\approx \frac{1}{2}\sigma_{inel}^{\pi N} \approx 10 mbarn . \]
The probabilities of these processes can be written as
\begin{equation}
\label{28}
V_k^h = \frac{n!}{(n-k)!k!\sigma_{prod}^{hA}}\int d^2 b e^{-(n-k)\sigma_q
T(b)}[1-e^{-\sigma_q T(b)}]^k ,
\end{equation}
where $k$ is the number of the interacting quarks, and $h$ is the incident
hadron consisting of $n$ quarks ($n=2$ for mesons and $n=3$ for baryons
([20],[21]). The function $T(b)$ is expressed in terms of the nucleon
distribution density in the nucleus:
\begin{equation}
\label{29}
T(b) = A \int_{-\infty}^{\infty}dz \rho(b,z) .
\end{equation}
For $\rho(r=\sqrt{b^2+z^2})$, the Fermi parametrization,
\begin{equation}
\label{30}
\rho(r) = \frac{1}{1+e^{[(r-c_1)/c_2]}}, \qquad
 4\pi \int_0^{\infty}\rho(r)r^2 dr = 1
\end{equation}
is accepted. The parameters $c_1$ and $c_2$ are taken from the data on
$eA$ scattering ([38],[39]). The probabilities $\sigma_{prod}^{hA}$ for
the hadron - nucleus scatterings have the same meaning as the inelastic
hadron - hadron cross section with the production of at least one
secondary hadron. It is obtained from the normalization condition
$\sum_i V_i^h(A) = 1$:
\begin{equation}
\label{31}
\sigma_{prod}^{hA}\int d^2 b [1 - e^{-n_q T(b)}] .
\end{equation}
The values of the probabilities $V_i^h(A)$ which are calculated from
the nuclear density functions found in $eA$ scatterings for $p$, $\pi$
and $K$ beams (see [38]), are given in [22],[40]. For light nuclei
processes with the interaction of only one constituent quark are
dominating (Figs.19a, 19d). However, even for the nucleus of Be the
share of the process with two interacting constituent quarks (Fig.19b)
is not small (around 25\%). For nuclei $A \sim 100$ the probabilities of
all three processes of proton fragmentation are of the same order.

The calculation of the probability $V_i^h(A)$ allows us to obtain the
relation between multiplicities in different regions of hadron - nucleus
collisions.

We assume that the interacting dressed quarks produce secondary particles
independently of each other. Then in the central region the multiplicities
for the processes shown in Figs.19b, 19e is twice as large as in the
processes Figs.19a, 19c; the ratio of the multiplicities for processes
with three and one interacting quarks is three. Indeed, using (\ref{28}),
we can express the multiplicities $<n>_{pA}$ and $<n>_{\pi A}$ in the form
\[R\left(\frac{pA}{q A}\right) = \frac{<n>_{pA}}{<n>_{q A}} =
 \sum_{k=1}^3 kV_k^p = \frac{3}{\sigma_{prod}^{pA}}\int d^2
 b[1-e^{-\sigma(qN)T(b)}]     \]
\begin{equation}
\label{32}
R\left(\frac{\pi A}{q A}\right) = \frac{<n>_{\pi A}}{<n>_{q A}} =
 \sum_{k=1}^2 kV_k^{\pi} = \frac{2}{\sigma_{prod}^{\pi A}}\int d^2 b
 [1-e^{-\sigma(qN)T(b)}] .
\end{equation}
The ratio of multiplicities of the secondary particles in $pA$ and
$\pi A$ does not depend on $<n_{qA}$ and is
\begin{equation}
\label{33}
R\left(\frac{pA}{\pi A}\right) = \frac{<n>_{pA}}{<n>_{\pi A}} =
\frac{V_1^p(A) + 2V_2^p(A) + 3V_3^p(A)}{V_1^{\pi}(A) + 2V_2^{\pi}(A)} =
\frac{3}{2}\frac{\sigma_{prod}^{\pi A}}{\sigma_{prod}^{p A}} .
\end{equation}
For heavy nuclei $\sigma_{prod}^{\pi A}/\sigma_{prod}^{p A} \approx 1 $,
and we obtain relation (\ref{2a}). The comparison of $R(pA/\pi A)$ with
the experimental data [22] is presented in Fig.20.

\begin{figure}
\centerline{\epsfig{file=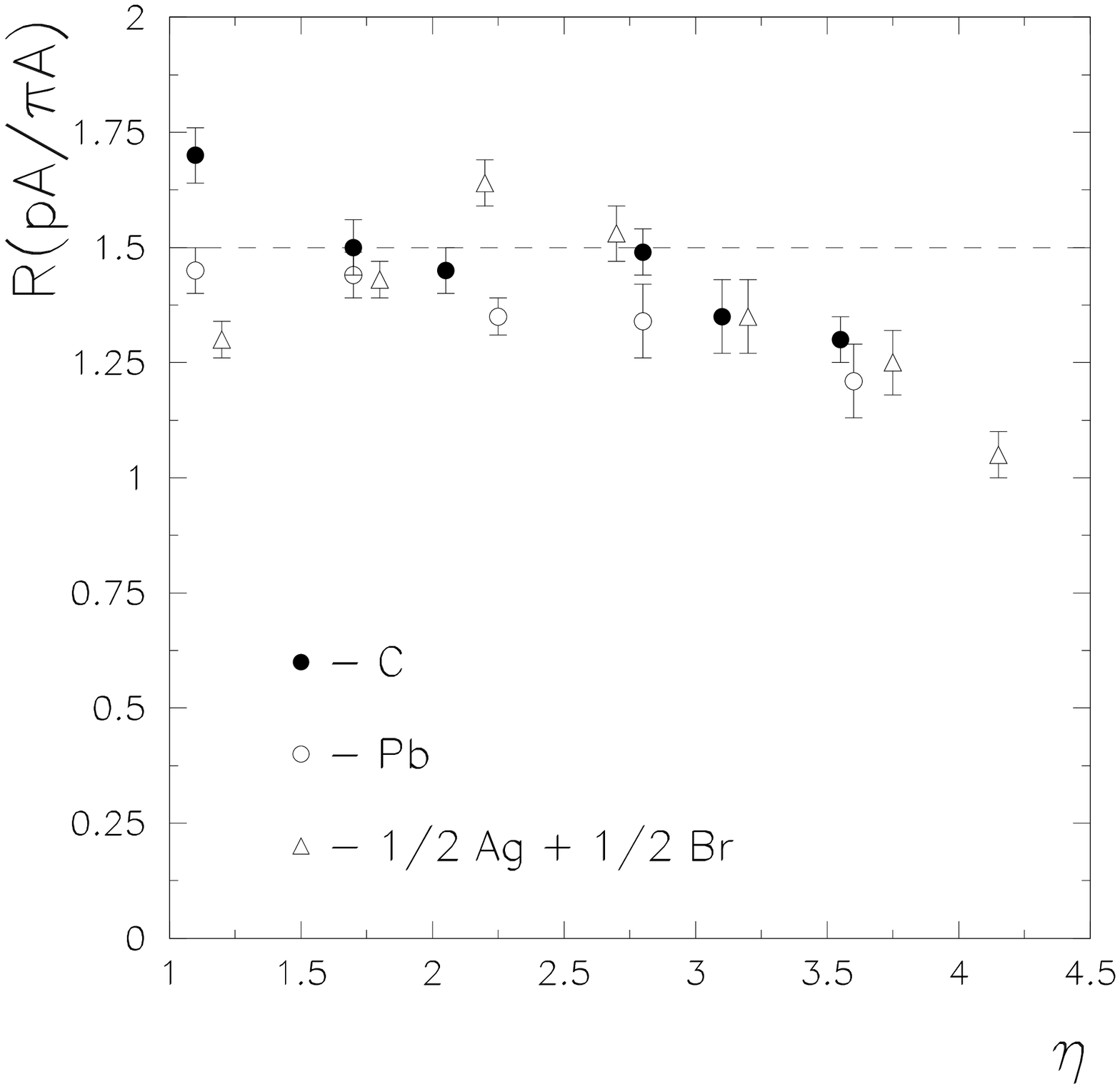,height=8.0cm}}
\vspace{1cm}
\centerline{Fig.20}
\end{figure}

The calculated value of $R\left(\frac{pA}{\pi A}\right)$ is in agreement
with experiment in the interval $1,5 < \eta < 3,5$ for the nuclei C(A=12)
and Pb(A=207) and for photoemulsion 1/2 Ag + 1/2 Br. The considered
region for the values of the quasirapidity $\eta = -\ln\tan\theta/2$
corresponds just to the central region of the collision processes.

Due to (\ref{32}), the ratios of the secondaries in $\pi A$ and $\pi p$
scatterings and in $pA$ and $pp$ scatterings depend on the ratios
$<n>_{qA}/<n>_{qq}$ (where $<n>_{qq}$ is the multiplicity in the
quark-quark collision):
\begin{equation}
\label{34}
R\left(\frac{\pi A}{\pi p}\right) = [V_1^{\pi}(A)+2V_2^{\pi}(A)]\frac
{<n>_{qA}}{<n>_{qN}} ,
\end{equation}
\begin{equation}
\label{35}
R\left(\frac{p A}{p p}\right) = [V_1^p(A)+2V_2^p(A)+3V_3^p(A)]\frac
{<n>_{qA}}{<n>_{qN}} .
\end{equation}
The relation (\ref{33}) is reasonably well satisfied experimentally, and
we can take $<n>_{qA}\simeq <n>_{qN}$.

Further, the multiplicities of secondary hadrons in the fragmentation
region are calculated as functions of the atomic number $A$ of the
target. The values of $V_1^p(A)$, $V_2^p(A)$, $V_3^p(A)$ and
$\sigma_{prod}^{pA}$ are shown in Fig.21. The proton-nucleus
cross section increases as $A^{2/3}$, in full accordance with
expectations.

\begin{figure}
\vspace{1cm}
\centerline{\epsfig{file=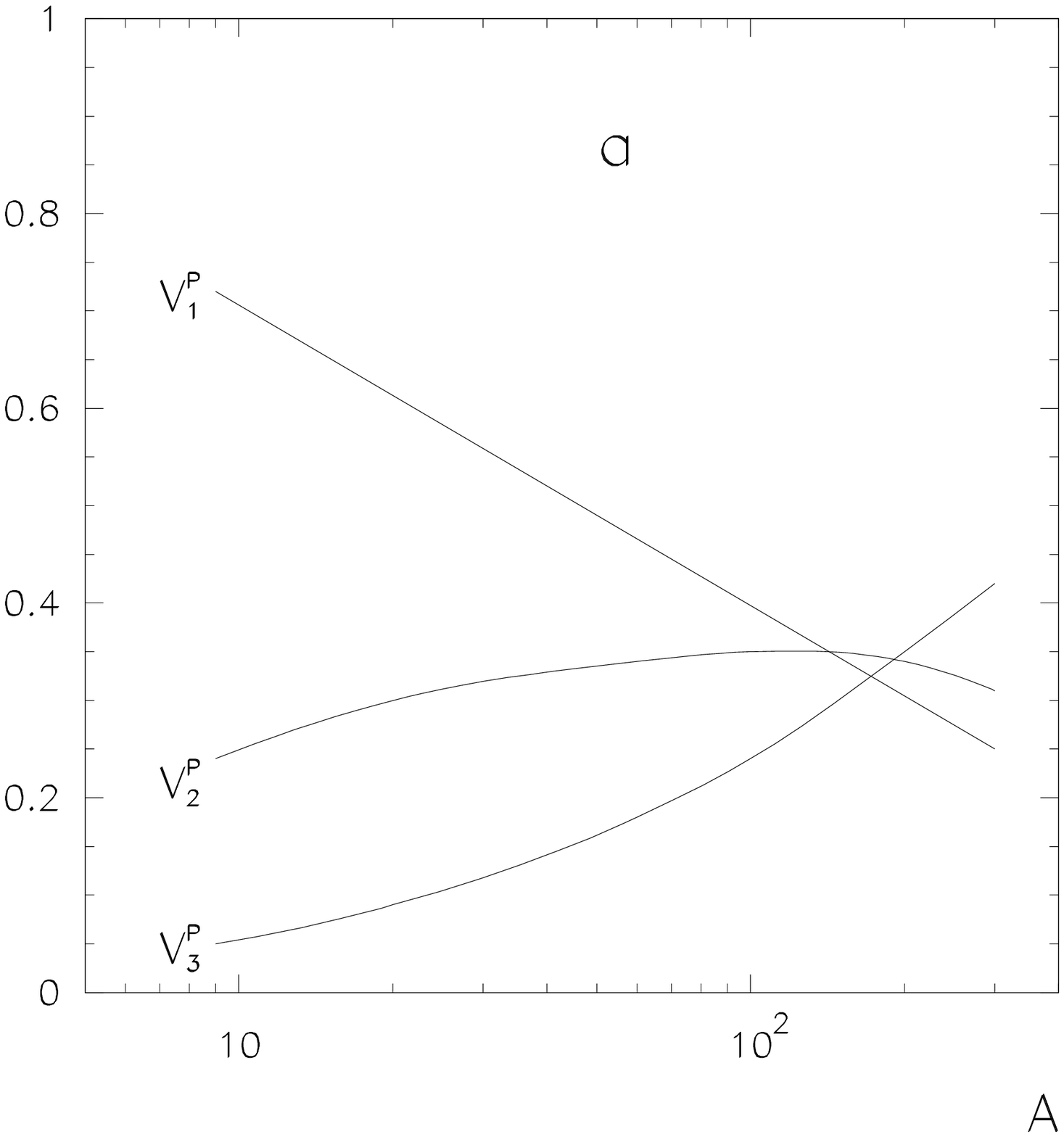,height=8.0cm}\hspace{1cm}
            \epsfig{file=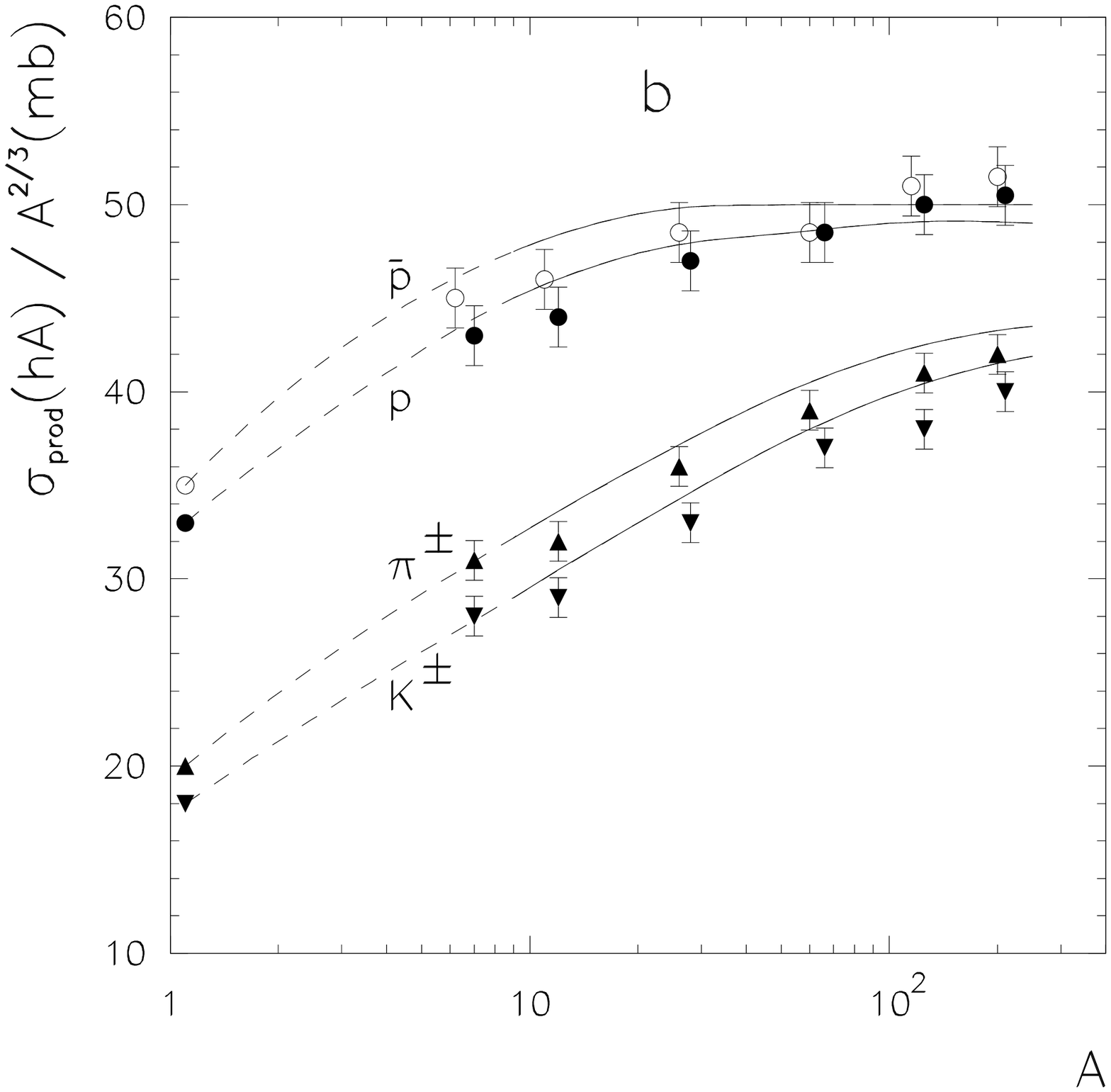,height=8.0cm}}
\centerline{Fig.21}
\end{figure}

As already said, the model with three spatially separated quarks enables
us to express the multiplicity of a fast secondary baryon with $x\simeq
2/3$ for proton - nucleus collisions in terms of the multiplicity for
$pp$ interactions. In both cases a fast baryon is produced by picking up
a newly made quark of the sea by the two non-interacting spectators.
The upper vertices in Figs.9b and 19a coincide, so they cancel in the
ratio of the cross sections or multiplicities. Therefore the ratio of
the inclusive cross sections must not depend on $x$ in a region near
$x=2/3$. This independence on $x$ provides a test of the hypothesis that
the three constituents in a nucleon are spatially separated, whatever
the formation mechanism of the secondaries is.

The calculated ratio of the absolute proton yields with nucleus and
proton targets is
\begin{equation}
\label{36}
\frac{\frac{d^2\sigma}{dpd\Omega}(pA \rightarrow pX)}{\frac{d^2\sigma}
{dpd\Omega}(pp \rightarrow pX)} = V_1^p(A)\frac{\sigma_{prod}^{pA}}
{\sigma^{pp}_{inel}}
\end{equation}
at $x\simeq 2/3$. The results of our calculations are displayed in
Fig.22a for the nuclei of Be, Al, Cu and Pb together with data obtained
at $19,2 GeV/c$ [39]. Theory and experiment are consistent in the wide
range $0,55 \leq x \leq 0,85$ where the experimental $x$-dependence of
the ratio (\ref{36}) is essentially flat. This indicates the absence
of a substantial spread in momenta of the constituents.

The experimental values of $V_1^p$ obtained from the data of [40]
are shown in Fig.22b to be consistent with our calculation.

The ratio of the meson yields near $x=1/3$ is obtained by using the
expressions (\ref{5}), (\ref{6}):
\begin{equation}
\label{37}
\frac{\frac{1}{\sigma_{prod}^{pA}}\frac{d^2\sigma}{dpd\Omega}(pA
\rightarrow MX)}{\frac{1}{\sigma_{inel}^{pp}}\frac{d^2\sigma}{dpd\Omega}
(pp \rightarrow MX)}=V_1^p(A) + \frac{4}{5}V_2^p(A) .
\end{equation}

\begin{figure}
\vspace{1cm}
\centerline{\epsfig{file=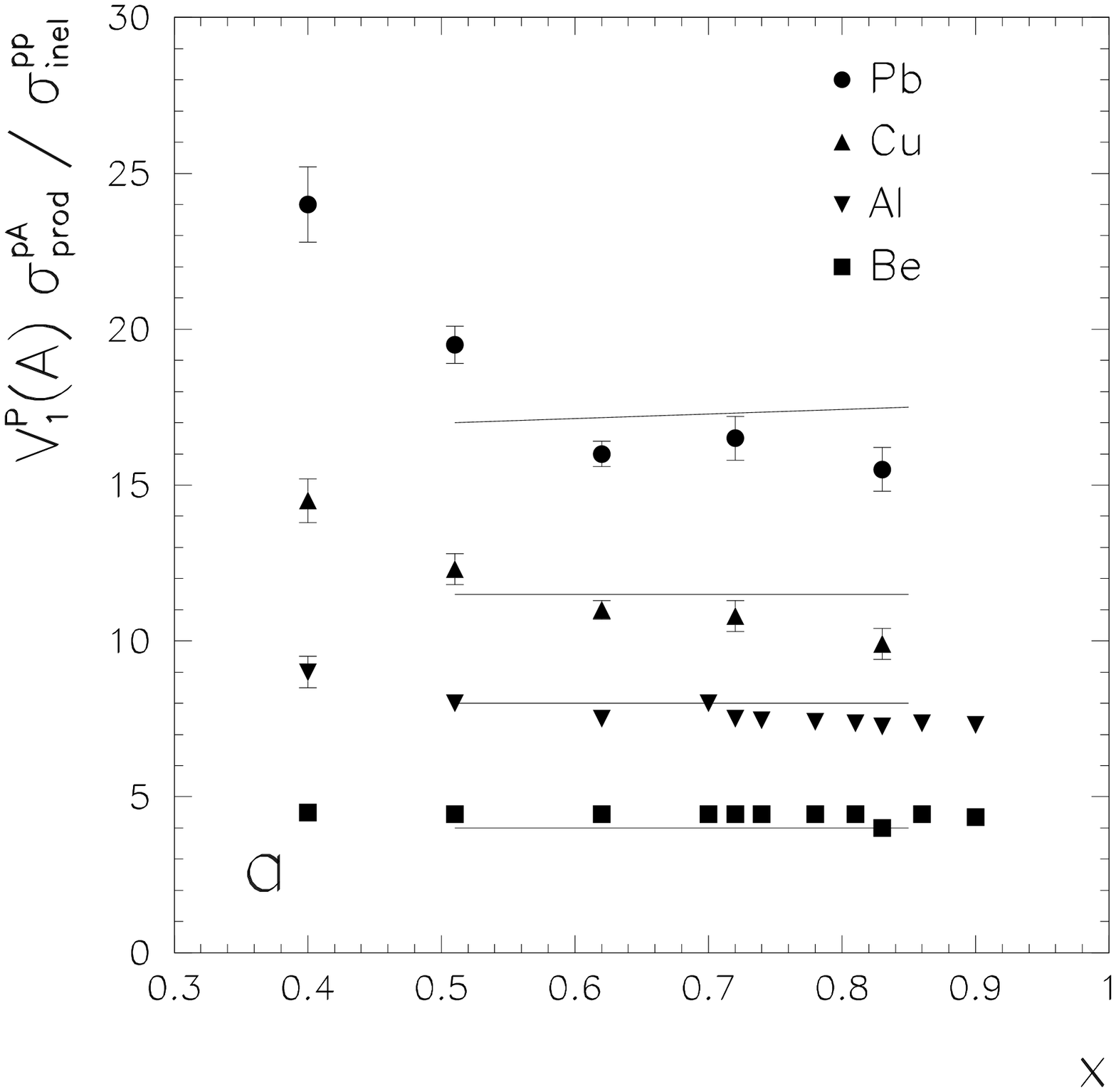,height=8.0cm}\hspace{1cm}
            \epsfig{file=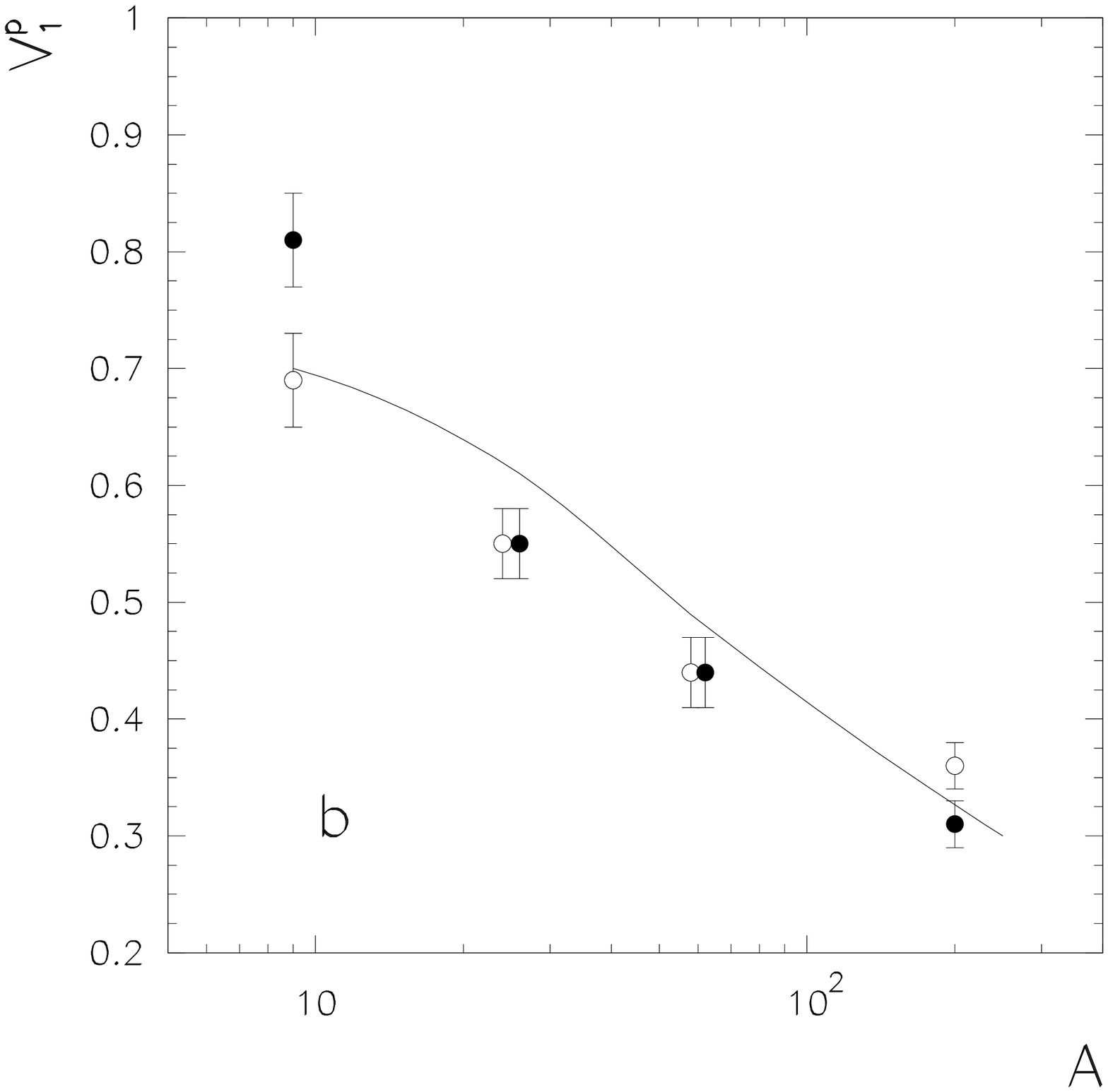,height=8.0cm}}
\centerline{Fig.22}
\end{figure}

\begin{figure}
\vspace{1cm}
\centerline{\epsfig{file=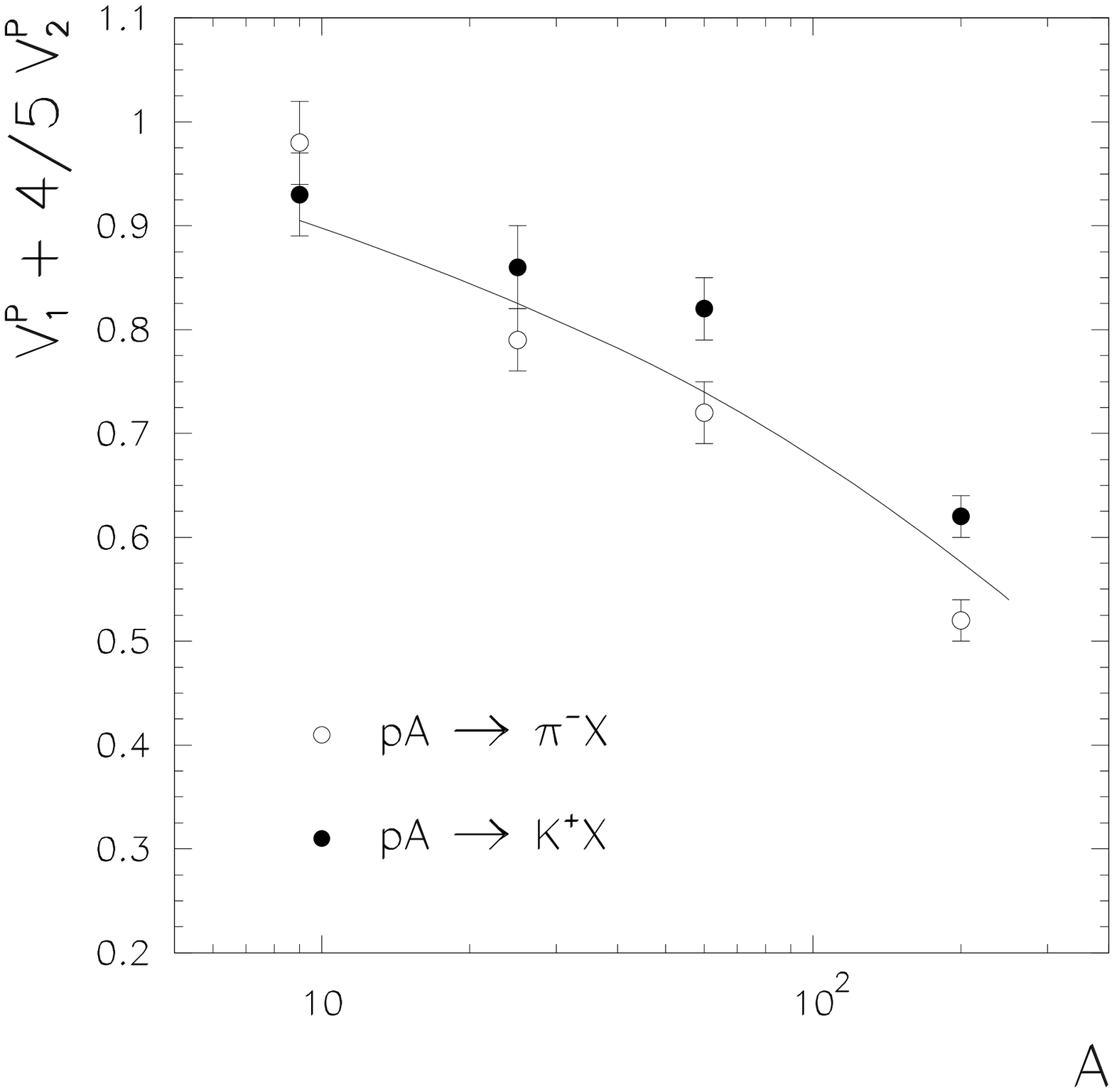,height=8.0cm}}
\centerline{Fig.23}
\vspace{1cm}
\end{figure}

In Fig.23 we plot the values of $V_1^p+4/5V_2^p$ calculated according
to Eq.(\ref{28}). Also shown are the experimental values of the
left-hand side of (\ref{37}), obtained from the data of [40] on
$\pi^-$ and $k^+$ yields at $p_{lab} = 19,2 GeV/c$, $\theta=12,5 mrad$
and $x=0,34$ for Be, Al, Cu and Pb nuclei. Agreement between theory
and experiment is quite good. The $\pi^-$ and $K^+$ mesons have been
chosen, since the chance of producing such particles near $x=1/3$ as
resonance decay products is negligible. The opposite case of $\pi^+$
production at $x\simeq 1/3$ is probably dominated just by the baryonic
resonance decays, and therefore we do not consider them here.

When a pion strikes a nucleus or a proton, the ratio of inclusive
spectra of the same fragments at $x\simeq 1/2$ containing one of the
pion quarks must be
\[ \frac{\frac{1}{\sigma_{prod}^{\pi A}}\frac{d^2\sigma}{dpd\Omega}
(\pi^-A\rightarrow hX)}{\frac{1}{\sigma_{inel}^{\pi p}}\frac{d^2\sigma}
{dpd\Omega}(\pi^-p \rightarrow hX)} \hspace{1cm}
h=\pi^-,\pi^0,p,n,\ldots  \]
independently of the kind of the secondary particle. Hence, the
single-hadron yield ratios like $\pi^-/K^-$, $\pi^-/p$ etc., must be
the same (at $x\simeq 1/2$) for all nuclei in $\pi^-A$ interactions.
The theoretical $A$ dependence of $V_1^{\pi}$ shown in Fig.22b can be
given as
\[  V_1^{\pi}(A) \simeq 1,75 A^{-0,24} \]
for $A>60$.

If the incident particle is a kaon, the production of a fragment
containing the strange quark is determined by the probability to absorb
the non-strange quark, $V_q^k$ . For instance, for the $K^-$ beam the
spectra of strange secondaries $K^-$, $\bar{K}^0$, $\Lambda$, $\Sigma$
etc. must be in the ratio
\[ \frac{\frac{1}{\sigma_{prod}^{KA}}\frac{d^\sigma}{dpd\Omega}(K^-A
\rightarrow h_sX)}{\frac{1}{\sigma_{inel}^{K p}}\frac{d^2\sigma}
{dpd\Omega}(K^-p \rightarrow h_sX)} = V_q^K(A)\left(1+\frac{\sigma_s}
{\sigma_q}\right) , \hspace{1cm} h_s=K^-,\Lambda, \Sigma, \ldots \]
According to Fig.22b, $V_q^K(A) \simeq 0,82 A^{-0,15}$ for $A>30$.
On the other hand, the ratio of the spectra of non-strange fragments
like $\pi^0, \pi^-, \bar{N}$ etc. is determined by the probability to
absorb the strange quark:
\[ \frac{\frac{1}{\sigma_{prod}^{KA}}\frac{d^\sigma}{dpd\Omega}(K^-A
\rightarrow hX)}{\frac{1}{\sigma_{inel}^{K p}}\frac{d^2\sigma}
{dpd\Omega}(K^-p \rightarrow h_sX)} = V_s^K(A)\left(1+\frac{\sigma_s}
{\sigma_q}\right) , \hspace{1cm} h=\pi^-,\pi^0, \bar{p},\bar{n} \ldots \]

In the case of a hyperon beam ($\Lambda$ or $\Sigma$) near $x=2/3$ the
multiplicity ratio for the baryons, containing the strange quark, is
again determined by the probability of absorbing a non-strange quark,
say $V_{1q}^{\Lambda}(A)$. On the other hand, a similar ratio for the
non-strange baryons. As can be seen, the difference in the $A$ dependences
of these quantities is very small. Experimental observation of the
predicted decrease of the multiplicity ratio for strange and non-strange
hadrons near $x=1/2$ in the case of a kaon beam would be a verification
of the hypothesis of the small cross section for a strange quark
interacting with a nucleon.

Similarly to the hadron-hadron interactions, in the hadron-nucleus
interaction processes we can observe the production of fast secondary
hadrons. Due to the presented mechanism of the interaction, we have
to consider those cases, when one or two constituents of the incident
baryon ($x \sim 2/3$ and $x \sim 1/3$, respectively) and one constituent
of the incident meson ($x \sim 1/2$) participate in the interaction.
Using the expressions (\ref{5}) and (\ref{6}), we have for the
baryon-nucleus collision
\begin{eqnarray}
\label{38}
\lefteqn{V_1^b(A)(q_iq_j+q,\bar{q}-sea)+V_2^b(A)(q_i+q,\bar{q}-sea)}
\nonumber\\
& \rightarrow & V_1^b\left(\frac{1}{2}B_{ij} + \frac{1}{12}(B_i+B_j) +
 \frac{5}{12}(M_i + M_j)\right) \nonumber\\
& + & V_2^b\left(\frac{1}{3}B_i + \frac{2}{3}M_i \right) .
\end{eqnarray}
In addition, the following distribution functions have to be introduced:
$f_{ij}(x,p_{\perp}^2)$ for $B_{ij}$, $f_i(x,p_{\perp}^2)$ for $B_i$ and
$\varphi_i(x,p_{\perp}^2)$ for $M_i$. We assume $f_{uu}=f_{ud}=f_{dd}$;
$\varphi_u=\varphi_d$, $f_u=f_d$. Instead of (\ref{34}) we then have
\[ V_1^b(A)\left[ \frac{1}{2}f_{ij}(x)B_{ij} + \frac{1}{12}\left(f_i(x)B_i
 + f_j(x)B_j\right) + \frac{5}{12}\left(\varphi_i(x)M_i+\varphi_j(x)M_j
 \right)\right] \]
\begin{equation}
\label{39}
+ V_2^b(A)\left(\frac{1}{3}f_i(x)B_i +\frac{2}{3}\varphi_i(x)M_i\right) .
\end{equation}
The meson-nucleus collision can be described as
\[ V_1^m(A)(q_i+q,\bar{q}-sea)\;\; \rightarrow \;\; V_1^m(A)\left(\frac{1}
 {3}B_i + \frac{2}{3}M_i \right) \]
\begin{equation}
\label{40}
\rightarrow \;\; V_1^m(A)\left(\frac{1}{3}f_i(x)B_i + \frac{2}{3}
\varphi_i(x)M_i \right) .
\end{equation}
Similarly to the hadron-hadron collisions case, one can easily get the
secondary particles produced in $pA$, $\Lambda A$, $\Sigma A$, $\pi A$
etc. processes.

\section*{Acknowledgements}

The author would like to thank V.V. Anisovich and V.N. Nikonov for many
helpful discussions during the preparation of this review.

\section*{References}

\begin{description}
\item[1] M. Gell-Mann, Phys.Lett. {\bf 8} 214 (1964)
\item[2] G. Zweig, CERN report 8419/TH 412 (1964)
\item[3] G. Morpurgo, Physics {\bf 2} 95 (1965)
\item[4] R.H. Dalitz, Proceedings of the Berkeley Conference (1966)
\item[5] E.M. Levin, L.L. Frankfurt, JETP Pisma {\bf 3} 652 (1965)
\item[6] H.J. Lipkin, F. Scheck, Phys. Rev. Lett. {\bf 16} 71 (1966)
\item[7] J.J.J. Kokkedee, L. Van Hove, Nuovo Cim. {\bf 42} 711 (1966)
\item[8] V.N. Gribov, Eur. Phys. J. {\bf C10} 71 (1999), e-print Archive
         hep-ph/9807224; Eur. Phys. J. {\bf C10} 91 (1999),
         e-print Archive hep-ph/9902279
\item[9] V.V. Anisovich, M.N. Kobrinsky, J. Nyiri, Yu.M. Shabelski,
         World Scientific, Singapore (1985)
\item[10] V.V. Anisovich, Proceedings of the 9th LNPI Winter School on
         Nuclear and Elementary Particle Physics, Leningrad (1974)
\item[11] V.V. Anisovich, Proceedings of the 14th LNPI Winter School on
         Nuclear and Elementary Particle Physics, Leningrad (1979)
\item[12] V.N. Gribov, autumn session of the Academy of Sciences USSR,
         Moscow (1977)
\item[13] E.V. Shuryak, Phys.Rept. {\bf 115} 151 (1984)
\item[14] E.M. Levin, M.G. Ryskin, Yad. Fiz. {\bf 21} 1072 (1975)
\item[15] N. Cabibbo, R. Petronzio, CERN preprint TH 2440 (1978)
\item[16] E.M. Levin, V.M. Shekhter, Proceedings of the 9th LNPI Winter
         School on Nuclear and Elementary Particle Physics, Leningrad
         (1974)
\item[17] J.P. Burq et al., Nucl. Phys. B {\bf 217} 285 (1983)
\item[18] S. Bondarenko, E.Levin, J.Nyiri, TAUP-2700-2002 (2002),
          e-print Archive hep-ph/0204156
\item[19] V.N. Gribov, Proceedings of the 8th LNPI Winter School on
         Nuclear and Elementary Particle Physics, Leningrad (1973)
\item[20] V.V. Anisovich, V.M. Shekhter, Nucl. Phys. {\bf B55}, 455 (1973)
\item[21] J.D. Bjorken, G.E. Farrar, Phys. Rev. {\bf D9}, 1449 (1974)
\item[22] V.V. Anisovich, F.G. Lepekhin, Yu.M. Shabelsky, Yad. Fiz.
          {\bf 27} 1639 (1978)
\item[23] V.V. Anisovich, Yu.M. Shabelsky, V.M. Shekhter, Nucl. Phys.
          {\bf B133} 477 (1978)
\item[24] F.E. Close, Yu.L. Dokshitzer, V.N. Gribov, V.A. Khoze and
\item[25] V.N. Gribov, Effective Theories and Fundamental Interactions,
          Proceedings of the International School of Subnuclear Physics,
          Erice {\bf 34}, 42 (1996)
\item[26] V.V. Anisovich, Yad. Fiz. {\bf 28} 761 (1978)
\item[27] V.V. Anisovich, M.N. Kobrinsky, J. Nyiri,
          LNPI preprint 631 (1980); Yad. Fiz. {\bf 34} 195 (1981)
          (Sov. J. Nucl. Phys. {\bf 34} 111 (1981))
\item[28] V.V. Anisovich, J. Nyiri, Yad. Fiz. {\bf 30} 539 (1979)
          (Sov. J. Nucl. Phys. {\bf 30} 279 (1980))
\item[29] ALEPH Collaboration, R. Barate et al., Phys. Rep. {\bf 294} 1
          (1998)
\item[30] K. B\"{o}ckmann, Invited Talk at the Meeting on Multiparticle
          Production Processes and Inclusive Reactions at High Energies,
          Serpukhov (1976)
\item[31] A. Bia{\l}as, Invited Talk at the First Workshop on
          Ultra-Relativistic Nuclear Collisions, LBL (1979);
          Fermilab-Conf. 79/35-TH4 (1979)
\item[32] V.A. Nikonov, J. Nyiri, e-print Archive hep-ph/0006219 ,\\
          V.V. Anisovich, V.A. Nikonov, J. Nyiri, e-print Archive
          hep-ph/0008163; Phys. Atom. Nucl. {\bf 64} 812 (2001);
          Yad. Fiz. {\bf 64} 877 (2001)
\item[33] S.S. Gershtein, A.K. Likhoded, Yu.D. Prokoshkin, Z. Phys. {\bf
          C24}, 305 (1984)
\item[34] C. Amsler, F.E. Close, Phys. Rev. {\bf D53}295 (1996)
\item[35] V.V. Anisovich, Phys. Lett. {\bf B364} 195 (1995)
\item[36] Yu.L. Dokshitzer, XIth Rencontres de Blois (1999)
\item[37] H.\"{U}berall, Electron Scattering from Complex Nuclei, part A,
          New York -- London (1971)
\item[38] L. Elton, Nuclear sizes, Oxford (1961)
\item[39] Yu.M. Shabelski, Yad.Fiz. {\bf 33} 1379 (1981)
\item[40] J.V. Allaby et al., Preprint CERN 70-12 (1970)
\item[41] H. Kichimi et al., Lett. Nuovo Cim., {\bf 24}, 129 (1979);
          Phys. Rev. {\bf D20}, 37 (1979). \\
          A. Suzuki et al., Lett. Nuovo Cim., {\bf 24}, 449 (1979)
\item[42] C. Cochet et al., Nucl. Phys. {\bf B155}, 333 (1979)
\item[43] V.M. Shekhter and L.M. Shcheglova, Yad. Fiz. {\bf 27} 1070
          (1978); Sov. J. Nul. Phys. {\bf 27} 567 (1978)
\item[44] V.V. Anisovich, M.N. Kobrinsky, J. Nyiri, Phys. Lett. {\bf B102}
          357 (1981)
\item[45] M.A. Voloshin, Yu.P. Nikitin and P.I. Porfirov, Sov. J. Nucl.
          Phys. {\bf 35} 586 (1982)
\item[46] Yi-Jin Pei, Z. Phys. {\bf C72} 39 (1996)
\item[47] P.V. Chliapnikov, CERN-EP/99-142 (1999); Phys. Lett. {\bf B462}
          341 (1999)
\item[48] Particle Data Group, C. Caso et al., Eur. Phys. J. {\bf C3} 1
          (1998)
\item[49] C.A. Baker, C.J. Batty, P. Bl\"{u}m et al., Phys. Lett. {\bf B449}
          114 (1999)
\item[50] A.V. Anisovich, C.A. Baker, C.J. Batty et al., Phys. Lett.
          {\bf B452} 187 (1999); {\it{ibid,}} {\bf B452} 173 (1999);
          {\it{ibid,}} {\bf B452} 180 (1999); Nucl. Phys. {\bf A651}
          253 (1999)
\item[51] A.V. Anisovich, V.V. Anisovich, Yu.D. Prokoshkin and
          A.V.Sarantsev, Z. Phys {\bf A357} 123 (1997) ; \\
          A.V. Anisovich, V.V. Anisovich and A.V.Sarantsev, Z. Phys
          {\bf A359} 173 (1997); {\it{ibid,}} Phys. Lett. {\bf B395} 123
          (1997)
\item[52] V.V. Anisovich, D.V. Bugg and A.V.Sarantsev, Phys. Rev.
          {\bf D58}: 111503 (1998)
\item[53] L3 Collaboration, M. Acciari et al., Phys. Lett. {\bf B393}
          465 (1997); {\it{ibid,}} {\bf B407} 389 (1997)
\item[54] DELPHI Collaboration, P. Abreu et al., Phys. Lett. {\bf B475}
          429 (2000); {\it{ibid,}} {\bf B449} 364 (1999)
\item[55] OPAL Collaboration, K. Ackerstaff et al., Eur. Phys. J. {\bf C4}
          19 (1998); G. Alexander et al., Z. Phys. {\bf C73} 569, 587
          (1997)
\item[56] E.A. Kuraev, L.N. Lipatov and V.S. Fadin, Sov. Phys. JETP {\bf
          44} 443 (1976); Ya.Ya. Balitsky and L.N. Lipatov, Sov. J. Nucl.
          Phys. {\bf 28} 822 (1978)
\item[57] L.N. Lipatov, Sov. Phys. JETP {\bf 63} 904 (1986)
\item[58] V.V. Anisovich, D.I. Melikhov and V.A. Nikonov, Phys. Rev.
          {\bf D52} 5295 (1995); {\bf D55} 2918 (1997)
\item[59] V.V. Anisovich, D.I. Melikhov, B.Ch. Metsch and H.R. Petry,
          Nucl. Phys. {\bf A563} 549 (1993)
\item[60] T. Barnes, in Hadron Spectroscopy, AIP Conference Proceedings,
          Woodbury, New York (1998)
\item[61] V.V. Anisovich, M.G. Huber, M.N. Kobrinsky and B.Ch. Metsch,
          Phys. Rev. {\bf D42} 3045 (1990); V.V. Anisovich and B.Ch.
          Metsch, Phys. Rev. {\bf D46} 3195 (1992)
\item[62] ALEPH Collaboration (D. Busculic at al.), Z. Phys. {\bf C69}
          393 (1996)
\item[63] DELPHI Collaboration (P. Abreu et al.), Z. Phys. {\bf C68}
          353 (1995)
\item[64] L3 Collaboration, M. Acciari et al., Phys. Lett. {\bf B345}
          353 (1995);
\item[65] OPAL Collaboration, K. Ackerstaff et al., Z. Phys. {\bf C74}
          413 (1997)
\item[66] ALEPH Collaboration (R. Barate at al.), Preprint CERN/EP/99-94
\item[67] DELPHI Collaboration (P. Abreu et al.), Preprint CERN/EP/99-66
\item[68] OPAL Collaboration, K. Ackerstaff et al., Eur. Phys. J. {\bf C5}
          1 (1998)
\item[69] D.B. Leinweber et al., Phys. Rev. {\bf D58} 031501 (1998)
\item[70] K.M Watson, Phys. Rev. {\bf 88} 1163 (1952); A.B. Migdal,
          ZETF {\bf 28} 10 (1955)
\item[71] M.A. Voloshin, Yu.P. Nikitin and P.I. Porfirov, Sov. J. Nucl.
          Phys. {\bf 35} 586 (1982)

\item[72] K. Peters and E. Klempt, Phys. Lett. {\bf B352} 467 (1995)
\item[73] A.V. Anisovich, V.V. Anisovich and A.V. Sarantsev,
          e-Print Archive hep-ph/0003113, Phys. Rev. {\bf D62} 051502 (R)
\end{description}

\end{document}